\RequirePackage{fix-cm}
\documentclass[smallextended]{svjour3}  

\smartqed  
\usepackage[utf8]{inputenc}
\usepackage[T1]{fontenc}
\usepackage{graphicx}
\usepackage{amsmath, bm} 
\usepackage[bookmarks=true]{hyperref}
\usepackage{caption}
\usepackage{fancyhdr}
\usepackage{emptypage}
\usepackage{xcolor}
\usepackage[notlof,notlot,nottoc]{tocbibind}
\usepackage{natbib}

\usepackage{color}
\definecolor{red}{RGB}{255, 0, 0}
\definecolor{myyellow}{RGB}{255, 210, 0}
\definecolor{mygreen}{RGB}{0, 150, 46}
\definecolor{myblue}{RGB}{0, 216, 255}
\definecolor{mymagenta}{RGB}{255, 0, 255}
\definecolor{azzurro}{RGB}{20,233,240}

\usepackage{lipsum}

\newcommand{\apj}{The Astrophysical Journal}
\newcommand{\apjl}{The Astrophysical Journal Letters}
\newcommand{\apjs}{The Astrophysical Journal Supplement}
\newcommand{\nat}{Nature}
\newcommand{\ssr}{Space Science Reviews}
\newcommand{\jgr}{Journal of Geophysical Research}
\newcommand{\prl}{Physical Review Letters}

\newcommand{\solphys}{Solar Physics}
\newcommand{\grl}{Geophysical Research Letters}
\newcommand{\aap}{Astronomy and Astrophysics}

\newcommand{\planss}{Planetary and Space Science}
\newcommand{\apss}{Astrophysics and Space Science}
\newcommand{\mnras}{Monthly Notices of the Royal Astronomical Society}
\newcommand{\cpam}{Communications on Pure and Applied Mathematics}
\newcommand{\physrep}{Physics Reports}

\newcommand{\de}{\partial}
\newcommand{\divg}{\bm{\nabla} \cdot}  

\newcommand{\grad}{\bm{\nabla}}
\newcommand{\g}{\textbf}

\newcommand{\quotes}[1]{``#1''}

\newcommand{\BB}{\g{B}}
\newcommand{\EE}{\g{E}}
\newcommand{\bb}{\g{b}}
\newcommand{\VV}{\g{V}}
\newcommand{\UU}{\g{U}}

\newcommand{\xx}{\g{x}}

\newcommand{\fig}{Figure~\ref}
\newcommand{\eq}[1]{Eq.~\eqref{#1}}

\journalname{Space Science Reviews}

\begin{document}

\title{Current sheets, plasmoids and flux ropes in the heliosphere
}
\subtitle{Part II: Theoretical aspects.}

\author{O. Pezzi \and
        F. Pecora \and 
        J. le Roux \and
        N.~E. Engelbrecht \and
        A. Greco \and
        S. Servidio \and
        H.~V. Malova \and
        O.~V. Khabarova \and
        O. Malandraki \and
        R. Bruno \and
        W.~H. Matthaeus \and
        G. Li \and
        L.~M. Zelenyi \and
        R.~A. Kislov \and
        V.~N. Obridko \and
        V.~D Kuznetsov
}

\authorrunning{O. Pezzi et al.} 

\institute{O. Pezzi \at
    Gran Sasso Science Institute (GSSI), Viale F. Crispi 7, 67100 L’Aquila, Italy \\
    INFN, Laboratori Nazionali del Gran Sasso (LNGS), I-67100 Assergi, L'Aquila, Italy \\
    Istituto per la Scienza e Tecnologia dei Plasmi, CNR, Via Amendola 122/D, I-70126 Bari, Italy \\
    \email{oreste.pezzi@gssi.it}           
    \and
    A. Greco, F. Pecora, S. Servidio \at 
    Dipartimento di Fisica, Universit\`a della Calabria, I-87036 Rende (CS), Italy
    \and
    J. le Roux, G. Li \at
    Center for Space Plasma and Aeronomic Research (CSPAR) and Department of Space Science, University of Alabama in Huntsville, Huntsville, AL 35805, USA
    \and 
    N.~E. Engelbrecht \at
    Centre for Space Research, North-West University, Potchefstroom, 2522, South Africa 
    \and
    R. Bruno \at
    Istituto di Astrofisica e Planetologia Spaziali, Istituto Nazionale di Astrofisica (IAPS-INAF), Roma, Italy
    \and
    V.~D. Kuznetsov, V.~N. Obridko \at
    Pushkov Institute of Terrestrial Magnetism, Ionosphere and Radio Wave Propagation of the Russian Academy of Sciences (IZMIRAN), Moscow, 108840 Russia 
    \and
    O.~V. Khabarova, R.~A. Kislov \at
    Pushkov Institute of Terrestrial Magnetism, Ionosphere and Radio Wave Propagation of the Russian Academy of Sciences (IZMIRAN), Moscow, 108840 Russia \\
    Space Research Institute (IKI) RAS, Moscow, 117997 Russia
    \and
    H.~V. Malova \at
    Space Research Institute (IKI) RAS, Moscow, 117997 Russia \\
    Scobeltsyn Nuclear Physics Institute of Lomonosov Moscow State University, Moscow, 119991 Russia
    \and
    O. Malandraki \at
    IAASARS, National Observatory of Athens, Penteli, Greece 
    \and
    W.~H. Matthaeus \at
    Department of Physics and Astronomy, University of Delaware, Newark, DE 19716, USA
    \and 
    L.~M. Zelenyi \at
    Space Research Institute (IKI) RAS, Moscow, 117997 Russia
}

\date{Received: date / Accepted: date}

\maketitle

\begin{abstract}
Our understanding of processes occurring in the heliosphere historically began with reduced dimensionality - one-dimensional (1D) and  two-dimensional (2D) sketches and models, which aimed to illustrate views on large-scale structures in the solar wind. However, any reduced 
dimensionality vision of the heliosphere limits the possible interpretations of {\it in-situ} observations. Accounting for non-planar structures, e.g. current sheets, magnetic islands, flux ropes as well as plasma bubbles, is decisive to shed the light on a variety of phenomena, such as particle acceleration and energy dissipation. In part I of this review, we have described in detail the ubiquitous and multi-scale observations of these magnetic structures in the solar wind and their significance for the acceleration of charged particles. Here, in part II, 
we elucidate existing theoretical paradigms of the structure of the solar wind and the interplanetary magnetic field, with particular attention to the fine structure and stability of current sheets. Differences in 2D and 3D views of processes associated with current sheets, magnetic islands and flux ropes are discussed. We finally review the results of numerical simulations and {\it in-situ} observations, pointing out the complex nature of magnetic reconnection and particle acceleration in a strongly turbulent environment.


particle acceleration

\end{abstract}

\section{Introduction}
\label{sect:intro}
The heliosphere is a highly structured medium, characterized by the presence of a variety of plasma structures observed over a wide range of scales, from the energy-containing, large scales, to kinetic \citep{malandraki2019current}. As described in Part I of this review, understanding physical processes related to the variety of large-scale plasma structures observed in the solar wind and the magnetosphere implies analyzing their fine structure. The latter, in turn, requires a comprehensive analysis of properties of current sheets (CSs), flux ropes (FRs), and plasmoids. The dipolar nature of the main solar magnetic field and, consequently, of the interplanetary magnetic field (IMF) leads to the formation of the heliospheric current sheet (HCS), which is the largest CS in the heliosphere. Its configuration has historically been modelled as a waved ``ballerina skirt'' that follows the solar magnetic equator and the Parker IMF spiral shape \citep{wilcox1980origin, hoeksema1983structure}. Similarly strong but less long-lived CSs are formed in the solar wind at different helio-latitudes owing to the presence of higher harmonics of the solar magnetic field (see Part I, Section 2.1.2). Coronal Mass Ejections (CMEs), often associated with explosive phenomena triggered by magnetic reconnection, as well as their interplanetary counterpart, ICMEs, significantly perturb the heliosphere and its magnetic field, producing shock waves, dubbed interplanetary shocks (ISs) at which strong CSs may occur. ISs can also be produced by long-lived Corotating Interaction Regions (CIRs) or less stable Stream Interaction Regions (SIRs) formed when a fast flow from a coronal hole overtakes the surrounding slow solar wind \citep{heber1999corotating}. Notably, large-scale structures, such as SIRs, ICMEs and the HCS, coexist with structures of much smaller scales. For example, reconnecting CSs and FRs or plasmoids -- whose two-dimensional (2D) counterparts are magnetic islands (MIs)-- are often observed not only at the edges but also within complex ICMEs and SIRs/CIRs (see \citet{xu2011observations,khabarova2016small,khabarova2017energetic} and Part I, Section 2.1.2). Furthermore, the HCS is often rippled and surrounded by the much wider heliospheric plasma sheet (HPS) in which numerous secondary reconnection regions produce a sea of thin CSs (TCSs), FRs as well as plasma bubbles/blobs/plasmoids (see \citet{adhikari2019role} and Part I, Section 2.1). FRs or plasmoids have been also found in the inner heliosphere \citep{zhao2020identification}, in the Earth's magnetosheath at kinetic scales \citep{yao2020kinetic}, and in laboratory plasmas \citep{gekelman2019spiky}. This ensemble of structures have a significant impact on the topology of the surrounding IMF, which -in turn-- regulates the propagation of charged particles of both heliospheric and galactic origin (see  \citet{d13,battarbee2017solar,bat18,Eea19} and Part I, Section 3).

Another keystone of {\it in-situ} measurements regards the pervasive significance of both plasma turbulence and magnetic reconnection in governing small-scale processes that occur in space plasmas (e.g. \citet{matthaeus2011needs,Zank2014particle,matthaeus2015intermittency,servidio2015kinetic,LeRoux2015kinetic}). In Part I, we discussed the fact that magnetic reconnection and turbulence are always linked. An analogous argument holds for CSs, the occurrence of which always suggests the formation of FRs/plasmoids/magnetic islands and vice versa. The solar wind, which embeds these magnetic and plasma structures, is a strongly turbulent and intermittent medium, characterized by a complex interplay of different phenomena \citep{servidio2009magnetic,matthaeus2015intermittency, bruno2016turbulence}. The energy of the magnetic field and bulk speed fluctuations injected at large scales is cross-scale transferred towards smaller scales, at which Hall and kinetic effects can be significant \citep{ServidioEA07, servidio2015kinetic}. Spectral steepening \citep{Leamon98,alexandrova2008small,sahraoui2009evidence} and dispersive wave effects are routinely observed in the solar wind. These latter are compatible with either strongly turbulent fluctuations not described in terms of linear
modes \citep{alexandrova2008small} or kinetic Alfvén waves (KAWs) \citep{howes2008model, sahraoui2009evidence, salem2012identification} and whistler waves \citep{beinroth1981properties, gary2010whistler, vasko2020quasi}, although whistler modes possess a smaller power content with respect to the KAW branch \citep{chen2013nature}.

Signatures of kinetic effects are often found in the particle velocity distribution function (VDFs) that exhibit non-Maxwellian properties, e.g. temperature anisotropy, heat fluxes, beams and rings \citep{marsch2006kinetic,maruca2011relative,servidio2012local,servidio2015kinetic,valentini2016differential,perri2020deviation}. These non-equilibrium features may drive the onset of microinstabilities \citep{HellingerEA06,MatteiniEA13,bandyopadhyay2020interplay}, although understanding the significance of linear instabilities within a turbulent environment is still under debate \citep{qudsi2020intermittency}. The dissipation of turbulent energy is thought to locally occur at small scales \citep{osman2011evidence,osman2012intermittency,osman2012kinetic,matthaeus2015intermittency,vaivads2016turbulence}, thus ultimately heating the plasma. Fine velocity-space structures have also been recently observed in the magnetosheath \citep{servidio2017magnetospheric} and obtained in kinetic simulations performed within solar-wind-like conditions \citep{pezzi2018velocityspace,cerri2018dual} by means of the Hermite decomposition of the plasma VDF \citep{Grad49,tatsuno2009nonlinear,SchekochihinEA16}. This supports the idea that non-equilibrium features in particle VDF readjust in a very complex way, resembling the development of a enstrophy velocity-space turbulent spectrum \citep{servidio2017magnetospheric}. The presence of fine velocity-space structures may also enhance the dynamical role of inter-particle collisions \citep{pezzi2016collisional,pezzi2017solarwind,pezzi2019protonproton}.

Magnetic reconnection, during which a local breaking of the frozen-in law causes a rapid release of magnetic energy in both flow energy and heating, goes hand in hand with plasma turbulence \citep{lazarian1999reconnection,kowal2011mhd,LazarianEA15}. Indeed, CSs that separate vortices and magnetic islands in plasma turbulence often represent reconnection sites \citep{retino2007insitu,servidio2009magnetic,servidio2010statistics,haggerty2017exploring, phan2018electron}. Furthermore, reconnection exhausts and jets generated by magnetic reconnection can be turbulent themselves \citep{franci2017magnetic,pucci2017properties, pucci2018generation} and can also host secondary reconnection regions driven by nonlinear waves and/or instabilities \citep{lapenta2015secondary,lapenta2018nonlinear,wang2020direct}. Very recent observations conducted at an unprecedented high-resolution by the Magnetospheric Multi-Scale (MMS) mission \citep{burch2016magnetospheric, fuselier2016magnetospheric} allowed, for the first time, to investigate the electron-scale magnetic reconnection by explicitly showing the electron diffusion region and by addressing the relevant question about the physical mechanism that breaks the frozen-in law in a collisionless plasma \citep{burch2016electron, lecontel2016whistler, torbert2016estimates, torbert2018electronscale, breuillard2018new, chasapis2018insitu}. 

Magnetic reconnection is also a decisive phenomenon in astrophysical plasmas owing to its role in particle acceleration \citep{lyutikov2003explosive,uzdensky2011magnetic}. In the planetary magnetospheres and the solar wind, signatures of particle acceleration are observed in the particle energy distribution, that often shows non-Maxwellian higher energy tails described in terms of kappa  \citep{vasyliunas1968low, sarris1976location, christon1989spectral, collier1999evolution} or power-law distributions. Although acceleration mechanisms in the solar wind and magnetospheres have common features, they also present distinctive traits due to the different physical conditions. For example, the maximum particle energies that can be obtained in the Earth's magnetosphere do not exceed a few MeV \citep{zelenyi2007universal} since accelerated particles with gyroradii approximately equal to the magnetospheric scales leak out to the solar wind. In smaller-scale magnetospheres, the maximum acceleration energies are much smaller. For example, in Mercury's magnetosphere, due to repeating substorms and magnetotail dipolarization events, particle energies are estimated to be about $100 {\rm keV}$. In the corona and the solar wind, even single mechanisms act on scales many orders larger, and the mechanisms in combination can contribute to the particle energy gain up to GeV. At the same time, one should note that magnetic reconnection, wave propagation and turbulent processes widely present in the solar wind are limited by MeV energies (see Part I, Section 3).

Historically, seminal works by \citet{fermi1949origin, fermi1954galactic} proposed two general acceleration mechanisms based either on the stochastic interaction of particles with randomly moving magnetic clouds/inhomogeneities (second-order Fermi acceleration) or on the systematic acceleration of particles in the case of converging magnetic traps (first-order Fermi acceleration). Several acceleration mechanisms in space have been proposed and are found to be in accordance with {\it in-situ} observations. For example, the large-scale dawn-dusk electric field can accelerate quasi-adiabatic ions during the so-called Speiser acceleration \citep{speiser1965particle,lyons1982evidence,cowley1983current,ashour1993shaping,zelenyi2007universal}. Plasma particles can also be energized by Alfv\'en and cyclotron waves in the proximity of the magnetopause \citep{drake1994turbulence,johnson2001stochastic,panov2008high} and heated by ultralow frequency (ULF) waves \citep{glassmeier2003concerning,baumjohann2006magnetosphere}. In presence of a shock (e.g. magnetospheric bow shocks or ICME/CIR/SIR -- driven interplanetary shocks) \citep{thampi2019acceleration,slavin2008mercury}, nonadiabatic heating, diffusive shock acceleration (DSA), shock drift acceleration (SDA) and also stochastic shock drift acceleration (SSDA) (see \citet{katou2019theory} for a review on SDA) can occur. At variance with SDA, both DSA and SSDA assume the presence of fluctuations (e.g. pre-existing turbulence) on which particles diffuse in order to be accelerated. DSA, first introduced to address cosmic-ray acceleration at supernova remnants \citep{bell1978accelerationA,bell1978accelerationB,blandford1978particle}, has also been invoked to explain acceleration in solar wind \citep{zank2000particle}. However, it is not easy to interpret some {\it in-situ} observations through the original steady-state DSA model (see \citet{malandraki2019current} for a further discussion). At the same time, SDA and SSDA are efficient in the energization of electrons at planetary bow shocks \citep{leroy1984theory,burgess1987shock,giacalone1992shock,katou2019theory,amano2020observational}. 

Other mechanisms of acceleration are directly related to magnetic reconnection. Both reconnection events of solar origin \citep{higginson2018structured} and local magnetic reconnection \citep{khabarova2015small,khabarova2016small,khabarova2020counterstreaming,adhikari2019role,malandraki2019current} can lead to particle acceleration. Near the current-sheet X-line, the electric field produced by reconnection can accelerate charged particles \citep{matthaeus1986turbulent, Ambrosiano88test}. Moreover, \citet{Hoshino05} found that electrons, trapped by the polarization (Hall) electric field produced by reconnection, are efficiently accelerated owing to a surfing acceleration mechanism. Current sheets are often unstable due to tearing instability \citep{zelenyi1984dynamics,malara1992competition,malara1996parametric,tenerani2015tearing,primavera2019parametric}. Electric fields induced by i) the tearing instability, ii) the CS destruction \citep{lutsenko2005source}, iii) dipolarization processes in the terrestrial magnetosphere \citep{zelenyi1990generation, delcourt1994plasma, delcourt2002particle, slavin2004mercury, delcourt2005electron, apatenkov2007multi, zhou2010accelerated,  ukhorskiy2017ion, parkhomenko2019acceleration}, or iv) other fluctuations occurring in the turbulent environment with the presence of magnetic reconnection \citep{malara2019electron} are also able to accelerate charged particles. For example, in the magnetosphere, a power-law tailed energy distribution of energetic particles has been obtained by considering the model of the CS with electromagnetic fluctuations excited by electromagnetic wave ensembles. The occurrence of electromagnetic fluctuations close to the reconnecting CSs also allows explaining additional particle acceleration (additional to the acceleration due to the large-scale electrostatic $E_y$) often observed in velocity distributions at open field lines \citep{grigorenko2013current}. Temperature and density inhomogeneities across the reconnecting CS may also speed up the formation of secondary plasmoids, as has recently been shown using test-particle \citep{catapano2015current, catapano2016proton, catapano2017charge} and PIC simulations \citep{karimabadi2013coherent,lu2019turbulence} as well as spacecraft observations \citep{lu2019effects}.

Furthermore, numerical simulations show that particles can be efficiently energized due to the contraction of magnetic islands (MIs) \citep{drake2006electron} and their merging \citep{Drake2013power}. The combined role of these different mechanisms and their relative importance has been described by \citet{Zank2014particle} via analytical solutions, finding that MIs contraction and merging are the dominant acceleration mechanisms. Further models discussing the effect of multiple FRs and the combined role of magnetic reconnection-related acceleration mechanisms and DSA have been developed by \citet{LeRoux2015kinetic} and \citet{ZankEA15,leRoux2016combining}, respectively. \citet{leRoux2018self} self-consistently coupled energetic particles, for which the transport equation is solved, and MHD turbulence that controls the FRs dynamics. The role of small-scale FRs has been recently considered by \citet{leroux2019modeling,mingalev2019modeling}. Magnetic reconnection is thus an efficient mechanism able to accelerate particles locally, at least at levels comparable with DSA \citep{garrel2018diffusive}. An analysis of spacecraft observations also supports these findings. Local particle acceleration at the HCS has been reported, e.g., by  \citet{zharkova2012particle, zharkova2015additional}. \citet{khabarova2015small, khabarova2016small,khabarova2020counterstreaming,khabarova2017energetic,adhikari2019role,malandraki2019current} found observational evidence for local particle acceleration associated with the occurrence of reconnecting CSs and MIs. Recurrent (stochastic or turbulent) magnetic reconnection occurring at the HCS, MIs inside the rippled HCS and smaller CSs within the HPS presumably energizes charged particles via the mechanisms proposed by \citep{Zank2014particle,Zank2015particle,LeRoux2015kinetic,leRoux2016combining,leroux2019modeling}. Atypical energetic particle events (AEPEs), not easily explained in terms of the DSA mechanism, are instead well described by local particle acceleration in regions filled with small-scale MIs of width $<\sim 0.01$ AU \citep{khabarova2016small, khabarova2017energetic}. Prior findings regarding locally accelerated ions have recently been systematized by \citet{adhikari2019role, chen2019analysis}. A local origin of some suprathermal electrons observed in the solar wind, as reflected in the specific features of pitch-angle electron distributions, has also been suggested by \citet{khabarova2020counterstreaming}. 
Explosive particle acceleration \citep{Pecora18JPP,Pecora19SolPhys} that combines several types of acceleration mechanisms mentioned above can finally take place, e.g. in the solar corona. 

It is important to note that different mechanisms can co-act simultaneously in space plasmas. This can be clearly seen in observations (e.g., \citet{khabarova2016small, adhikari2019role} and Part I, Section 3.1, 3.2). However, this effect may considerably complicate the interpretation of the observed picture \citep{malandraki2019current}. Note that, in presence of turbulent fluctuations, stochastic acceleration of charged particles (of both Fermi types) \citep{milovanov2001strange,trotta2020fast,sioulas2020stochastic} and, in general, various acceleration mechanisms mentioned above can operate \citep{artemyev2009acceleration,kobak2000energetic,parkhomenko2019acceleration,ergun2020observations,ergun2020particle}. We also remark here that \citet{milovanov2001strange} generalized second order Fermi acceleration for systems with long-range correlations of spatial turbulence structures \citep{comisso2018,comisso2019interplay}. This process can be considered as a transport process in the velocity space. In this perspective, the ``random movement'' term is usually described by the Gaussian variance $\langle V^2(t)\rangle \sim t$ for velocities of scattered particles $V(t)$. Meanwhile, the Gaussian variance in principle ignores long-range dynamical correlations occurring in turbulent self-organized systems. The effect of correlations appears in multi-scale nonrandom acceleration events that do not comply with standard velocity diffusion. A fractional-dynamic approach solves the issue of describing stochastic acceleration in the presence of the long-range correlations \citep{milovanov1994development, metzler2000random, metzler2004restaurant, zimbardo2017fractional}. Fractional generalization of Einstein’s Brownian motion and the subsequent fractional kinetic equations are believed to provide a powerful framework that can be applied to many physical systems \citep{perri2007evidence,perri2012superdiffusive,zimbardo2013levy,zimbardo2015superdiffusive,zimbardo2017superdiffusive}. The fractional equation allows incorporating spatial and temporal effects of long-range correlations and is capable of describing both particle super-diffusion (i.e., persistent random walks), and sub-diffusion (anti-persistent random walks) \citep{zimbardo2006superdiffusive}, while the well-known Fermi acceleration dynamics scales as $V(t)\sim t^{1/3}$ in the true random classical case. 

These observations ultimately confirm the idea that heliospheric structures, such as CSs, FRs and plasmoids, have a complex topology and, in general, are far from being planar or spherically-symmetric. Due to both non-stationarity and non-uniformity of the solar wind, acceleration processes can take place everywhere where the electric field is present and the scales of magnetic inhomogeneities are of the order of the proton or electron gyroradii. Plasma turbulence complexly links stream propagation, magnetic reconnection and particle energization in a puzzling multi-scale and multi-process scenario. Owing to strong nonlinearities, numerical simulations are important to properly interpret {\it in-situ} observations. 

Obviously, fully three-dimensional (3D) models retain the complete physical picture of the described phenomena. Indeed, small-scale phenomena (e.g., secondary magnetic reconnection, waves/instabilities and ripples of the HCS) are hardly described within a 2D approach. However, 2D or quasi-2D models have been adopted owing to their capability to properly paint the picture of several physical phenomena, at least at a qualitative level. For example, the basic physics of magnetic reconnection is well depicted employing 2D models \citep{matthaeus1986turbulent, Ambrosiano88test}. On the other hand, \citet{kowal2011mhd} stressed the importance of considering the fully 3D space and the presence of a guide field. Moreover, \citet{kowal2012particle} pointed out that turbulent fluctuations superimposed to the reconnecting CSs can enhance the acceleration rate since magnetic reconnection becomes fast and a thick volume filled with several multiple reconnecting regions is formed, thus leading to a first-order acceleration mechanism. Other advanced studies have been focused on kinetic simulations, in which it is possible to describe the non-ideal mechanism leading to magnetic reconnection. Several studies, often focusing on the relativistic regime, confirmed that the first-order Fermi mechanism related to reconnection can accelerate particles to energies described by a power-law distribution \citep{guo2015particle,guo2016efficient,li2019particle,xia2018particle,xia2020particle}. The fully 3D space guarantees the interplay of several wave modes and instabilities, leading to turbulence \citep{huang2017development, lapenta2015secondary, lapenta2018nonlinear}. The contribution of several drifts has also been analyzed by \citet{li2017particle}, finding that the major energization is due to the particle curvature drift along the induced electric field. The particle energization created by FR merging has recently been investigated by \citet{du2018plasma}. 

It is important to emphasize that even while 3D models include  
distinctive effects not included in 2D, some very important effects occur in both geometries even if details may differ. 
For example, multiple secondary island formation occurs even in 2D incompressible MHD \citep{matthaeus1986turbulent,WanEA13-xpts}. 
Turbulence also elevates reconnection rates in 2D and 2.5D MHD and Hall MHD 
\citep{Matthaeus85,matthaeus1986turbulent,smith2004hall} as well 
as in the more realistic 3D geometry as discussed above. 
While it is almost always preferable to have a 3D picture of natural 
phenomena, often the 2D models are preferred because they can achieve larger systems sizes and higher Reynolds numbers, these also being important and even crucial parameters in describing observations of natural processes.

We finally remark that the description of solar-wind turbulence has faced similar debates about 2D vs 3D models. The strong wavevector anisotropy of MHD turbulence with respect to the guide field \citep{Shebalin83}, also supported by spacecraft observations revealing the ``Maltese Cross'' pattern in the two-dimensional correlation function measurements of solar wind fluctuations \citep{matthaeus90}, motivated the development of a two-component model, which is composed by fluctuations with wavevector parallel to the ambient magnetic field (slab) and by fluctuations with wavevector quasi-perpendicular to the ambient magnetic field (2D) \citep{b96,GhoshEA98} (see \citep{oughton2017reduced,oughton2020critical} for recent reviews on MHD turbulence in the solar wind). Moreover, numerical simulations performed in the kinetic regime showed that a 2.5D approach describes well solar-wind dynamics in the inertial range and at sub-proton scales \citep{servidio2014proton, servidio2015kinetic, wan2015intermittent, li2016energy, franci2018solarwind}, although it has been claimed that a fully 3D approach should be retained to properly model nonlinear couplings at large-scale in the incompressible limit \citep{howes2015inherently}. A contrasting, and even puzzling,
result in the realm of kinetic 
PIC simulations is that averages of the electromagnetic work conditioned on local values of the current density (a rough equivalent of an Ohm's law) 
behaves very similarly in 2.5D and in 3D,
and, perhaps surprisingly,  
both behave very much like collisional MHD 
\citep{WanEA16}. 

This part of the review follows Part I \citep{review1}, which, mainly focuses on providing the general information about the objects of this review and observations of non-planar magnetic structures and their association with particle acceleration. The structure of this part of the review is the following. In Sect. \ref{sect:olgaeug} we provide details on some models adopted in order to describe the dynamics of CSs and MIs, with particular emphasis on the formation of their fine structure. We also discuss their stability properties and overview theoretical and numerical models aimed at analyzing the impact of CSs, FRs and MIs on the transport of energetic particles, in particular on galactic cosmic rays in the heliosphere (HCS). Then, in Sect. \ref{sect:leroux} we speculate about the theoretical approaches usually employed to describe particle acceleration within contracting/merging FRs. Sect. \ref{sect:unicalturb} is dedicated to the description of recent advances in numerical studies of the evolution of plasma turbulence and, in particular, the appearance of numerous discontinuities representing potential reconnection and heating sites. The discussion of the results covers both high-resolution fluid (MHD and Hall MHD) and kinetic simulations. Particle acceleration in such complex environments is described in Sect.\ref{sect:unicalaccel}. Finally, we summarize the discussed results and make conclusions about the past and future development of studies of CRs, FRs/plasmoids and MIs in Sect. \ref{sect:concl}.

\section{Current sheets: their fine structure, stability and the ability to accelerate particles}
\label{sect:olgaeug}

As introduced above, FRs, blobs/plasmoids/bubbles, CSs and MIs are ubiquitous features in the dynamics of astrophysical plasmas. In this Section, we revisit the basic dynamics of current sheets by highlighting, in particular, the formation of their fine structure and their stability properties. 

One can roughly classify planar and conic current sheets as either ``thick'' or ``thin'' structures based on the ratio of their transverse scale to the proton gyroradius. In the thick CSs, electric currents are supported by plasma drift currents in circular magnetic field gradients, repeating the shapes of large-scale structures themselves. For CSs of this type, the presence of a neutral magnetic surface is not necessary and they are quite stable relative to the tearing mode. Meanwhile, crossings of thin CSs (TCSs) are characterized by the occurrence of the neutral line altogether with a strong jump of the magnetic field in their vicinity. These CSs seem to be a reservoir of free energy. Therefore, it is quite possible that these very common structures in space plasmas are responsible for the energy storage and release, magnetic reconnection, plasma acceleration and other important processes in planetary magnetospheres \citep{retino2007insitu,ergun2020observations}, in the solar wind \citep{zelenyi2010metastability,zelenyi2011thin,zelenyi2019current}, and the solar corona \citep{syrovatksi1971formation}.

As discussed in Part I of this review, current sheets also exhibit a fine structure. They are multi-layered, containing several TCSs embedded into the main CS of much wider thickness \citep{malova2017evidence}. Cascading or production of similar current sheets around the main one is a well-known process observed over a wide range of scales. In general, a conglomerate of current sheets formed around the mother current sheet and resembling the multi-scaled heliospheric plasma sheet (HPS) (see Part I) is observed. Below we describe the main physical mechanisms responsible for the formation of the fine structure of current sheets. By developing a quasi-adiabatic model, \citet{zelenyi2004nonlinear} demonstrated that CSs are multi-scaled and meta-stable structures that can be quite stable for a long time but then can spontaneously be destroyed by tearing instability. TCSs are also sensitive to the impact of a wide spectrum of kink and oblique waves, which leads to their instability in variable conditions. This process is followed by the explosive CS destruction and the release of the energy excess in a form of plasma acceleration and wave activity. Note that excited waves of different types can generally coexist also after the CS destruction. Electromagnetic turbulence can hence develop in the TCS-filled plasmas. As a result, plasma particles can be accelerated by plasma turbulence, and this mechanism is common for the magnetospheric and the solar wind plasmas. 

\subsection{Quasi-adiabatic model of thin current sheets in space plasmas}

It should be noted that the first analytical and mathematically-simple model of a thin one-dimensional (1D) current sheet was proposed by \citet{harris1962plasma}. In this model, only the tangential component of the magnetic field was taken into account. Further analysis of proton dynamics in the neutral plane with the reversed magnetic field that also considered a finite but small normal magnetic field was done by \citet{speiser1965particle}, who showed that in such CSs protons are demagnetized near the neutral plane and their quasi-adiabatic integral of transverse (along z coordinate) oscillations
\begin{equation}
    I_z = \frac{1}{2\pi} \oint p_z dz
\end{equation}
is approximately conserved.

A semi-analytical 1D model of the TCS in the collisionless space plasma was proposed by \citep{zelenyi2000thin,zelenyi2004nonlinear} based on the quasi-adiabatic description of ions with the plasma anisotropic pressure tensor and magnetized electrons in the fluid approximation. In this model, the general self-consistent system of Vlasov-Maxwell equations has the following simple form:
\begin{eqnarray}
\frac{d B_x}{dz} &&= \frac{4\pi}{c}\left(j_{i,y}(z)+ j_{e,y}(z)\right) \\
j_{i,y} &&= e \int v_y f_{i}(z,v) dv 
\end{eqnarray}
being $f_i$ the ion distribution function and $j_{i,y}$ the ion current density in the y-direction. The ion distribution function can be obtained at any location inside the current sheet by mapping the source distribution from the edges of the TCS to the neutral plane using Liouville's theorem; thus it can be re-written as a function of a quasi-adiabatic integral $I_z$. Electron current density $j_{e,y}$ can be calculated in the guiding center approach. The detailed description of the basic model and its modifications can be found in \citet{zelenyi2011thin} and references therein.

\begin{figure*}
\centering
\includegraphics[width=0.95\textwidth]{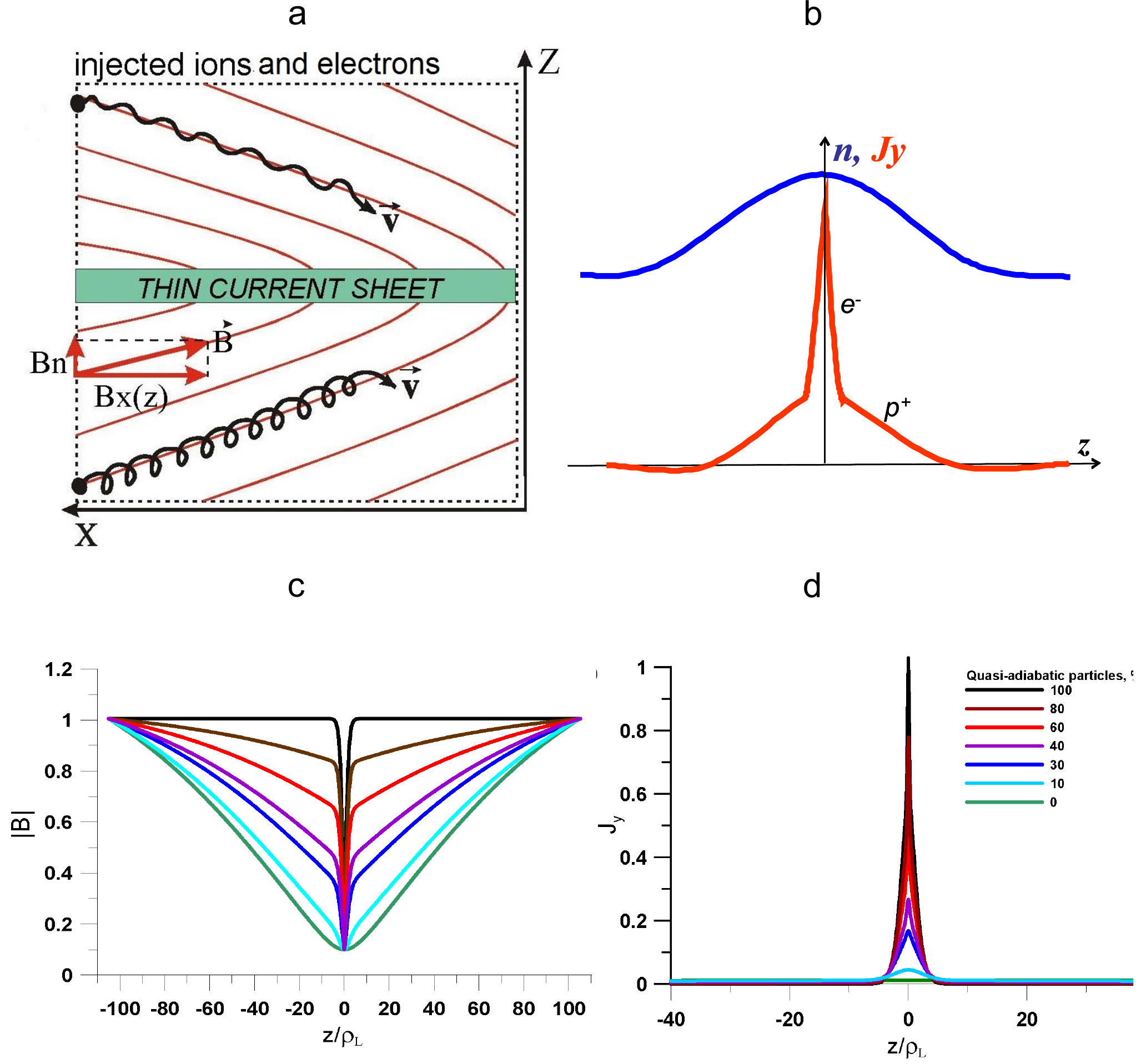}
\caption{(Color online) Multi-scaled structure of SCSs/HCS: (a) General scheme of the model (adapted from \citet{zelenyi2006matreshka}); (b) Schematic view of the embedding of current sheets (adapted from \citet{zelenyi2006matreshka}). Multiscale CSs are embedded in the thicker plasma sheet in which the plasma density tends to constant at the edges, i.e. $n(L)\rightarrow n_0$. Regions of differently-dominated plasma populations can be seen; (c) Self-consistent profiles $|{\bm B}(z/\rho_L)|$ (adapted from \citet{malova2017evidence}); (d) Current density profile $J_y=j_y/enV_{D1}$ within the current sheet (adapted from \citet{malova2017evidence}). Colored curves correspond to different densities of quasi-adiabatic particles (as a percentage of the total number of particles). Transverse coordinate $z$ showed in the abscissa is normalized to the proton gyroradius $\rho_L$, $|{\bm B}|$ and $n$ are normalized to their values at $100 z/\rho_L$.}
\label{fig:OLGAfig1}
\end{figure*}

Later, numerical models of TCS were adopted to verify the semi-analytical model \citep{zelenyi2004nonlinear}. Comparison of model results with experimental observations from the Cluster satellites in the Earth's magnetotail demonstrated a good agreement between them \citep{artemyev2009thin}. The application of the model to the case of the heliospheric current sheet (HCS) and similarly strong current sheets (SCSs) in the solar wind was made by \citet{malova2017evidence}. The general configuration of the model is shown in Fig. \ref{fig:OLGAfig1} (a). Particles move from current sheet edges toward the neutral plane. Within the CS plane, particles are demagnetized and move along serpentine-like orbits; afterwards, they go outward and carry the electric current across the current sheet. Since electrons are magnetized, their curvature electric current dominates in the plane of the neutral sheet. As a result, the electron-determined current flows within a thin current sheet is embedded in a thicker current sheet created by protons, as it is shown schematically in Fig. \ref{fig:OLGAfig1} (b). This is in good accordance with observations showing that TCSs are located within much thicker Harris-like background current sheets which can be called plasma sheets analogous to the HPS (see \citet{malova2018structure} and Figure 1 of Part I). 

Figure \ref{fig:OLGAfig1}(c) displays self-consistent solutions for the module of the tangential component of the magnetic field as a function of the $z$ coordinate. Here the parameter $n_r$ is the relative concentration of quasi-adiabatic particles; it varies from $1$ to $0$. The black curve corresponds to the case when all particles are quasi-adiabatic ($n_r=1$ or $100\%$) and the green curve corresponds to the opposite case of a pure Harris-like current sheet with the isotropic pressure distribution ($n_r=0$ or there is $0\%$ of quasi-adiabatic particles). From panels (b) and (d) of Fig. \ref{fig:OLGAfig1}, one can easily figure out that the background current sheet has a very smooth and wide profile. The corresponding green curve has a width of about $10^2$ proton gyro-radii, and the ordered thinner proton current peak is manifested in the central part of CS. The electron current is the narrowest, and at some parameters, it is seen in the region of the neutral plane [Fig. \ref{fig:OLGAfig1}(b)]. Panel (b) of Fig. \ref{fig:OLGAfig1} also shows that a very narrow $|\bf{B}|$ peak of the proton current becomes noticeable for $n_r\geq 0.3$ (see the case with $n_r = 0.3$, shown by the dark blue curve). 

One may conclude that in TCSs electrons may carry only a part of the azimuthal electric current due to curvature drifts, whereas the main part of the cross-sheet current is carried by demagnetized quasi-adiabatic protons \citep{zelenyi2011thin,malova2017evidence,malova2018structure}. As seen from Figs. \ref{fig:OLGAfig1} (b--c), the central part of the multi-scale current sheet is a very narrow structure with a thickness of several proton Larmor radii $\rho_L$, while the wider background part of the SCS with negligible current density has a thickness $L=100 \rho_L$. Therefore Fig. \ref{fig:OLGAfig1} (b--c) demonstrate the multiscale character of a thin current configuration embedded in a much thicker plasma sheet in which the plasma density tends to have an almost uniform distribution, while the current density vanishes in the same region. 

\subsection{Current sheet dynamics and formation of multiple current sheets in the solar wind}

The current sheet filamentation {\it via} tearing instability was proposed as a key factor of sub-storm activity in the Earth’s magneto-tail since the early papers (e.g., \citet{coppi1966dynamics}), in which the 1D Harris sheet equilibrium solution has been used for the analysis of current sheets stability. It further was shown that the existence of a finite but small normal component $B_z$ in the current sheet can lead to a strong stabilizing effect due to the electron compressibility (see, e.g. \citet{schindler1979theories} and references therein). Many other possible mechanisms of sub-storm triggering were proposed to solve this problem (see details in the review by \citet{zelenyi2010metastability}). In particular, the stochastic particles motion \citep{buchner1989regular,kuznetsova1991magnetic}, pitch angle scattering \citep{coroniti1980tearing}, transient electrons \citep{sitnov1997quasi,sitnov2002reconnection}, and current-driven instabilities \citep[and references therein]{lui1996current} were considered and analyzed in detail. 

Meanwhile, it was later shown that all these mechanisms cannot destroy the strong stabilization effect of electrons, at least in the Harris-like current-sheet case \citep{pellat1991does}.
Over time this paradoxical result had lead to the loss of interest of scientists in the tearing mode as a trigger of substorm onset \citep{zelenyi2011thin}. In two decades, experimental observations of the Earth’s magnetotail performed during the Cluster mission time made clear that magnetotail TCSs are principally different from well-known Harris-like configurations, therefore some additional factors may control the TCS stability \citep{runov2005electric,runov2006local,zelenyi2003splitting}.

\begin{figure*}
    \centering
    \includegraphics[width=0.95\textwidth]{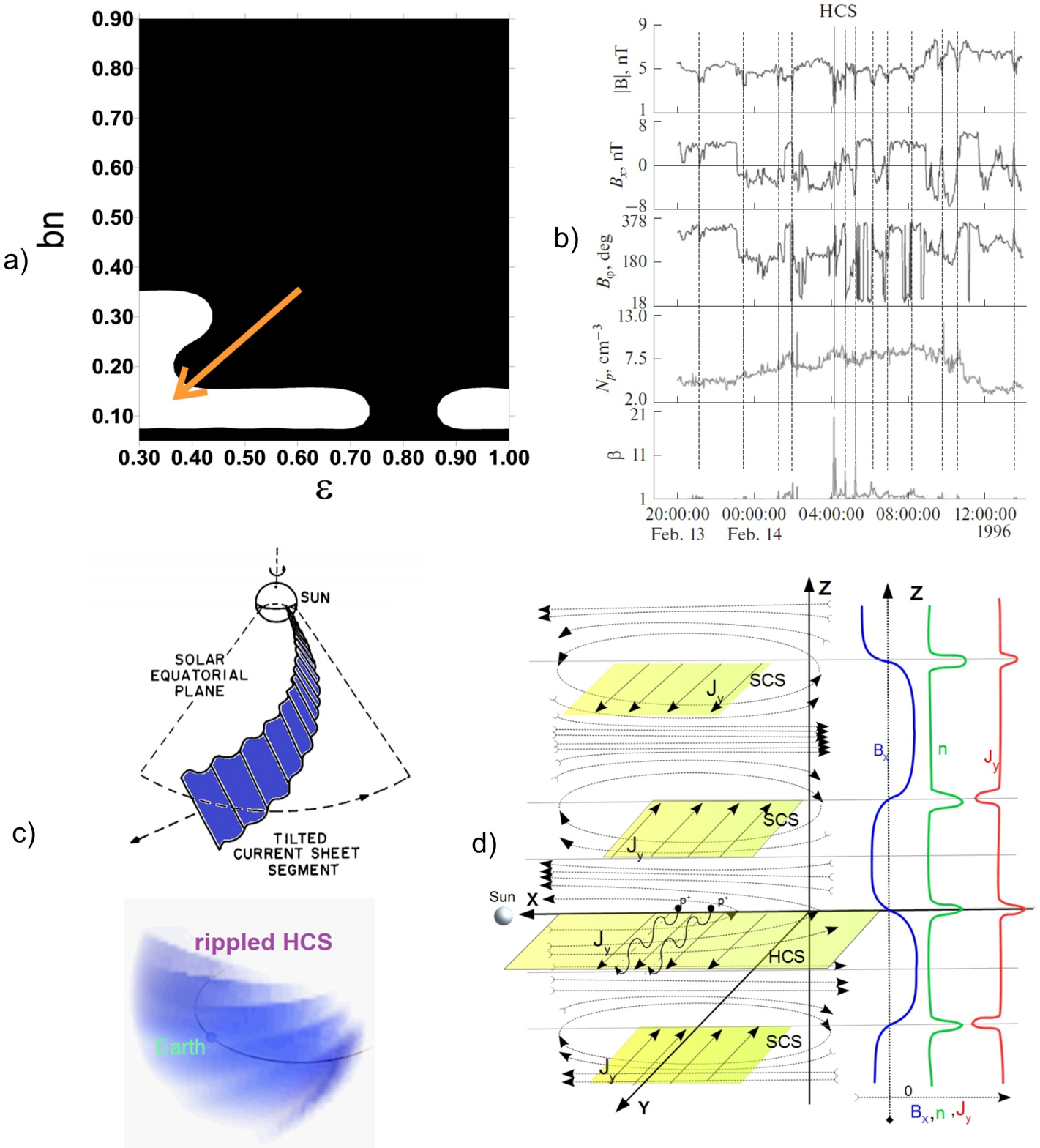}
    \caption{Current sheet dynamics and multi-sheet current configurations in the solar wind: observations and theoretical interpretations. a) Stable (black) and unstable (white) regions of a thin current sheet under the tearing mode, according to the model by \citet{zelenyi2008marginal} (adapted from \citet{zelenyi2008marginal}). Red arrow indicates a possible direction of the switch of the current sheet state from the stable to unstable. b) Example of observations of the prolonged HPS/HCS crossing at the Earth’s orbit (reproduced from \citet{malova2018structure}). From top to bottom: the magnitude of the IMF, the radial component of the IMF in the GSE coordinate system, the azimuthal angle of the IMF, the solar wind density, and the plasma beta. Multiple strong current sheets (indicated by vertical lines) are observed within the HPS. c) HCS ripples as predicted by \citet{behannon1981finescale} (upper panel) and observed in the solar wind via interplanetary scintillations (lower panel) (adapted from \citet{khabarova2015small}). d) Configuration of the magnetic field lines in the vicinity of the HCS (central yellow plane) surrounded by magnetic islands/blobs (left sketch), and corresponding reconstructed profiles of the magnetic field $B_x$, the electric current $j_y$ and the plasma density $n$.
    }
    \label{fig:OLGAfig2}
\end{figure*}

These results enabled a solution to the old problem, finally forming the paradigm that associates the substorm activity with the onset of tearing-type instability. The idea that magnetotail current sheets can be unstable only at some narrow regions in the parameter space was for the first time formulated by \citet{galeev1976tearing}. Further, \citet{zelenyi2008marginal} investigated this effect in details in the frame of a quasi-adiabatic model of TCS supported by satellite observations. It was concluded that the thinning of the magnetotail current sheet is followed by an increase of the anisotropy in the ion and electron distribution functions. Hence, finally, the magnetotail becomes a new metastable equilibrium with a TCS in the equatorial plane that can spontaneously be destroyed, being accompanied by the processes of fast magnetic reconnection. 

Both observational and theoretical studies allow suggesting that tearing instability and magnetic reconnection of thin current sheets can play a substantial role in the formation of a system of multiple current sheets in the solar wind (see, e.g., \citet{wang1998origin, reville2020slow} and references therein). More recently, the stability of tearing mode and the onset of fast magnetic reconnection have received a detailed attention \citep{tenerani2015tearing, tenerani2015magnetic, pucci2017fast, PucciEA18}.
	
As discussed in Part I, Section 2.3.1, crossings of sector boundaries are often associated with prolonged observations of multiple thin current sheets generally following the heliospheric plasma sheet (HPS) shape  (see also \citet{liu2014statistical}). These current sheets with a thickness of several proton gyroradii \citep{behannon1981finescale,khabarova2015small,xu2015angular} are often called strong current sheets (mentioned above in paragraph 2.1) (see \citet{malova2017evidence}). A possible theoretical interpretation of such events and a typical HPS/HCS crossing observed by a 1 AU spacecraft were discussed by \citet{malova2018structure}. The corresponding figures are reproduced here as Fig. \ref{fig:OLGAfig2} (a) and (b), respectively. 

Fig. \ref{fig:OLGAfig2} (a) shows stable and unstable regions in the parameter space, where $b_n$ is the transverse magnetic component normalized to the total magnetic field at current sheet edges, and $\varepsilon$ is the ratio of the thermal velocity to the drift flow plasma velocity. Stability properties of TCSs can be changed during their formation and possible evolution under changing external conditions. An example of this process is shown in Figure \ref{fig:OLGAfig2} (a) with the red arrow that indicates the direction of a possible switch of a TCS from the stable to the unstable state under evolving external conditions, following the decrease in the magnetic transverse component $b_n$ and the relative increase in the plasma flow velocity with respect to the thermal speed (i.e. the decreasing parameter $\varepsilon$, see, e.g., \citet{zelenyi2009tearing,zelenyi2011thin}).

Figure \ref{fig:OLGAfig2} (b) depicts numerous current sheets observed within the HPS at 1 AU. The beginning of the HPS crossing is characterized by a sharp change in the azimuthal IMF angle direction that was generally stable before. It further becomes very unstable and varies many times for hours along with the corresponding signatures of current sheet crossings in the IMF components and the plasma beta until the IMF direction changes to the opposite. It was acknowledged tens of years ago that studying the nature of multiple SCSs in the solar wind is very important for the understanding of current sheet dynamics in the heliosphere. \citet{crooker1993multiple,crooker2004largescale} proposed a mechanism of the formation of multiple current sheets due to the large-scale radial folding/bending of the HCS and magnetic tubes with a subsequent formation of giant loops, which, however, was not confirmed. Modern views on this point suggest that both the loops and HCS folding/rippling exist but at far smaller scales, produced by local dynamical processes (see Part I, Section 2.3.1).

Generally, two main hypotheses on the formation of a set of multiple current sheets in the solar wind near the HCS have been developed:
\begin{itemize}
    \item exogenous: the solar corona is the source of current sheets in the solar wind. 
    \item endogenous: wave and reconnection processes locally trigger multiple CSs in the HCS/HPS vicinity. 
\end{itemize}

Note that, the exogenous factor may lead, e.g., to the distortion and the consequent destruction of neutral surfaces originated from the top of coronal streamers (see, e.g., \citet{sanchezdiaz2019insitu}, and references therein). On the other hand, the two following mechanisms can be attributed to the endogenous factors:
\begin{enumerate}
    \item The development of wave processes as a result of the impact of non-stationary flows/streams, owing to which the HCS can possess a folded or highly rippled form with ripples crossed for tens of minutes at 1 AU \citep{behannon1981finescale,khabarova2015small}. The rippled HCS is shown in Figure \ref{fig:OLGAfig2}(c). The upper panel is adapted from \citet{behannon1981finescale}. They gave a correct explanation of the fine structure of the HCS/HPS, suggesting the existence of the pliss\'e-like shape of the HCS, crossing of which may be reflected in observations of numerous current sheets. The direction of ripples was suggested to be perpendicular to the radial (sunward) direction, which indeed can be the case either at far heliocentric distances where the Parker spiral twist angle is near $90^\circ$ or in front of/behind fast streams pushing the HCS (see Figure 22 and Figure 30 in Part I of the review). The lower panel in Figure \ref{fig:OLGAfig2}(c) shows observational evidence for the formation of HCS ripples. It is easy to see that the crossing of a ripple can be interpreted as an intersection of pairs of current sheets with the opposite direction of the flowing electric current  \citep{mingalev2019modeling}.
    \item The development of current sheet–associated instabilities (mainly, tearing instability) in the solar wind, accompanied by magnetic reconnection and the formation of plasmoids/magnetic islands (see \citet{gosling2007observations,ruffenach2012multi,khabarova2015small,tenerani2015magnetic,tenerani2015tearing,pucci2017fast,khabarova2017energetic,adhikari2019role} and Section 2.1.4. of Part I). The electric currents can flow at both internal and external surfaces of such closed structures. 
\end{enumerate}

It is important to note that the origin of magnetic islands and plasmoids, as well as blobs and bubbles surrounding the HCS, may be of the combined exogenous-endogenous origin (see Part I and \citet{janvier2014arethere}). The largest of these magnetic structures can be formed due to the splitting of streamer current sheets and subsequent magnetic reconnection during the extension and evolution of streamers in the solar wind \citep{somov2013solarcorona}. It is shown that multi-sheet magneto-plasma configurations are unstable in principle \citep{dahlburg1995triple}, i.e. magnetic fields tend to reconnect, which leads to the formation of magnetic islands contributing to changes of the magnetic surface topology and stratification of the current system within the HCS \citep{zelenyi2004fractal,sanchezdiaz2019insitu}. An idea about magnetic reconnection of coronal streamers above loops appeared in 1990th (see \citet{mccomas1991observations,antiochos1999role,wang2000dynamical,somov2013solarcorona}). Later, it was suggested that coronal streamers can eject a series of sequentially arranged magnetic blobs adjacent with magnetic flux ropes into the solar wind. In the plane perpendicular to the neutral surface at the cusp of a streamer, these structures represent a chain of blobs or 2D magnetic islands that can be observed not only in-situ but also in white light (e.g., \citet{wang2000dynamical,song2009quasi,wang2018gradual}). This process has been modelled as well (e.g., \citet{higginson2018structured}). Owing to the impact of the dynamical processes occurring in the surrounding plasma and the development of tearing and other instabilities, magnetic bubbles/blobs originated from the corona can be distorted, split or merged. As a result, all those structures create an HCS turbulent/intermittent environment consisting of multiple current sheets occurring within the system of numerous magnetic islands. In this perspective, it is clear that a fully 3D approach is quite important in order to properly understand the dynamics of these structures.

Figure \ref{fig:OLGAfig2}(d) schematically shows a possible configuration of the HCS surrounded by several magnetic islands. On the right of the sketch, the corresponding profiles of the magnetic field, the plasma and electric current density are shown as a function of the transversal $z$-coordinate. Such current structures are relatively thin and can be comparable by width with the proton gyroradii. In such current configurations, electrons can be magnetized while protons move along quasi-adiabatic (serpentine-like) orbits \citep{zelenyi2004nonlinear,zelenyi2011thin,malova2017evidence}. Two protons on quasi-adiabatic trajectories are shown in Fig. \ref{fig:OLGAfig2}(d) in the HCS plane.

Therefore, multiple current sheets embedded in the HPS are self-consistent structures with characteristic peaks of the plasma density, the plasma beta and the alternating direction of the electric current in the adjacent current sheets. Owing to their small thicknesses and large spatial scales, such magnetic configurations can be described in the frame of an almost one-dimensional quasi adiabatic model (see \citep{zelenyi2004nonlinear}) in which spatial characteristics in the direction transverse to the current sheet midplane are most important.

\subsection{Oblique low-frequency electromagnetic modes and TCS stability}
As described above, TCSs can become unstable when they are influenced by the electromagnetic modes propagating along the magnetic field. Meanwhile, modes with the wave vector ${\bf k}=k \bf{e}_x$ in the GSM system of coordinates represents a limiting case of a more general situation when modes propagate in the arbitrary direction with respect to the magnetic field (e.g., for ${\bf k}=k \bf{e}_y$, these are well-known kink and sausage modes \citep{lapenta1997kinetic,daughton1999unstable,buchner1999sausage}). These fluctuations could be either symmetric or asymmetric with respect to $z=0$ plane. A lot of research were devoted to the analysis of all possible instabilities both analytical \citep{lapenta1997kinetic,daughton1999unstable,yoon2001drift} and numerical \citep{pritchett1996three,zhu1996tearing,buchner1999sausage,karimabadi2003ionion}.

All these studies were limited to the case of 1D Harris equilibrium with $B_z\ne0$. Meanwhile, it is known that real TCSs cannot be described by this model (see above). Some alternative approaches were suggested by \citet{sitnov2004current}, who studied lower-hybrid drift perturbation in the y-direction, although considering electrons as a cold neutralizing background. \citet{silin2002instabilities} studied electromagnetic perturbations in current sheets, in the limit of an infinitely thin 1D current sheet (so-called Syrovatsky’s sheet). Moreover, \citet{zelenyi2009low} showed that different values of the growth rate are possible in TCSs as a function of angle $\theta$ with respect to the direction of magnetic field lines. This implies that different types of wave modes can exist at the neutral plane of the magnetotail simultaneously to form a complex turbulent structure \citep{milovanov2001geometric}.

\citet{zelenyi1998multiscale} suggested analyzing the problem from another perspective. In particular, they studied the state of the magnetotail after that all linear CS modes grew up and nonlinearly interacted with each other (see \citet{zelenyi1998multiscale} for details about such interaction). It is important to point out that the resulting CS state keeps an intrinsic variability, i.e. modes may grow and decay, but on average, the system remains steady. Such a system is, hence, in the so-called non-Equilibrium Steady State (NESS). In the NESS state, particles (e.g., ions as suggested in \citet{zelenyi1998multiscale}) supporting the cross-tail current in the current sheet drift through current sheet-associated magnetic turbulence and are scattered by magnetic fluctuations. The fluctuating part of the electric current is determined by this scattering and controls the parameters of magnetic fluctuations via Maxwell equations. Magnetic fluctuations, in turn, govern peculiarities of ion scattering. Therefore, the fluctuating part of the electric current and magnetic fluctuations are coupled self consistently, while the average current should have an exact value supporting the current sheet magnetic field reversal considered as a boundary condition.

Technically, methods of fractal geometry appear to be very effective to describe particle scattering in the ensemble of multi-scale magnetic fluctuations with certain fractal properties. This analysis uniquely defines the fractal measures of turbulence necessary for the self-consistency of a system. The final task is to convert these measures to the observable standard characteristics of turbulence, such as Fourier spectra. \citet{zelenyi1998multiscale} adopted a simple assumption, namely, the Taylor hypothesis that magnetic structures are frozen into the bulk of moving plasma and their characteristics are determined by the Doppler effect in the frame of an observer/spacecraft. Bringing all these considerations together, \citet{zelenyi1998multiscale} obtained the so-called universal shape of the spectra of magnetic fluctuations: $\delta B/B\sim k^{-7/3}$, where $k$ is the wavenumber.  Certainly, such spectral indexes are expected to exist as ``universal'' only within the frequency (wavelength) domain in which the Taylor hypothesis operates. A theoretical analysis predicted that there should be well-defined spectral breaks at large- and small $k$ boundaries of the universal interval.  An analysis of numerous satellite observations brought together by \citet{zelenyi1998multiscale} confirmed that the spectral index of 7/3 is indeed quite common for magnetospheric processes in numerous cases.

\subsection{Effects of the occurrence of current sheets, flux ropes and magnetic islands for the transport of energetic particles in the heliosphere}

Beyond the importance to accelerate particles, that will be the content of next section, drift along HCS as well as drift due to gradients in and the curvature of the heliospheric magnetic field (HMF) play an integral role in the modulation of galactic cosmic rays, as evinced by more than half a century of neutron monitor \cite[e.g.][and references therein]{cl2019} and spacecraft observations \cite[e.g.][]{jan}, of cosmic-ray intensities. The characteristic 22-year cycle of alternating peaks and plateaux corresponding respectively to negative and positive heliospheric magnetic field polarities seen from neutron monitor observations can in principle be explained by invoking the drift patterns of charged particles in the heliosphere, where, in the former case positively charged cosmic rays drift Earthwards along with the heliospheric current sheet, and in the latter from the polar regions \cite[see, e.g.,][]{j77,j81,burger1985,p13}. This has led to implementing these effects, especially CSs drift, in cosmic ray transport codes since earlier works by \citet{j79} and \citet{k83} as well as in subsequent studies. Implementing drift effects due to the HCS in such a code is, however, not a straightforward endeavour, and several approaches to this problem have been implemented in the past. For example, the approach of \citet{k83} numerically calculated the required drift velocities, while studies like those of \citet{strauss} and \citet{Pei2012} numerically calculate perpendicular particle distances from a modelled HCS, using this distance as an input for an approximate expression for the current sheet drift speed which is calculated assuming a locally flat sheet (see also \citet{burger1985}). The approach of \citet{Burger2012} calculates drift velocities directly by assuming a transition function over the HCS and has also been successfully implemented in CR modulation studies \cite[e.g.][]{EB2015}. All of these approaches, however, treat the HCS as a differentially thin surface, doubtlessly motivated by the fact that high energy galactic cosmic rays have large Larmor radii, this being the characteristic length scale associated with drift processes in a scatter-free environment \cite{f74}, which are assumed to be much greater than the thickness of the current sheet. This quantity is observed to be approximately $10^{4}$~km at $1$~au, and to increase with heliocentric radial distance \cite[see, e.g.,][]{Smith01}. Turbulence in the HMF does, however, act so as to reduce the drift scale, and hence drift effects, as has been shown theoretically \cite[e.g.][]{Bieber97}, from numerical CR modulation studies \cite[e.g.][]{2003Ferreira_etalAnGeo,burger2008} and from numerical test-particle simulations of diffusion and drift coefficients \cite[e.g.][]{minnie2007b,ts2012}. Various approaches have been taken to model this effect \cite[for a review, see][]{Burger2010}, which imply that the assumption that drift processes would be unaffected by physical processes taking place within the finite thickness of the HCS may not always be correct. To investigate this, the turbulence-reduced drift length scale proposed by \citet{E17} is used in what is to follow, as it yields results for the turbulence-reduced drift coefficient in reasonable agreement with numerical simulations of that quantity in various turbulence scenarios, and has been used successfully in \textit{ab initio} CR modulation studies \cite[e.g.][]{kat}. This length scale is given by 
\begin{equation}
\lambda_{A}=R_{L}\left[1 + \frac{\lambda_{\perp}^{2}}{R_{L}^{2}}\frac{\delta B_{T}^2}{B_0^2}\right]^{-1},
\label{eq:la}
\end{equation}
where $R_L$ denotes the maximal Larmor radius to which this drift scale reduces to for low levels of turbulence, $\delta B_{T}^2$ the total transverse magnetic variance, $B_0$ the HMF magnitude, and $\lambda_{\perp}$ the perpendicular mean free path. The inherent three-dimensionality of heliospheric magnetic turbulence has long been taken into account in the various scattering theories used to describe the diffusion of charged particles. An example of this is the study of \citet{b94}, who invoke the observed composite slab/2D nature of turbulence at Earth \cite[see, e.g.,][]{M90,b96} to resolve the then long-standing issue of parallel mean free paths derived from the quasilinear theory of \citet{Jokipii66} for magnetostatic turbulence being considerably smaller than expected from simulations \citep{p82}. Since then, more advanced theories of particle scattering take this geometry into account, and require also as key inputs information as to the behaviour of turbulence power spectra \cite[see, e.g.,][]{matthaeusetal2003,Shalchibook2009,Ruffolo12}, being sensitive in particular to the behaviour of the 2D turbulence spectrum at low wavenumbers \citep{shalchietal2010,EB2015} (See \citet{dundovic2020novel} for a recent review of theoretical models and test-particle simulations). For demonstration, as input for Eq.~\ref{eq:la} an expression of the perpendicular mean free path derived by \citet{EB2015} from a scattering theory proposed by \citet{Shalchi2010} will be employed, given by
\begin{equation}
\lambda_{\perp}\approx a^{2}C_{-1}\lambda_{\parallel}\frac{\delta B^{2}_{2D}}{B_{0}^{2}}\log\left[\frac{3\lambda_{out}+2\sqrt{\lambda_{\parallel}\lambda_{\perp}}}{3\lambda_{2D}+2\sqrt{\lambda_{\parallel}\lambda_{\perp}}}\right],
\label{eq:kperpnlgc4}
\end{equation}
with $C_{-1}$ a normalisation constant for the 2D turbulence power spectrum employed in that study, $a^{2}=1/3$ a constant chosen based on the results of numerical test particle simulations of the perpendicular diffusion coefficient by \citet{matthaeusetal2003}, while the parallel mean free path $\lambda_{\parallel}$ is modelled using the quasilinear theory result employed by \citet{burger2008}. The quantity $\lambda_{2D}$ denotes the length scale at which the break between the energy-containing and inertial ranges occurs on the 2D turbulence power spectrum, and $\lambda_{out}$ the length scale at which the energy range begins. The spectrum employed by \citet{EB2015} to derive Eq.~\ref{eq:kperpnlgc4} is assumed to decrease with decreasing wavenumber at scales larger than $\lambda_{out}$, following physical considerations as discussed by \citet{matthaeusetal2007}. Observationally, it is unclear how to model this latter quantity, given uncertainties as to observations of turbulence quantities \cite[see, e.g.,][]{gr99}, and various estimates for this quantity, which can have a considerable effect on perpendicular mean free paths, have been proposed. \citet{ad17} argue that the largest turbulent injection scale should correspond with the solar rotation rate, while \citet{EB2013} found that scaling of this quantity being proportional to the 2D correlation scale, as modelled using the \citet{so2011} turbulence transport model, would lead to the differential galactic CR proton intensities computed with their ab initio modulation code to be in reasonable agreement with observations. The quantity $\lambda_{out}$ can, however, be indirectly calculated if it is assumed that magnetic island sizes can give one a measure of the 2D ultrascale \cite{matthaeusetal2007}. \citet{E19} does this by making use of observed magnetic island sizes reported by \citet{khabarova2015small} and calculating corresponding values for $\lambda_{out}$ from an expression for the 2D ultrascale derived from an observationally-motivated form for the 2D turbulence power spectrum (the same as that used to derive Eq.~\ref{eq:kperpnlgc4}). The results are quite close to the scaling used by \citet{EB2013} for this quantity, and are used as inputs for Equations~\ref{eq:kperpnlgc4} and~\ref{eq:la}.
\begin{figure}
\centering
  \includegraphics[width=0.5\textwidth]{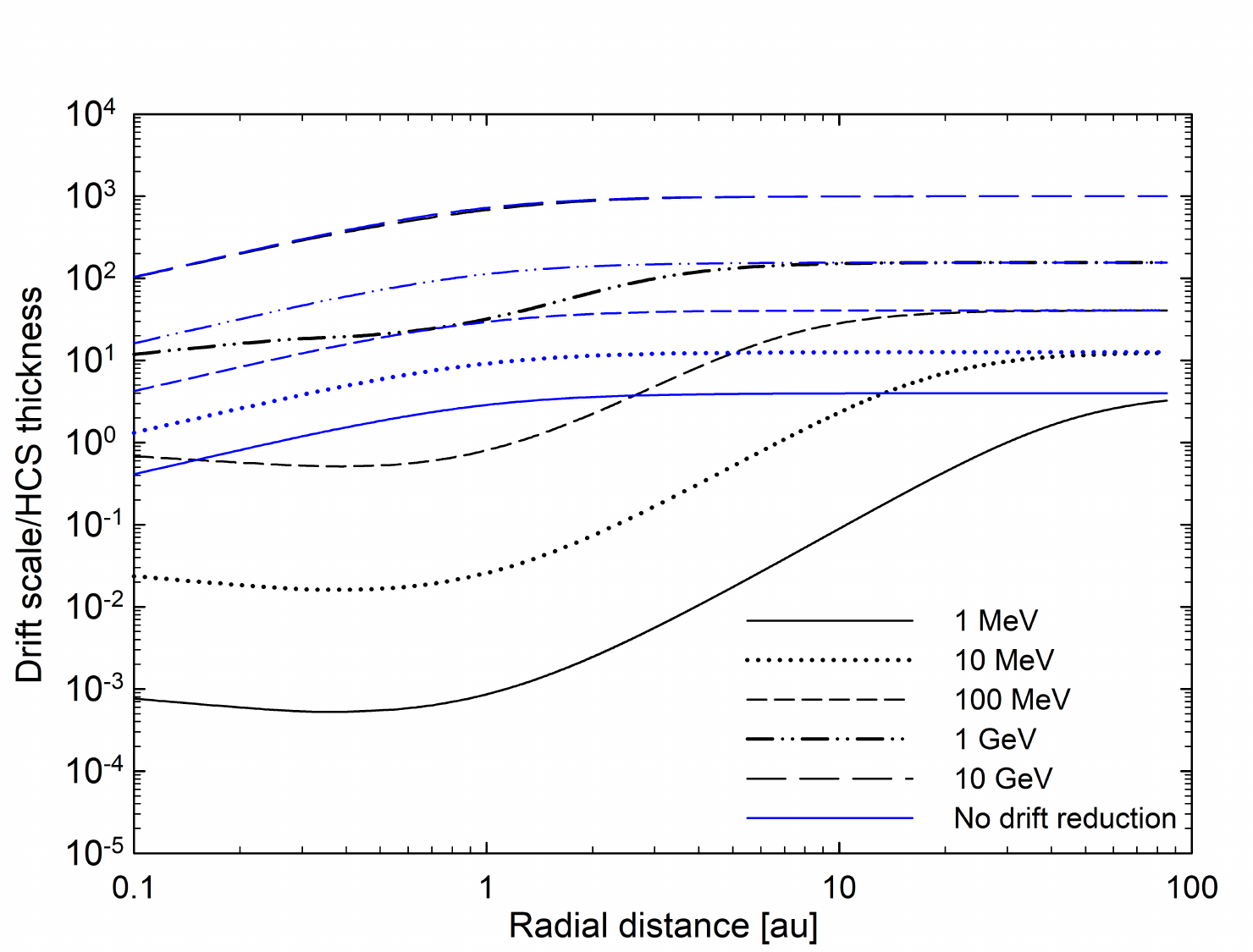}
\caption{Ratio of proton turbulence-reduced drift scale (black lines) and Larmor radius (blue lines) at various energies to the HCS thickness, as a function of heliocentric radial distance in the solar ecliptic plane.}
\label{fig:ratios}       
\end{figure}

To roughly ascertain the validity of the assumption of a negligibly thin HCS, the ratio of the proton drift scale to the HCS thickness is plotted in Figure~\ref{fig:ratios} as a function of heliocentric radial distance, for the purposes of simplicity in the solar ecliptic plane, and assuming a Parker HMF. For the purposes of comparison, the current sheet thickness is assumed to scale as $\sim r$, with a value of $10^4$~km at Earth \citep{Smith01}. Turbulence quantities are modelled using the simple power-law scalings employed by \citet{E19b} (based on the turbulence transport model results of \citet{chhiber}). The ratios are also only evaluated to $85$~au, as these assumptions are no longer valid in the heliosheath. Fig.~\ref{fig:ratios} shows ratios of the drift scale to HCS thickness for both the weak scattering (blue lines) and turbulence-reduced (black lines) cases at various energies. A ratio considerably larger than one would imply that physical processes occurring within the current sheet could safely be neglected when the transport of particles is modelled. For the weak-scattering drift length scale, this ratio is greater than unity for all proton energies considered, barring for a small distance in the very inner heliosphere at $1$~MeV. When the effects of turbulence on the drift scale are taken into account, the picture changes considerably. At the highest energies, corresponding to galactic CR protons above $100$~MeV, and beyond $\sim 1-10$~au, the ratio is very large. Below this energy, and in the very inner heliosphere, it quickly drops below unity, and considerably so at the lowest energies shown. The implication is that the finite thickness of the current sheet, as well as the detailed physics thereof, need to be taken into account when the transport of such particles is being studied. This is especially relevant given the recent interest in the effects of drift on solar energetic particles \cite[see, e.g.,][]{d13,ma13}, and has to some degree been taken into account \cite[e.g.][]{bat18,Eea19}. It is nevertheless also surprising that relatively low-energy ($\leq 100$~MeV) galactic protons may also be influenced by these effects, and that the detailed physics of the HCS might also need to be incorporated in CR transport models.

\section{Kinetic Transport Theory of Energetic Particle Acceleration by Small-Scale Flux Ropes}
\label{sect:leroux}

Irrespective of whether small-scale magnetic flux ropes (SMFRs) originate in a turbulent plasma through magnetic reconnection at large-scale (primary) current sheets or because of the presence of a significant guide/background magnetic field in the turbulent plasma, they form a dynamic turbulence component that generates electric fields through processes like contraction and merging (reconnection) of neighbouring SMFRs that can accelerate particles. Evidence from simulations suggest an efficient acceleration of charged particles traversing regions filled with dynamic SMFRs that can result in power-law spectra for energetic particles as first pointed out by \citet{Matthaeus1984particle} and \citet{Ambrosiano88test} and confirmed later by many others such as \citet{Gray1992MHD, Dmitruk2004test, drake2006electron, drake2010magnetic, Drake2013power, Li2015nonthermally, li2017particle, Li2018roles}. 

\begin{figure}
    \centering
    \includegraphics[width=0.7\textwidth]{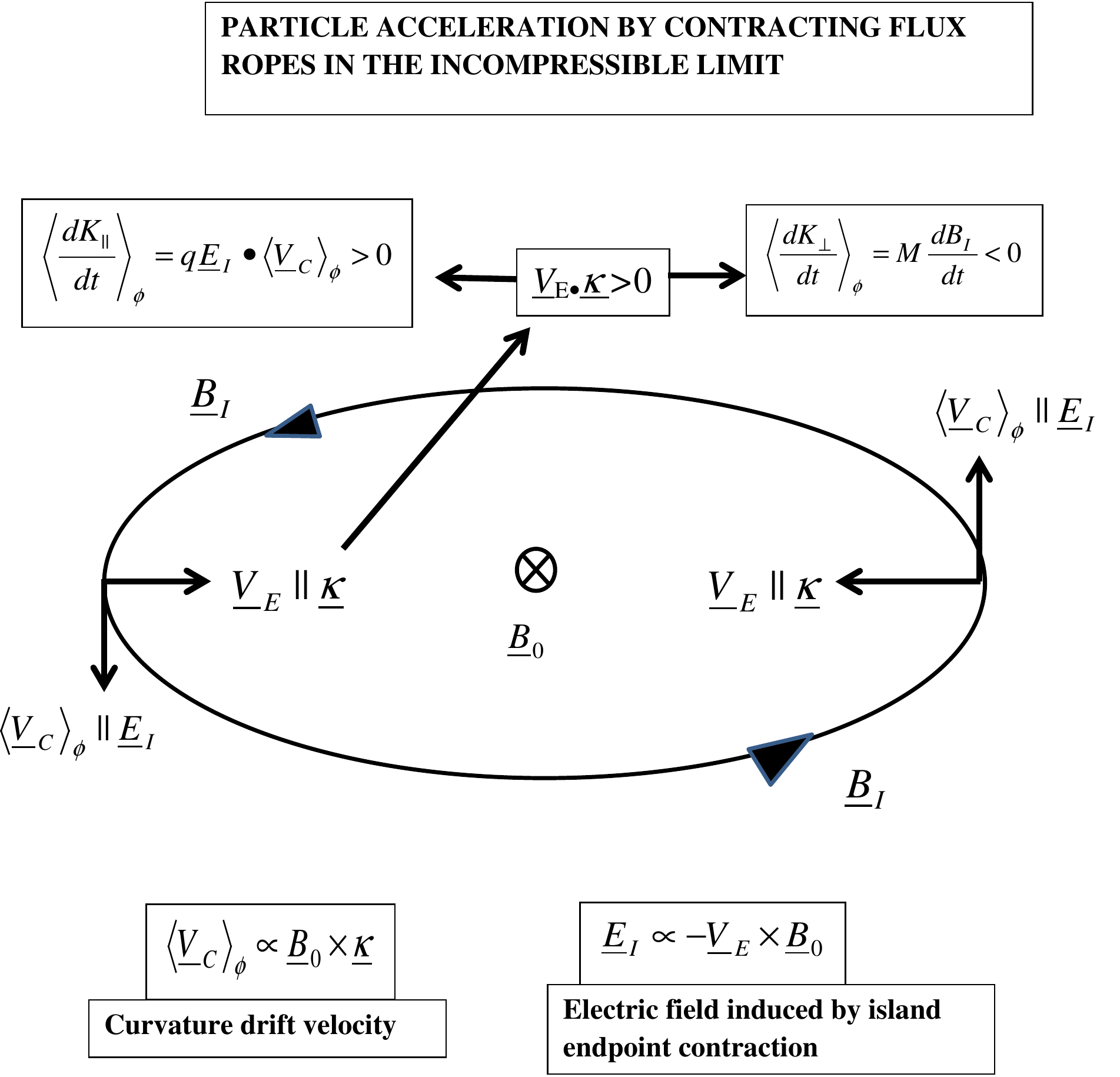}
    \caption{A schematic diagram of energetic ion acceleration by a contracting quasi-2D SMFR in the incompressible limit (reproduced from \citet{leRoux2018self}). Shown is the island (twist) magnetic field component $\BB_I$ of the SMFR structure in the 2D plane perpendicular to the locally uniform guide field (axial) component $\BB_0$ of the SMFR pointing into the page. $\VV_E$ is the contraction velocity (plasma drift velocity) at the endpoints of the island, and $\bm{\kappa} = ( \bb \cdot \grad ) \bb $ is the magnetic curvature vector which points in the same direction as the contraction velocity. Thus, for a contracting island $\VV_E\cdot \bm{\kappa} > 0$. This ensures parallel kinetic energy gain from curvature drift acceleration by the in-plane electric field $\EE_I \approx - \VV_E \times \BB_0 $ induced by contraction because $q\EE_i \cdot \langle \VV_C \rangle_\phi \propto \VV_E \cdot \bm{\kappa} > 0$, where $\langle \VV_C \rangle_\phi$ is the curvature drift velocity, but perpendicular kinetic energy loss from Lagrangian betatron acceleration (combination of betatron and grad-B drift acceleration) because $M dB/dt \propto -\VV_E \cdot \bm{\kappa} < 0$, where M is the magnetic moment (see \eq{eq:leRoux1} and its discussion).}
    \label{fig:leRoux1}
\end{figure}

Theoretical explanation of the main SMFR acceleration mechanisms in the simulations often rely on kinetic transport theories constructed based on the first and second adiabatic invariants combined with magnetic flux conservation (e.g., \citet{drake2006electron, Drake2013power, Zank2014particle}), guiding center kinetic transport theory (e.g., \cite{Dahlin2016parallel, Dahlin2017role}), and the closely related focused transport theory \citep{LeRoux2015kinetic,leRoux2018self,Li2018roles}. These theoretical approaches, besides providing familiar non-resonant acceleration concepts for SMFRs, also show promise in reproducing simulation results of acceleration on macro scales (scales larger than the magnetic island size) where the limitation of nearly gyrotropic particle phase angle distributions inherent in these approaches is less problematic (e.g., \cite{Li2018roles}). This opened up the possibility that such kinetic transport theories can be used to model particle acceleration by SMFRs on large scales in the solar wind, which is computationally beyond the capability of full kinetic particle simulations of acceleration especially.

\begin{figure}
    \centering
    \includegraphics[width=0.7\textwidth]{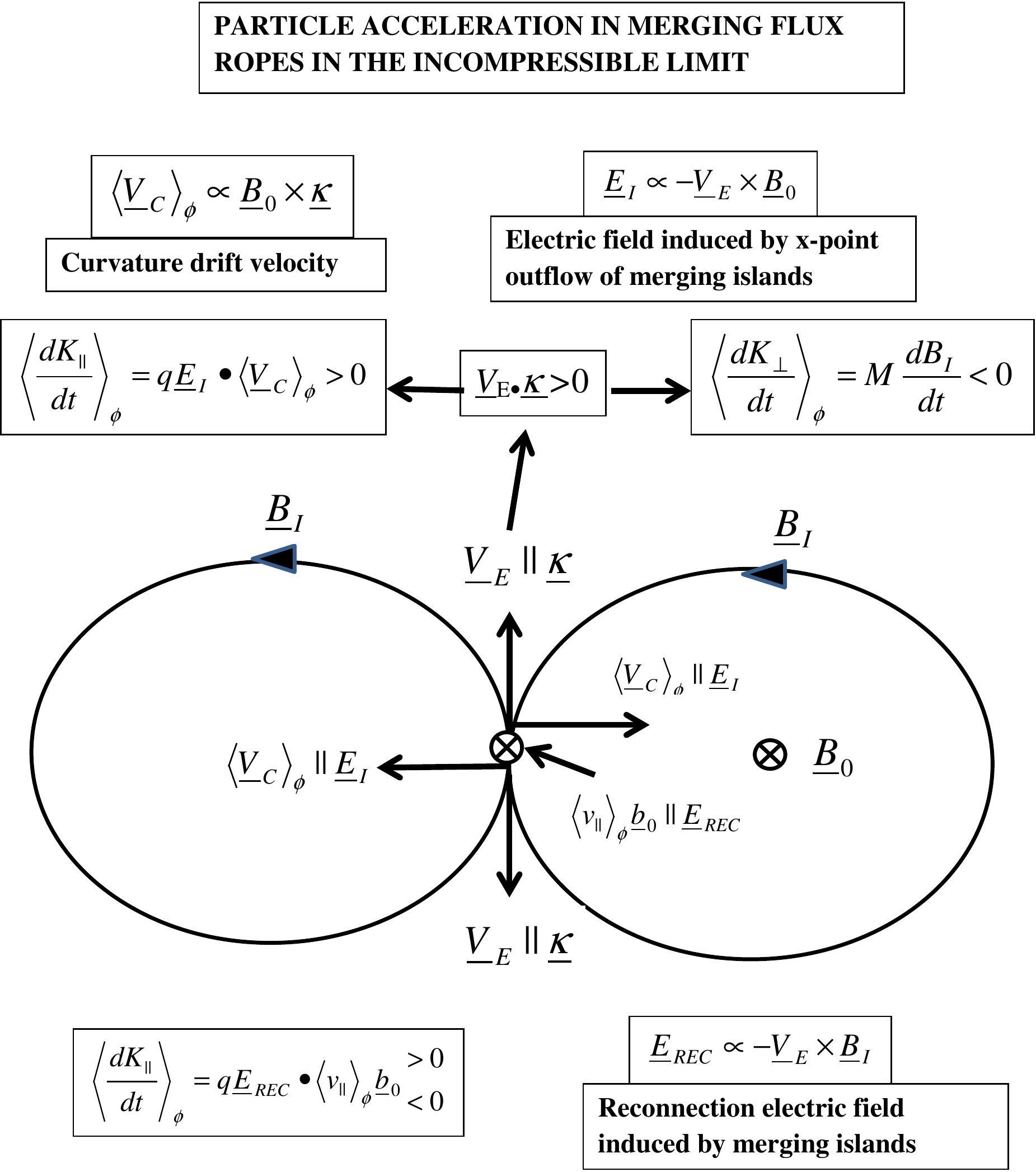}
    \caption{A schematic diagram of ion acceleration by two merging (reconnecting), quasi-2D flux ropes in the incompressible limit (reproduced from \citet{leRoux2018self}). Shown is the island magnetic field $\BB_I$ in the 2D plane perpendicular to a uniform guide field component $\BB_0$ pointing into the page. $\VV_E$ is the x-point plasma outflow drift velocity in the merging area at the center of the merging magnetic islands pointing in the same direction as the magnetic curvature vector $\bm{\kappa} = ( \bb \cdot \grad ) \bb $. Thus, in the merging region (reconnecting area)  $\VV_E \cdot \bm{\kappa} >0$, resulting in parallel kinetic energy gain from curvature drift acceleration and perpendicular kinetic energy loss from Lagrangian betatron acceleration (see \eq{eq:leRoux1} and the caption of \fig{fig:leRoux1}). In the center of the merging area, the reconnection electric field $\EE_{REC} = -\VV_E \times \BB_I$ points into the page. Energetic particle guiding center motion along/against $\BB_0$ results in parallel kinetic energy gain/loss from the reconnection electric field.}
    \label{fig:leRoux2}
\end{figure}

\begin{figure}
    \centering
    \includegraphics[width=0.7\textwidth]{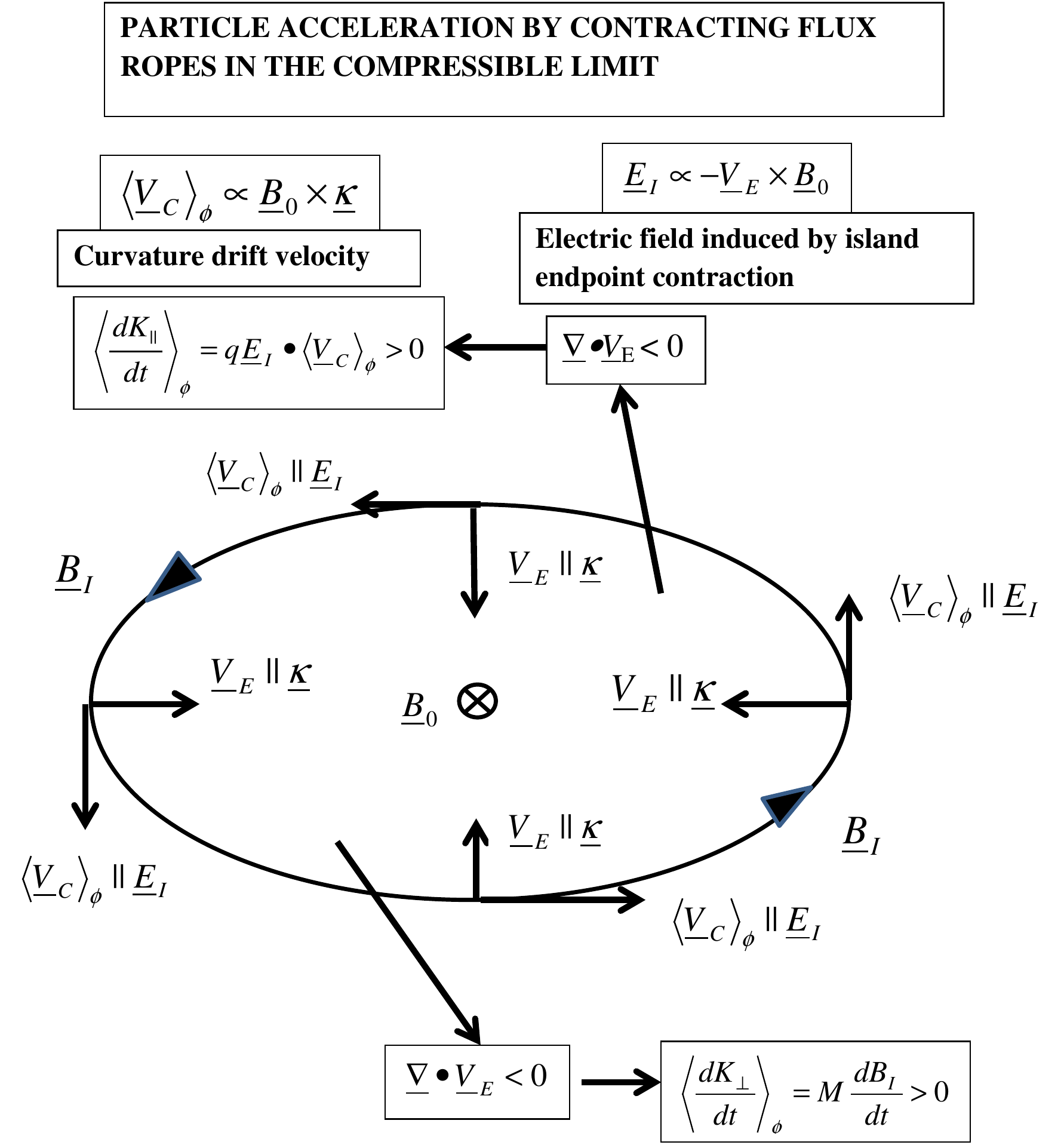}
    \caption{A schematic diagram of ion acceleration in a compressible contracting quasi-2D SMFR (reproduced from \citet{leRoux2018self}). Shown is the island magnetic field (twist component) $B_I$ in the 2D plane perpendicular to a uniform guide field component $B_0$ pointing into the page. In the compressible limit the divergence of the contraction velocity $\divg \VV_E <0$ results in perpendicular kinetic energy gain of energetic particles from the Lagrangian betron acceleration term  $ MdB/dt \propto -\divg \VV_E > 0 $ (see \eq{eq:leRoux1} and its discussion) because the island magnetic field strength increases with time (\eq{eq:leRoux1}). Since $\VV_E\cdot \bm{\kappa} > 0$ for compressible contraction, there is also parallel kinetic energy gain from curvature drift acceleration, but because $|\divg \VV_E | \gg \VV_E \cdot \bm{\kappa} > 0$ in the compressible limit, perpendicular kinetic energy gain from Lagrangian betratron acceleration is expected to be dominant.}
    \label{fig:leRoux3}
\end{figure}

This promise prompted \cite{Zank2014particle} and \cite{LeRoux2015kinetic,leRoux2018self} to develop kinetic focused transport theories that unify the main non-resonant SMFR acceleration mechanisms identified in simulations. Expressed in terms of guiding center kinetic theory, the main acceleration mechanisms are: (i) parallel guiding center motion acceleration by the parallel reconnection electric field generated when neighboring SMFRs form secondary (small-scale) reconnecting current sheets between them to merge (e.g., \cite{Oka2010electron}), (ii) curvature drift acceleration by the motional electric field generated when SMFRs contract or merge  (e.g., \cite{drake2006electron, Drake2013power, li2017particle, Li2018roles}), (iii) Lagrangian betatron acceleration which involves magnetic moment conservation when the magnetic field strength in the plasma drift flow frame slowly varies in time and space. This mechanism includes grad-B drift acceleration by the motional electric field generated in contracting and merging SMFRs (e.g., \cite{Dahlin2016parallel, Dahlin2017role}). In focused transport theory the same acceleration mechanisms manifest in terms of non-uniform plasma flow effects in SMFRs. In this context the main SMFR acceleration mechanisms identified in kinetic particle simulations are SMFR compression acceleration, SMFR incompressible parallel shear flow acceleration referring to the SMFR shear flow tensor in the limit of a SMFR flow with a zero divergence, while acceleration by the parallel reconnection electric field is the same as in guiding center kinetic theory \citep{Zank2014particle,LeRoux2015kinetic,leRoux2018self}. Having studied how the SMFR acceleration mechanisms in focused transport theory and guiding center kinetic transport theory are connected, \cite{leRoux2018self} concluded that: (i) SMFR compression acceleration can be viewed as combining curvature drift momentum gain from the motional electric field  induced by magnetic island compression with Lagrangian betatron momentum gain (momentum gain from unified betatron and grad-B drift acceleration) due to the increasing magnetic field strength resulting from magnetic island compression (magnetic-island-containing area shrinks as a result). This mechanism is illustrated diagrammatically in Figure \ref{fig:leRoux3}. (ii) SMFR incompressible parallel shear flow acceleration can be interpreted as combining curvature drift momentum gain in the motional electric field generated by magnetic island contraction or merging with competing Lagrangian betatron momentum loss from the decreasing magnetic field strength resulting from magnetic island contraction or merging (magnetic-island-containing area is conserved). For illustration of shear-flow acceleration in contracting magnetic islands, see Figure \ref{fig:leRoux1}, and in merging magnetic islands, see Figure \ref{fig:leRoux2}. Figure \ref{fig:leRoux2} also illustrates parallel guiding center motion acceleration by the parallel reconnection electric field generated in the merging process.  (iii) Focused transport theory also includes an acceleration mechanism connected to parallel guiding center motion momentum gain or loss from the non-inertial force associated with the parallel component of the acceleration of the plasma flow in SMFRs \citep{leRoux2018self}. This mechanism can be traced back as being part of the term in guiding center kinetic theory describing the effect of the electric field on drift inertia that also is the source of the curvature drift acceleration mechanism.  From the SMFR focused transport equations, diffusive Parker type transport equations were derived assuming that on large spatial scales in the solar wind pitch-angle scattering will inevitably result in near-isotropic accelerated energetic particle distributions (the diffusive approximation). Accordingly, the distributions were expanded out to the second anisotropic moment in pitch-angle space using Legendre polynomials \citep{Zank2014particle,LeRoux2015kinetic}. By inspecting the SMFR focused transport equation it is clear that SMFR compression acceleration, first advocated by \cite{Zank2014particle} and later also by \cite{LeRoux2015kinetic, leRoux2018self, Li2018roles, du2018plasma}, can be considered as the only true 1st order Fermi SMFR acceleration mechanism because it involves energy gain without competing energy losses.  In the Parker transport equation limit, however, SMFR compression acceleration appears to be a 1st order Fermi acceleration only in terms of the dominating isotropic part of the distribution function because SMFR compression acceleration also contributes to 2nd order Fermi acceleration through second anisotropic moment of the particle distribution \citep{leRoux2018self}. Because the second moment of the distribution function is small compared to the dominating isotropic part of the distribution function, the net compression acceleration for the total distribution function still yields 1st Fermi acceleration mechanism consistent with the focused transport equation. In contrast, SMFR incompressible parallel shear flow acceleration in the Parker transport equation is intrinsically a second order Fermi acceleration mechanism during significant pitch-angle scattering with net acceleration originating from the second anisotropic moment in the particle distribution expansion.  In the focused transport equation this mechanism acts like a 1st order Fermi acceleration mechanism when pitch angle scattering is weak because shear-flow acceleration tends to beam particles along the magnetic field so that energy losses become negligible. This is consistent with discussions by \cite{drake2010magnetic} of incompressible magnetic island contraction leading to 1st order Fermi acceleration because the betatron energy losses are negligible compared to curvature drift acceleration when pitch-angle scattering is weak.  SMFR parallel guiding center motion acceleration by both the parallel reconnection electric field force and by the parallel component of the non-inertial force associated with the acceleration of the plasma flow also yield 2nd order Fermi acceleration in the Parker transport equation limit of focused transport theory, but with the difference that net acceleration comes from the 1st anisotropic moment of the particle distribution function expansion. Thus, for a purely isotropic particle distribution only first Fermi SMFR acceleration will be operative (see also, \cite{drake2010magnetic}), while the other mechanisms need a distribution with a pitch-angle anisotropy to yield a net acceleration effect. 

\subsection{The Small-scale Flux Rope Acceleration Mechanisms}

\subsubsection{A Guiding center kinetic theory perspective}

SMFRs detected near Earth have cross sections of $L_I \sim 0.01 - 0.001 AU $ \citep{Cartwright2010heliospheric, khabarova2015small} where energetic protons, e.g., have gyro-radii $r_g \ll L_I$ for a wide range of energies that easily includes energies in the MeV range. Thus, standard guiding center kinetic theory, which is restricted to gyro-radii much less than scale of the electromagnetic field in the plasma, is well suited for modeling energetic particle acceleration in SMFR regions up to several MeV as observed (e.g., \cite{khabarova2017energetic}). In guiding center kinetic theory, the gyro-phase-averaged rate of change in kinetic energy for energetic charged particles interacting with dynamic SMFRs can be expressed in different ways: 

\begin{equation} \label{eq:leRoux1}
\begin{split}
    \left\langle \frac{dK}{dt} \right\rangle_\phi & \approx \left[ q\EE \cdot \left( v_\parallel \BB + mv_\parallel \bb \times \frac{d\bb}{dt} + \frac{mv_\parallel^2}{qB} \bb \times \bm{\kappa} \right) \right]_\parallel  \\
    & + \left[ M \frac{\de B}{\de t} + q\EE \cdot \left( \frac{M}{q} \frac{\BB \times \grad \BB}{B^2} + \frac{M}{q} \left( \grad \times \bb \right)_\parallel \bb \right) \right]_\perp \approx \\
    & \left[ qE_\parallel v\mu + mv\mu \left( \VV_E \cdot \frac{d\bb}{dt} \right) + mv^2 \mu^2 \left( \VV_E \cdot \bm{\kappa} \right) \right]_\parallel  \\
    & + \left[ \left( 1-\mu^2 \right) M' \left( \frac{\de B}{\de t} + \left( \VV_E \cdot \grad \right) B \right) + (1-\mu^2)B \frac{dM'}{dt} \right]_\perp \approx \\
    & \left[ qE_\parallel v\mu - mv\mu \left( \frac{d\VV_E}{dt}\cdot \bb \right) + mv^2\mu^2 \left( \VV_E \cdot \bm{\kappa} \right) \right]_\parallel  \\
    & + \left[ -mv^2 \frac{1}{2}\left( 1-\mu^2\right) \left[ \left( \VV_E\cdot \bm{\kappa} \right) + \left( \divg \VV_E \right) \right] \right]_\perp
\end{split}
\end{equation}

where $q$ is the net particle charge, $v_\parallel$ is the parallel guiding velocity component, $\bb$ is the unit vector along the SMFR magnetic field, $\bm{\kappa} = ( \bb \cdot \grad ) \bb $ is its magnetic curvature, $M$ is the magnetic moment of a gyrating particle, $\EE \cdot \bb = E_\parallel$ is the parallel reconnection electric field component associated with merging SMFR pairs, $\mu$ is the cosine of the particle pitch angle, $\VV_E$ is the electric field drift (plasma drift) velocity (velocity at which the SMFR magnetic field is contracting or merging), and $M'$ is the maximum magnetic moment (magnetic moment when $\mu = 0$).

In \eq{eq:leRoux1}, first and second line, the basic SMFR acceleration mechanisms are grouped in terms of parallel kinetic energy changes (terms in first square bracket) and perpendicular kinetic energy changes (terms in second square bracket). The mechanisms are: (1) Parallel guiding center motion acceleration by the parallel reconnection electric field component $E_\parallel$ generated in reconnection regions between merging neighbouring SMFRs (first term in first square bracket), (2) the effect of the electric field force on the drift inertia term minus curvature drift acceleration (second term in first square bracket in which $d/dt = \de / \de t + (\VV_E \cdot \grad )$), (3) curvature drift acceleration by the motional electric field induced by contracting or merging SMFRs (third term in the first square bracket), (4) betatron acceleration due to time variations in the SMFR magnetic field strength (first term in the second square bracket), (5) grad-B drift acceleration by the motional electric field induced by contraction and merging of SMFRs (second term in the second square bracket) and, (6)  parallel drift acceleration by $E_{REC\parallel}$ (the last term in second square bracket). Direct comparison of the terms in the first two lines of \eq{eq:leRoux1} with the corresponding terms in the third and fourth lines reveals that one can express the curvature drift acceleration term in terms of $ \VV_E \cdot \bm{\kappa} $ (advection of curved SMFR magnetic field at plasma drift velocity), and the grad-B drift acceleration term in terms of  $(\VV_E\cdot \grad )B$ (advection of the perpendicular gradient in SMFR magnetic field strength at the plasma drift velocity). This version of the grad-B drift acceleration term can be combined with the betatron acceleration term into a Lagrangian betatron acceleration expression $M'dB/dt = M'(\de B/\de t + (\VV_E\cdot \grad ) B)$ that tracks the time and spatial variations in the SFMR magnetic field strength $B$ following the SMFR flow  due to contraction or merging processes (first two terms in second square bracket, in line three of \eq{eq:leRoux1}) (e.g., \cite{Dahlin2016parallel, Dahlin2017role}). Comparison of the last terms in line two and line four indicates that approximate conservation of the magnetic moment $M$ requires a small $E_\parallel$ value. Furthermore, comparing the terms in the fourth line with those in the sixth line of \eq{eq:leRoux1} reveals that the Lagrangian betatron acceleration expression can be related to a combination of the $\VV_E\cdot\bm{\kappa}$ and the $\divg \VV_E$ terms, assuming approximate magnetic moment conservation (i.e. a small $E_\parallel$-value). Thus, Lagrangian betatron acceleration is a competition between incompressible SMFR contraction or merging $(\VV_E\cdot \bm{\kappa} > 0)$ resulting in energy loss because $dB/dt < 0$ and compressible contraction or merging $( \divg \VV_E < 0 )$ resulting in energy gain because $dB/dt > 0$. Inspection of the middle term in the first square bracket and the corresponding middle terms in lines 3 and 5 of \eq{eq:leRoux1} reveals how the drift inertia term in line one was converted into a term modelling parallel guiding center motion acceleration by the parallel component of the non-inertial force associated with the acceleration of the plasma drift velocity $dV_E/dt$.

Combining the first two terms of the fifth line and separately the last term in the fifth line with the first term in the sixth line of \eq{eq:leRoux1}, results in the expression 

\begin{equation}\label{eq:leRoux2}
\begin{split}
\left\langle \frac{dK}{dt} \right\rangle_\phi \approx & \; v\mu \left( q\EE_{REC} - m \frac{d\VV_E}{dt} \right) \cdot \bb +  mv^2\frac{1}{2}\left( 3\mu^2-1 \right) \left( \VV_E \cdot \bm{\kappa} \right)  \\
 & - mv^2 \frac{1}{2}\left(1-\mu^2\right) \left( \divg \VV_E \right) 
\end{split}
\end{equation}

The combined term containing $\VV_E \cdot \bm{\kappa}$  in \eq{eq:leRoux2} unifies curvature drift acceleration with Lagrangian betatron and parallel drift acceleration, but it seems that the last term associated with $\divg \VV_E$ combines Lagrangian betatron acceleration with parallel drift acceleration only without a contribution from curvature drift acceleration. The latter raises a question because this contradicts the standard Parker cosmic-ray transport equation in which the $\divg \VV_E$-term has been shown to combine curvature drift acceleration with Lagrangian betatron and parallel drift acceleration \citep{Kota1997energy, Webb1981scatter}. By introducing the shear flow tensor into the bottom line of equation \eq{eq:leRoux1} using the relationship 

\begin{equation}\label{eq:leRoux3}
   \VV_E \cdot \bm{\kappa} = -\bb \cdot \left( \bb \cdot \grad \right) \VV_E = - \left[ b_ib_j\sigma_{ij} + \frac{1}{3} \left( \divg \VV_E \right) \delta_{ij} \right]
\end{equation}

where we relate the magnetic curvature advection term, $\VV_E \cdot \bm{\kappa}$, to the parallel shear-flow term, $\bb \cdot \left( \bb \cdot \grad \right) \VV_E$, which in turn is expressed in terms of the shear-flow tensor $\sigma_{ij}$

\begin{equation}\label{eq:leRoux4}
    \sigma_{ij} = \frac{1}{2}\left[ \frac{\de V_{Ei}}{\de x_j} + \frac{\de V_{Ej}}{\de x_i} - \frac{2}{3} \left( \divg \VV_E \right) \delta_{ij}\right]
\end{equation}

one finds that

\begin{equation}\label{eq:leRoux5}
\begin{split}
\left\langle \frac{dK}{dt} \right\rangle_\phi \approx & \; v\mu \left( q\EE_{REC} - m \frac{d\VV_E}{dt} \right) \cdot \bb - mv^2 \left( 3\mu^2 - 1 \right)\bb \cdot \left( \bb \cdot \grad \right) \VV_E  \\ 
& -\frac{1}{3}mv^2\left( \divg \VV_E \right) + \frac{1}{3} mv^2 \frac{1}{2} \left( 3\mu^2 - 1 \right) \left( \divg \VV_E \right)
\end{split}
\end{equation}

In \eq{eq:leRoux5} the $\bb \cdot \left( \bb \cdot \grad \right) \VV_E$-term can be viewed as a combination of curvature drift acceleration with Lagrangian betatron and parallel drift acceleration acting collectively as plasma drift parallel shear flow acceleration in SMFRs. The first $\divg \VV_E$-term in \eq{eq:leRoux4} is exactly the $\divg \VV_E$-term appearing in the standard Parker cosmic-ray transport equation that has been shown before to combine curvature drift acceleration with Lagrangian betatron and parallel drift acceleration. The last $\divg \VV_E$-term in \eq{eq:leRoux5} depends on the factor $\left( 3\mu^2 - 1 \right)$ just like the $\bb \cdot \left( \bb \cdot \grad \right) \VV_E$-term, providing evidence that it too unifies curvature drift acceleration with Lagrangian betatron and parallel drift acceleration. The difference between the two $\divg \VV_E$-terms is that the first term accelerates the isotropic part of the distribution function while the second term affects the energy of particles belonging to the anisotropic part of the particle distribution as discussed above (for more details, see \cite{leRoux2018self}).

\subsubsection{The Connection Between Guiding Center Kinetic and Focused Transport Theory}

By doing the substitution $\VV_E = \UU_\perp$ (which follows from specifying the macroscale electric field in contracting and merging SMFRs as the induced motional electric field $\EE = - \UU \times \BB$ where $\UU$ is the plasma flow velocity in SMFRs) in \eq{eq:leRoux5}, we get the expression

\begin{equation}\label{eq:leRoux6}
\begin{split}
\left\langle \frac{dK}{dt} \right\rangle_\phi = & \; v\mu \left( q\EE - m \frac{d\UU_\perp}{dt} \right) \cdot \bb - \frac{1}{2}\left( 1-\mu^2 \right)\left( \divg \UU_\perp \right)  \\
& - \frac{1}{2}\left( 3\mu^2 - 1 \right) \bb \cdot \left( \bb \cdot \grad \right) \UU_\perp
\end{split}
\end{equation}

This expression reveals the close connection between standard guiding-center kinetic theory and focused transport kinetic theory that we use to model particle acceleration by dynamic small-scale flux ropes because, according to focused transport theory, the relative momentum rate of change is given by 

\begin{equation}\label{eq:leRoux7}
\begin{split}
\frac{1}{p}\left\langle \frac{dp}{dt} \right\rangle_\phi = & \; \mu \left( \frac{q\EE}{p} - \frac{1}{v} \frac{d\UU}{dt} \right) \cdot \bb - \frac{1}{2}\left( 1-\mu^2 \right)\left( \divg \UU \right) \\
& - \frac{1}{2}\left( 3\mu^2 - 1 \right) \bb \cdot \left( \bb \cdot \grad \right) \UU
\end{split}
\end{equation}

The main difference is that guiding-center kinetic theory describes particle acceleration in terms of the non-uniform plasma drift velocity $\VV_E = \UU_\perp$, whereas focused transport theory does it in terms of the total non-uniform plasma flow velocity in SMFRs. Thus, from the perspective of focused transport theory, we have four SFMR acceleration mechanisms. From left to right in \eq{eq:leRoux7} they are acceleration by the parallel reconnection electric field force, the parallel non-inertial force due to the acceleration of the SMFR flow, compression of the SMFR flow, and parallel SMFR shear flow. These mechanisms form the basis of our focused and Parker transport theories for energetic particle acceleration by and propagation through a field of dynamic SMFRs.   

\subsection{Focused and Parker Transport Equations for Particle Acceleration by SMFRs}

By applying standard perturbation analysis to the focused transport equation containing \eq{eq:leRoux7} we derived a modified focused transport equation that models how energetic particles respond in a statistical average sense on large spatial scales to both the non-uniform solar wind flow and interplanetary magnetic field, and to a collection of dynamic SMFRs advected with the solar wind flow \citep{LeRoux2015kinetic,leRoux2018self}. The derivation was done assuming that SMFR dynamics occur mainly in a 2D plane perpendicular to the guide field (axial) component $B_0$ of the SMFR assuming a strong guide field \citep{Birn1989, Dmitruk2004test}. The 2D plane contains the magnetic island (twist) component $B_I$ of the SMFR and the SMFR flow $U_I$ determining contraction or merging of the magnetic island. In the inner heliosphere, it appears that assuming a strong guide field so that $B_I/B_0 \ll 1$ is reasonable \citep{smith2006turbulent}. For simplicity, no distinction is made in the derivation between the guide field and the background magnetic field. This approach has some support from observational evidence that the SMFR guide field is aligned with the solar wind spiral magnetic field \citep{Zheng2018observational}. The modified focused transport equation is presented compactly as

\begin{equation}\label{eq:leRoux8}
\begin{split}
\left( \frac{df}{dt} \right)_{SW} = & - \divg \left[ \left< \frac{d\xx}{dt}\right>^I \left( \varepsilon_I \right) f \right] - \frac{1}{p^2}\frac{\de }{\de p} \left[ p^2 \left< \frac{dp}{dt} \right>^I \left( \varepsilon_I \right) f \right]  \\
& - \frac{\de}{\de \mu} \left[ \left< \frac{d\mu}{dt}\right>^I \left( \varepsilon_I \right) f \right] \\
& + \frac{\de}{\de \mu} \left[ D^I_{\mu\mu} \left( \varepsilon_I \right) \frac{\de f}{\de \mu} + D^I_{\mu p}\left(\varepsilon_I \right) \frac{\de f}{\de p} \right]  \\
& + \frac{1}{p^2}\frac{\de}{\de p} \left[ p^2 \left( D^I_{p\mu} \left( \varepsilon_I \right) \frac{\de f}{\de \mu} + D^I_{pp} \left( \varepsilon_I \right) \frac{\de f}{\de p} \right) \right]
\end{split}
\end{equation}
where $f\left(x,p,\mu,t\right)$ is the energetic charged particle distribution. On the left-hand side of this equation, $\left( df / dt \right)_{SW}$ represents the standard focused transport equation for energetic particle transport in the non-uniform solar wind flow and interplanetary magnetic field (e.g., \citet{isenberg97hemispherical}). On the right-hand side, there are additional terms to model the interaction of energetic particles with dynamic SMFRs. In the top line $ \left< dp/dt \right>^I$ models the average energetic particle momentum rate of change in response to average SMFR quantities (the average SMFR compression, parallel shear flow, parallel reconnection electric field due to SMFR merging, and the parallel acceleration of the SMFR flow). In the absence of pitch-angle scattering, the acceleration is coherent but can become stochastic if particles undergo significant pitch-angle scattering. In the bottom line, $D_{pp}^I$ signifies the variance in the energetic particle rate of momentum change due statistical fluctuations in the SMFR quantities expressed in terms of the variance in the SMFR compression, parallel shear flow, parallel reconnection electric field, and the parallel acceleration of the SMFR flow (for more details, see \cite{LeRoux2015kinetic, leRoux2018self}). Recently, a simplified telegrapher type Parker transport equation was derived from the modified focused transport equation by assuming that on large spatial scales the energetic particle distribution will inevitably become nearly isotropic due to significant levels of pitch-angle scattering \citep{leroux2019modeling}. This was accomplished by expanding the energetic particle distribution function $f\left(x,p,\mu,t\right)$ out to the second moment with respect to $\mu$ with aid of Legendre polynomials and deriving the zeroth, first and second moments of the modified focused transport equation (see also \cite{Zank2014particle,LeRoux2015kinetic,leRoux2018self}). The first and second moment equations were simplified to enable a closed evolution equation for the isotropic part of the distribution function $f_0\left(x,p,t\right)$ in the form of a telegrapher type Parker transport equation. Discussion of the simplifications can be found in \cite{leroux2019modeling}.  The telegrapher type Parker transport equation we derived is given by

\begin{equation}\label{eq:leRoux9}
\begin{split}
& \frac{3\kappa_\parallel^I}{v^2}\frac{\de}{\de t}\left[ \frac{\de f_0}{\de t} + \left( \UU^{coh}\cdot\grad \right) f_0 - \left(\divg \UU^{coh} \right) \frac{p}{3}\frac{\de f_0}{\de p}  \right. \\
& \left. - \frac{1}{p^2}\frac{\de}{\de p}\left( p^2 D_{pp}^{Istoch}\frac{\de f_0}{\de p} \right) \right]  \\
& + \frac{\de f_0}{\de t} + \left[ \UU^{coh} - \frac{1}{3p^2} \frac{\de }{\de p} \left( p^3 \left( U_{EA}^{coh} - U^{Istoch} \right) \bb \right) \right] \cdot \grad f_0  \\
& - \left[ \left( \divg \UU^{coh} \right) + \left( \divg \left( U_{EA}^{coh} + U^{Istoch} \right) \bb \right) \right]\frac{p}{3}\frac{\de f_0}{\de p} \\
& = \divg \left( \kappa_\parallel^I \bb \bb \cdot \grad f_0 \right) + \frac{1}{p^2}\frac{\de }{\de p} \left[ p^2 \left( D_{pp}^{coh} + D_{pp}^{Istoch} \right) \frac{\de f_0}{\de p} \right]  \\
& + \frac{2}{3} p U_{EA}^{coh} \left( \bb \cdot \grad \right) \frac{\de f_0}{\de p}
\end{split}
\end{equation}

In \eq{eq:leRoux9} the superscript 'coh' refers to a combination of background solar wind and average SMFR quantities, whereas the superscript 'Istoch' indicates the variance of fluctuating SMFR quantities only. Accordingly,  $U^{coh}$ models the net advection effect on energetic particles stemming from the combination of the solar wind velocity with the mean plasma flow velocity in dynamic SMFRs, $U_{EA}^{coh}\bb$ represents the net parallel advection effect on energetic particles from the average parallel component of the electric field and of the acceleration of the plasma flow in the background solar wind and in SMFRs, and $U^{Istoch}\bb$  refers to the average, field-aligned advection effect produced by the variance of statistical fluctuations in SMFR fields. $\kappa_\parallel^I$ denotes the parallel diffusion coefficient as a consequence of particle pitch-angle scattering by random magnetic mirroring forces in SMFRs (for more information, see \cite{LeRoux2015kinetic,leRoux2016combining,leRoux2018self}). Furthermore, there are two categories of second-order Fermi acceleration in \eq{eq:leRoux9}.  $D_{pp}^{coh}$ models second-order Fermi acceleration when particles undergoing pitch-angle scattering on macro scales respond to the average parallel electric field and acceleration of the plasma flow, the parallel shear flow tensor, and the flow compression for both the background solar wind and SMFRs, whereas $D_{pp}^{Istoch}$ describes second-order Fermi acceleration when particles experience the variance effects from fluctuations in the same SMFR quantities. The more familiar diffusive Parker transport equation can be recovered by neglecting the additional transport terms in the top line of \eq{eq:leRoux9}. Besides the well-known telegrapher term (first term in line one of \eq{eq:leRoux9} containing the second-order time derivative, there are also less familiar additional telegrapher terms involving second and third-order partial derivatives (rest of terms in line one and two of \eq{eq:leRoux9}). The telegrapher Parker transport equation addresses a deficiency in the diffusive Parker transport equation where some particles propagate to larger distances than physically possible by restoring causality. However, the causality in particle transport is only restored for leading edge particle pulses that are nearly isotropic, that is, for particles that experience significant pitch-angle scattering.  This kind of telegrapher equation is less accurate when it comes to model unscattered particle escape from the acceleration site during early times, resulting in a cutoff in the particle distribution spatially that is too abrupt so that the maximum distance particles reach as a function of time is underestimated (e.g., \cite{Effenberger2014diffusion, Malkov2015cosmic}). Since we model energetic particle acceleration inside the SMFR acceleration region on macro scales, where particles are expected to be scattered by low-frequency wave turbulence and by fluctuating magnetic mirroring forces inside these structures to maintain near-isotropic distributions, the telegrapher Parker equation should be applicable.

\subsection{Solutions of the Diffusive Parker Transport Equation for Acceleration by SMFRs}

Various analytical steady-state solutions for spatial transport in planar geometry, but also some in spherical geometry, were found for simplified versions of the diffusive Parker transport equation, that is, \eq{eq:leRoux9} neglecting telegrapher terms in the top line (e.g., \cite{Zank2014particle,Zank2015particle,LeRoux2015kinetic,leRoux2016combining,Zhao2018unusual,leroux2019modeling}).  The solutions either emphasized first-order Fermi due to the average SMFR compression and/or acceleration by the mean parallel reconnection electric field in the mixed-derivative transport term (see last term in \eq{eq:leRoux9}) generated during SMFR merging \citep{Zank2014particle,Zank2015particle,Zhao2018unusual}, or second-order Fermi acceleration involving the variance in the SMFR compression/parallel shear flow and/or 1st order Fermi acceleration \citep{LeRoux2015kinetic,leRoux2016combining,leroux2019modeling}. Irrespective of whether first-order or second-order Fermi SMFR acceleration in the solutions was emphasized, two key predictions were made that was verified by spacecraft data analysis: (1) the accelerated particle flux form spatial peaks in the SMFR acceleration region so that the flux amplification factor downstream of the particle injection point increases with particle energy, and (2) the accelerated particle spectra evolve through the SMFR region by becoming harder downstream of the injection location \citep{Zank2015particle,leRoux2016combining,Zhao2018unusual,adhikari2019role,leroux2019modeling,khabarova2016small}. In the new work \cite{leroux2019modeling} present both time-dependent and steady-state analytical solutions of \eq{eq:leRoux9} in planar geometry in which all the basic SMFR acceleration mechanisms present in the underlying focused transport theory (see \eq{eq:leRoux7}) are unified. The solutions are based on solving a simplified version of the telegrapher Parker transport equation \eq{eq:leRoux9} in which the solar wind flow velocity is defined in the x-direction as $\UU_0 =U_0\bm{e}_x$ and the background magnetic field is specified in the x-z-plane according to the expression $\BB_0 = B_0 \left( \cos (\psi) \bm{e}_x + \sin(\psi) \bm{e}_z\right)$ where $\psi$ is the spiral magnetic field angle (the angle between $\BB_0$ and $\UU_0$). The transport equation is

\begin{equation}\label{eq:leRoux10}
\begin{split}
& \tau_{SC}^I \frac{\de^2f_0}{\de t^2} + \frac{\de f_0}{\de t} + U_0^I \frac{\de f_0}{\de x} + \left\langle \nu_{COM}^I \right\rangle \frac{p}{3}\frac{\de f_0}{\de p} = \\
& = \kappa_{xx}^I \frac{\de^2 f_0}{\de x^2} + \frac{1}{p^2}\frac{\de }{\de p} \left( p^2 D_{pp}^I \frac{\de f_0}{\de p} \right) + \frac{2}{3} U_E^I \frac{\de }{\de x} \left( p \frac{\de f_0}{\de p} \right) - \frac{f_0}{\tau_{esc}} + \\
& + \frac{ dN / dt }{4\pi p_0^2} \delta\left(x-x_0\right)\delta\left(p-p_0\right) 
\end{split}
\end{equation}

where in the first term, the telegrapher term, $\tau_{SC}$, is the time scale for particle pitch-angle scattering by random magnetic mirroring forces in SMFRs. For simplicity, the other telegrapher terms in the top line of \eq{eq:leRoux9} were neglected. In the third term in \eq{eq:leRoux10}, the advection term, the effective advection velocity in the x-direction $U_0^I = U_0 - U_E^I$, where $U_0$ is the background solar wind flow velocity, and $U_E^I = \left( 3\kappa_\parallel^I / v \right) \left< \nu_{REC} \right> \cos(\psi)$ is an advective velocity along the background magnetic field $\BB_0$ projected in the x-direction using the factor $\cos(\psi)$. The $U_E^I$ advection effect is associated with particle interaction with the mean parallel reconnection electric field in numerous merging SMFRs generating an average relative momentum rate of change $\left< \nu_{REC}^I \right>$. The fourth term in \eq{eq:leRoux10} contains the mean SMFR compression rate $\left< \nu_{COM}^I \right>$ to model first-order Fermi SMFR acceleration, while the first term on the right hand side of \eq{eq:leRoux10} includes the diffusion coefficient in the x-direction $\kappa_{xx}^I$, which is related to the parallel diffusion coefficient $\kappa_\parallel^I$ according to the expression $\kappa_{xx}^I = \kappa_\parallel^I \cos^2(\psi)$. In the second term on the right of \eq{eq:leRoux10} we find the total momentum diffusion coefficient for second-order Fermi acceleration by SMFRs $D_{pp}^I = p^2 D_0^I = D_{pp}^{Icoh} + D_{pp}^{Istoch}$ where $D_{pp}^{Icoh}$ refers to second-order Fermi acceleration when particles undergo pitch-angle scattering on macro scales in response to the average parallel electric field and acceleration of the plasma flow, the parallel shear flow tensor, and the flow compression in SMFRs, whereas $D_{pp}^{Istoch}$ describes second-order Fermi acceleration when particles experience the variance effects from fluctuations in the same four SMFR quantities. This is followed by the third mixed derivative transport term which determines particle acceleration by the mean parallel reconnection electric field in numerous merging SMFRs. The second last term determines the rate of particle escape from the SMFR region which depends on the escape time scale $\tau_{esc}$ (see also, \cite{Zhao2018unusual}) whereas the last term represents a particle source in which particles in a thin spherical momentum shell with radius $p_0$ are continuously injected at a rate $dN/dt$ in the SMFR region at location $x = x_0$.
As an example, we present the steady-state solution of \eq{eq:leRoux10} which is

\begin{equation}\label{eq:leRoux11}
\begin{split}
f_0(x,p) = \frac{1}{2\pi}\left( \frac{dN/dt}{4\pi p_0^3} \right) \sqrt{\frac{1}{\Bar{D}_0^I\kappa_0^I}} e^{ \frac{1}{2} \frac{U_0^I + qU_E^I/3}{\kappa_0^I} \left( x - x_0 \right) } \left( \frac{p}{p_0} \right)^{-\frac{q}{2}} K_0 \left( 2 \sqrt{\alpha\beta} \right)
\end{split}
\end{equation}

where $K_0$ is the modified Bessel function of the first kind and

\begin{equation}\label{eq:leRoux12}
\begin{split}
& \Bar{D}_0^I = D_0^I \left[ 1 - \frac{1}{9} \frac{\left( U_E^I \right)^2}{\kappa_0^I D_0^I} \right] \\
& q = 3 \left[ 1 - \frac{1}{9} \frac{ \left< \nu_{COM}^I \right> - U_0^I U_E^I / \kappa_0^I }{ D_0^I } \right] / \left[ 1 - \frac{1}{9} \frac{\left( U_E^I \right)^2}{\kappa_0^I D_0^I} \right] \\
& \alpha = \frac{1}{4\kappa_0^I} \left[ \left(x-x_0\right)^2 + \frac{\kappa_0^I}{D_0^I} \left[ \ln \left( \frac{p}{p_0} \right) - \frac{1}{3} \left( \frac{U_E^I}{\kappa_0^I} \right) \left( x - x_0 \right) \right]^2 \right] \\
& \beta = \left( \frac{1}{2}\frac{U_0^I}{\kappa_0^I} \right)^2 \kappa_0^I + \left( \frac{q}{2} \right)^2 \Bar{D}_0^I + \frac{1}{\tau_{esc}}
\end{split}
\end{equation}

The solution includes all four SMFR acceleration mechanisms listed in the underlying focused transport equation (see \eq{eq:leRoux7}) that appears in the Parker transport equation as first-order Fermi acceleration, second-order Fermi acceleration, and acceleration by the mixed-derivative transport term. By taking limits of the solution, the basic characteristics of observations of energetic particle acceleration and transport through a dynamic SMFR region in the solar wind, namely, energetic particle distribution spatial peak formation and spectral hardening as discussed above. If $\ln^2\left(p/p_0 \right) \gg \left( x - x_0 \right)^2$ in the parameter $\alpha$ in $K_0 \left( 2 \sqrt{\alpha\beta} \right)$, one finds that

\begin{equation}\label{eq:leRoux13}
\begin{split}
f_0(x,p) \propto e^{ - \frac{U_0}{2\kappa_0^I}(x-x_0) }\left( \frac{p}{p_0} \right)^{-\frac{1}{2}\left[ q + |q| \sqrt{ 1 + (2/q)^2 \tau_{D_0^I} \left[ 1/\tau_{\kappa_0^I} + 1/\tau_{esc} \right]  } \right]}
\end{split}
\end{equation}

In the opposite limit $\ln^2\left(p/p_0 \right) \ll \left( x - x_0 \right)^2$

\begin{equation}\label{eq:leRoux14}
\begin{split}
f_0(x,p) \propto e^{ -\frac{1}{2} \left[  \left|  \frac{U_0}{\kappa_0^I}(x-x_0) \right|  \sqrt{ 1 + 4\tau_{\kappa_0^I} \left[ (q/2)^2/\tau_{D_0^I} + 1/\tau_{esc} \right] } - \frac{U_0}{\kappa_0^I}(x - x_0) \right] / \kappa_0^I } \left( \frac{p}{p_0} \right)^{-\frac{q}{2}}
\end{split}
\end{equation}

where $\tau_{\kappa_0^I} = \kappa_0^I / U_0^2$ is the diffusion time scale and $\tau_{D_0^I} = 1 / D_0^I$ is the time scale for second-order Fermi acceleration. In \eq{eq:leRoux13} and \eq{eq:leRoux14} we let $U_E^I = 0$, thus removing the effects of average parallel reconnection electric field in the mixed-derivative transport term and in spatial advection ($U_0^I = U_0$) in \eq{eq:leRoux10}, but not in second-order Fermi acceleration. The mixed-derivative transport term was found to counteract spatial peak formation by forming a plateau in the accelerated particle distribution contrary to observations \citep{Zank2014particle}. Both solution limits \eq{eq:leRoux13} and \eq{eq:leRoux14} indicate that, because energetic particle diffusive transport occurs against the solar wind flow upstream of the particle injection point at $x = x_0 \left(x < x_0 \right)$, the particle distribution decays exponentially with increasing upstream distance from the injection point (thus no spatial peak in the particle distribution). Since diffusive transport unfolds in the direction of the solar wind flow downstream of the injection point $\left(x > x_0 \right)$ during acceleration, the particle distribution increases exponentially with increasing distance downstream when sufficiently close to the injection point because then limit \eq{eq:leRoux13} applies. However, sufficiently far downstream of the injection point, the particle distribution at lower energies decays first with increasing distance because then limit \eq{eq:leRoux14} is applicable. The decay at higher energies occurs progressively further downstream of the particle source when \eq{eq:leRoux14} becomes applicable at those distances. Thus, peaks form in the accelerated downstream distribution that shifts increasingly to larger distances downstream with increasing particle energy. Consider the accelerated particle spectra. Close to the injection point the particle spectra form power laws steeper than $f_0(p) \propto (p/p_0)^{-q/2}$ at most energies above the injection energy $(p > p_0)$ because \eq{eq:leRoux13} is valid. With increasing distance from the injection point expression in \eq{eq:leRoux13} holds progressively at increasingly high particle energies only while at lower energies (\eq{eq:leRoux14}) applies where the spectrum approaches the harder power law $f_0(p) \propto (p/p_0)^{-q/2}$ for a growing energy interval. Thus, with increasing distance downstream of the injection point the accelerated particle spectrum becomes increasingly hard on average while assuming a more exponential character as it bends over more strongly at lower energies. Inspection of \eq{eq:leRoux13} also reveals that more efficient particle escape results in a steeper spectrum \citep{Zhao2018unusual} and a larger spatial diffusion coefficient produces a harder spectrum as particles sample more SMFRs in a given time interval.

\subsubsection{Second Order Fermi SMFR Acceleration}

\begin{figure}
    \centering
    \includegraphics[width=0.45\textwidth]{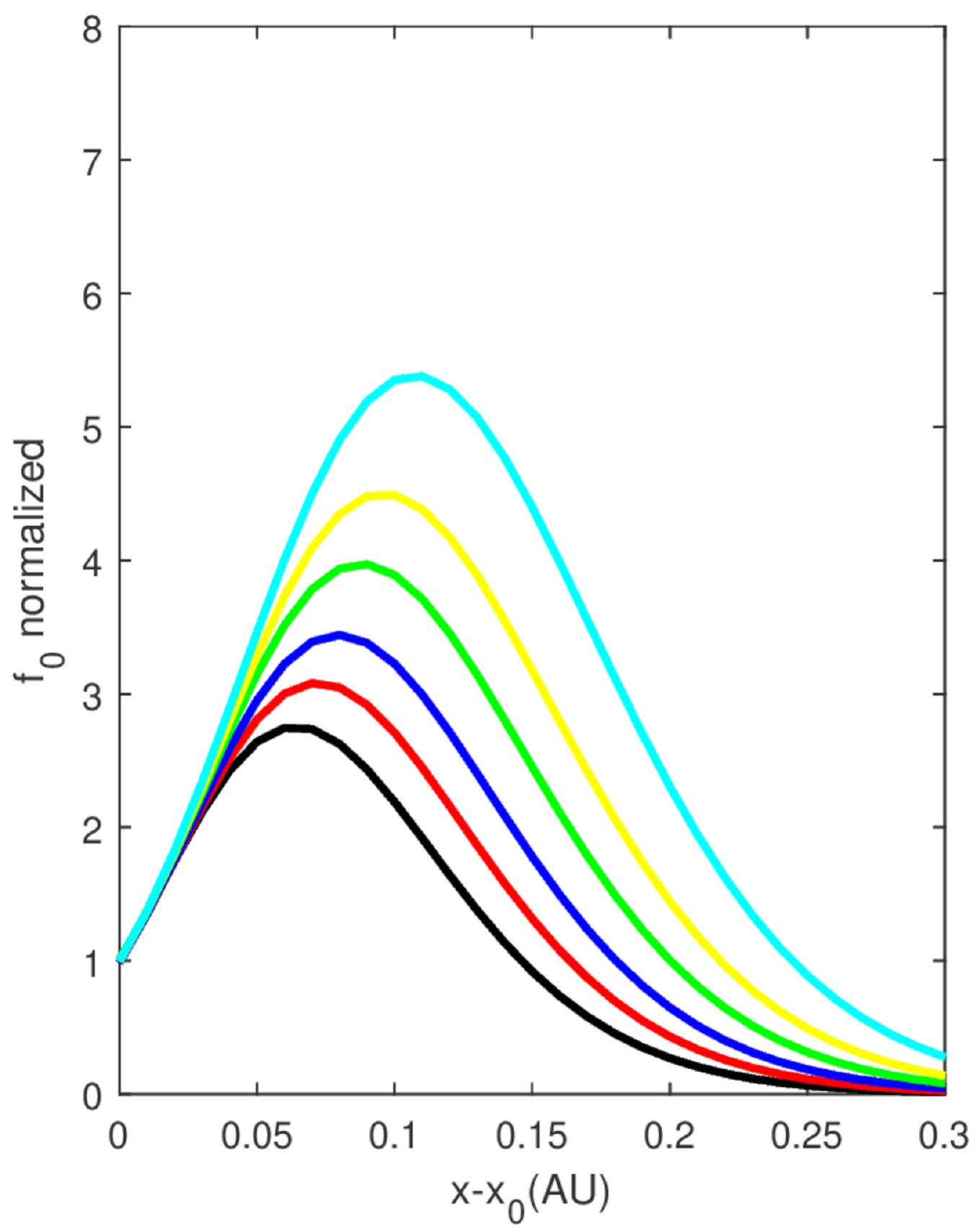}
    \includegraphics[width=0.45\textwidth]{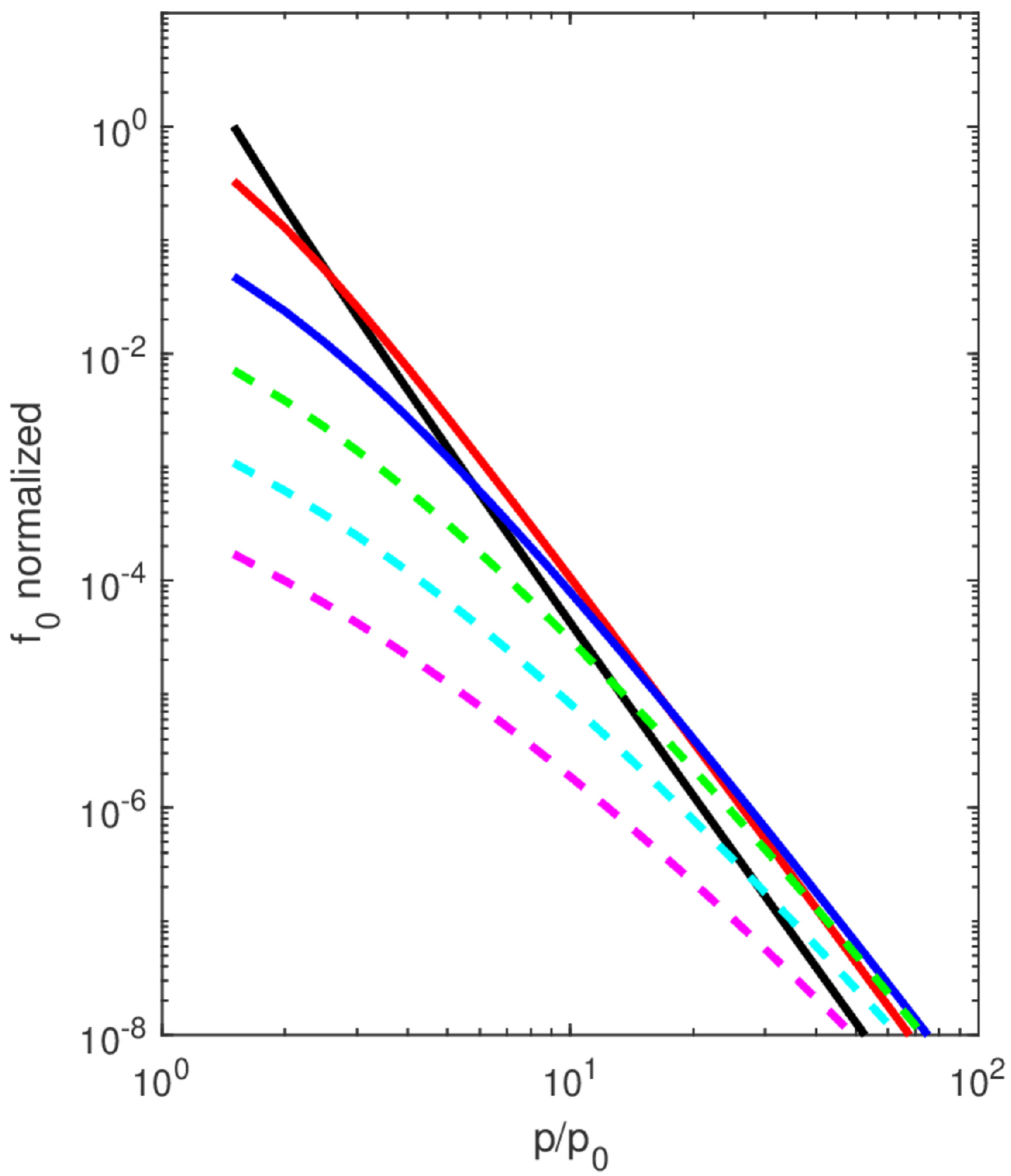}
    \caption{The 1D steady-state analytical solution (\eq{eq:leRoux11}) for energetic proton 2nd order Fermi acceleration by SMFRs in a uniform SMFR region in the vicinity of 1 AU (reproduced from \citet{leroux2019modeling}). The solution combines second order Fermi acceleration due the variance in the SMFR compression rate and in the SMFR incompressible parallel shear flow. Left panel: The direction-averaged proton distribution function $f_0(x,p)$ as a function of distance in AU relative to the particle injection position $x_0$ ranging from $x - x_0 = 0-0.3$ AU in the downstream direction (in the direction of the solar wind flow) for the particle energies 144 keV (black), 256 keV (red), 0.44 MeV (blue), 0.81 MeV (green), 1.44 MeV (yellow), and 3.31 MeV (cyan). The energy intervals were chosen to fall inside the energy intervals in the observed energetic flux enhancements at 1 AU in Fig. 10 of \citet{khabarova2017energetic}. The different curves for $f_0(x)$ were normalized to a value of 1 at $x-x_0 = 0$ AU to reveal the amplification factor of $f_0$ from the particle injection location to the peak in $f_0(x)$ further downstream for each particle energy. Since the differential intensity as a function of kinetic energy T is $j_T(T) = p^2 f_0(p)$, the amplification factor for $f_0$ also serves as the amplification factor for $j_T$. Right panel: Normalized $f_0(x,p)$ as a function particle momentum $p$ in the solar wind frame normalized to $p_0$ (the injection momentum). The spectra are displayed for the following values of $x-x_0$: 0 AU (solid black), 0.05 AU (solid red), 0.1 AU (solid blue), 0.15 AU (dashed green), 0.2 AU (dashed cyan), and 0.25 AU (dashed magenta). The curves were multiplied with the same factor so that the black curve has a value of one at the minimum momentum $p/p_0 = 1.5$. At $p/p_0 = 1$ the proton kinetic energy $T\approx 1$ keV while at the maximum momentum $p/p_0=100$, $T \approx 10$ MeV.}
    \label{fig:leRoux4}
\end{figure}

\cite{Zhao2018unusual} and \cite{adhikari2019role} had great success in reproducing the observed features of accelerated ions in SMFR regions at 1 AU and 5 AU in the equatorial plane using an analytical steady-state solution of their Parker transport equation in which 1st order Fermi SMFR compression acceleration appears to be the dominant acceleration mechanism, but without considering second-order Fermi SMFR acceleration or connecting the acceleration and transport time scales to SMFR properties. In \fig{fig:leRoux4} we illustrate steady-state solution (\eq{eq:leRoux11}) in the limit where second-order Fermi SMFR acceleration due to the variance in the SMFR compression and parallel shear flow is the dominant acceleration mechanism. The left panel in \fig{fig:leRoux4} shows the spatial variation in the accelerated proton distribution function amplification factor for different particle energies downstream of the injection point at $x-x_0 > 0$ by choosing an amplification factor of one for all energies at the injection location. The theoretical maximum amplification factor varies between $\sim 2.8-5.4$ for proton energies in the range $0.144 - 3.31$ MeV, thus increasing with energy. Figure 10 of \citet{khabarova2017energetic}, based on a superposed epoch analysis of energetic ion flux enhancements in the vicinity of 126 thin primary reconnecting current sheet events at 1 AU, suggests that the average intensity maximum amplification factor of energetic ion flux varies between $\sim 4.5-7.5$ in the energy range $0.112 - 4.75$ MeV (LEMS30 detector of EPAM instrument on the ACE spacecraft), and between $\sim 3.5-5.5$ in the energy range $0.066 - 4.75$ MeV (LEMS120 detector of EPAM) with the largest amplification factor occurring at the highest energies. The latter range of maximum amplification factors agrees best with the analytical results. The analytical solution also predicts a systematic shift in the position of the maximum amplification factor. It varies from $\sim 0.05-0.1$ AU from low to high particle energies downstream of the injection point that does not appear to be present in the observations averaged over many events in Fig. 10 of \citet{khabarova2017energetic}. Although such a shift does not appear to be present in Fig. 10 of \citet{khabarova2017energetic}, it is present in energetic ion observations of SMFR acceleration behind an interplanetary shock from the Ulysses spacecraft at $\sim 5$ AU reported by \cite{Zhao2018unusual}, and in anomalous cosmic ray observations from the Voyager 2 spacecraft behind the solar wind termination shock \citep{Zank2015particle}.

In \fig{fig:leRoux4}, right panel, we present the corresponding accelerated energetic proton spectra from our analytical solution for second-order Fermi acceleration at the particle injection point (solid black curve), and at increasing distances further downstream of the injection location: 0.05 AU  (solid red curve), 0.1 AU (solid blue curve), 0.15 AU (dashed green curve), 0.2 AU (dashed cyan curve), and 0.25 AU (dashed magenta curve). The spectra are normalized so that the distribution function at the injection point has a value of one at the lowest momentum shown which is $p/p_0 = 1.5$, where $p_0$ is the injection momentum specified to be at a suprathermal proton kinetic energy of $T\approx 1$ keV. At the particle injection location the accelerated spectrum is close to a power law, being slightly softer than $f_0(p) \propto p^{-5}$ (in terms of differential intensity it is somewhat harder than $j_T(T) \propto T^{-1.5}$) except at the lowest momenta where the spectrum steepens somewhat. Inspection of the spectral evolution with increasing distance downstream of the injection point reveals that the spectra become progressively harder and more exponential so that spectra at low energies are considerably harder compared to high energies. If one would fit a power law to the exponential spectrum at 0.2 AU downstream of the injection location (dashed cyan curve), the spectrum would be approximately a $f_0(p) \propto p^{-4} \left( j_T(T) \propto T^{-1} \right)$ above $\sim 100$ keV ($p/p_0 > 10$). This basic trend of spectral hardening and increasing exponential nature of accelerated proton spectra produced by SMFRs with increasing distance inside the SMFR region is consistent with SMFR acceleration events at 1 AU reported by \cite{adhikari2019role}. The variation in the power-law index through the SMFR region from $\sim -1.5$ to $\sim -1$ for particle energies $\sim 100$ keV $< T < 1$ MeV in the second event discussed by \cite{adhikari2019role} is close to the result reported here. A similar hardening trend in the energetic particle spectra through an SMFR at $\sim 5$ AU was detected in Ulysses data as reported by \cite{Zhao2018unusual}.

We conclude that one can potentially reproduce the observed flux amplification of energetic ions as a function of particle energy and the evolution of the accelerated spectra through SMFR regions at 1 AU by focusing solely on second-order Fermi SMFR acceleration of energetic ions in response to statistical fluctuations (variance) in the compression and the parallel shear flow in SMFRs, and we found that this can be accomplished with reasonable SMFR parameters (for more detail, see \cite{leroux2019modeling}). Based on the SMFR parameters that we used we found that stochastic acceleration by the variance in the parallel reconnection electric field is the dominant second-order Fermi acceleration mechanism. Unfortunately, we were unable to model particle acceleration for this mechanism using the solution of \eq{eq:leRoux11} because this analytical solution only holds for $D_0^I = D_{pp}^I / p^2 $ being a constant while for this mechanism $D_0^I$ depends on particle speed. Second-order Fermi acceleration involving the variance in the SMFR parallel shear flow was the second most efficient, while second-order Fermi acceleration due to the variance in SMFR compression was the least efficient. However, due to our limited knowledge of the SMFR parameters that enter into the acceleration expressions, and because of limitations of the analytical solutions, it is difficult to draw definitive conclusions about the ranking of the different second-order Fermi acceleration mechanisms associated with the variance in SMFR fields. Further progress requires intensifying data analysis of SMFR acceleration events, while at the same time increasing the sophistication of the solutions. 

\subsubsection{First Order SMFR Fermi Acceleration}

\begin{figure}
    \centering
    \includegraphics[width=0.45\textwidth]{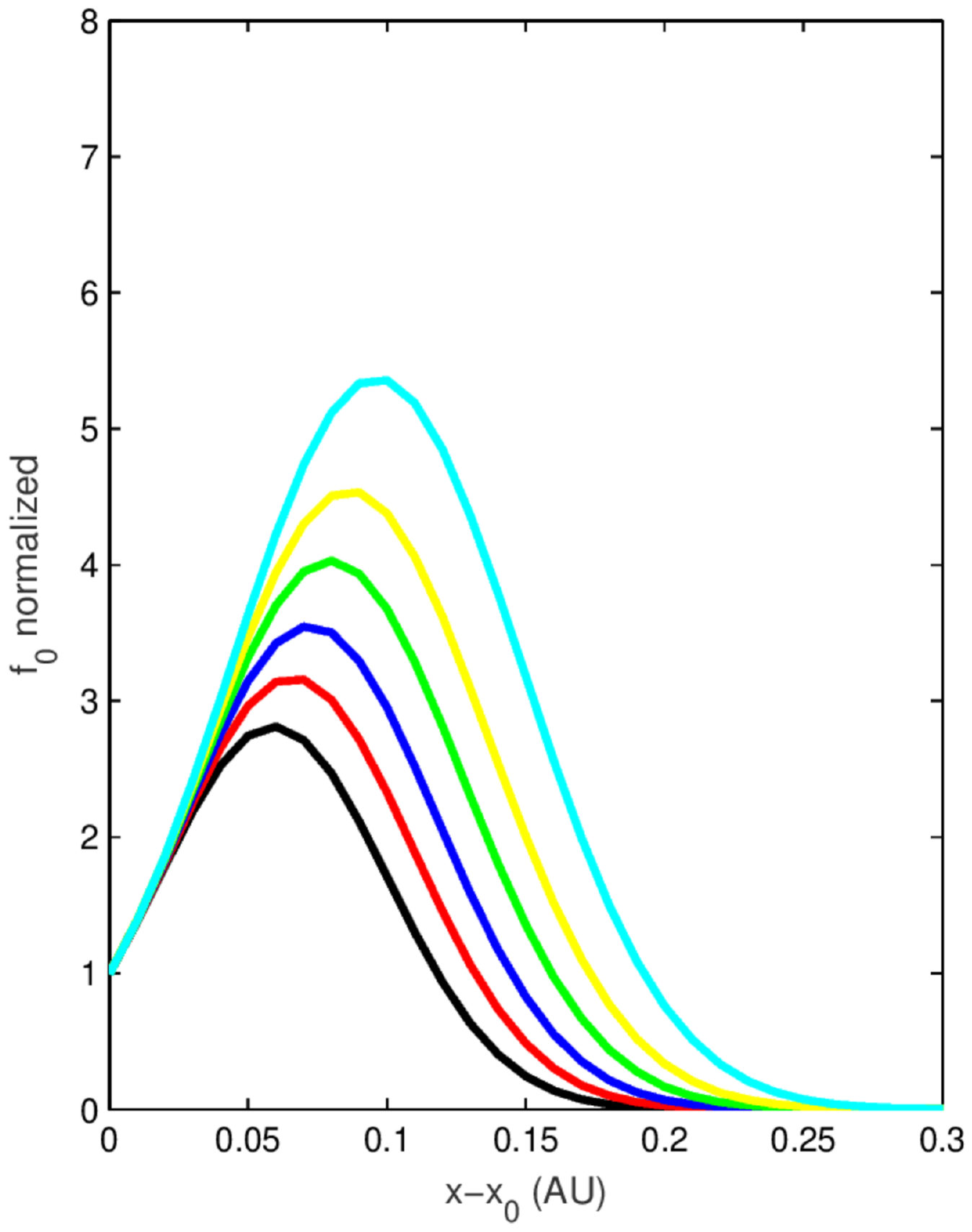}
    \includegraphics[width=0.45\textwidth]{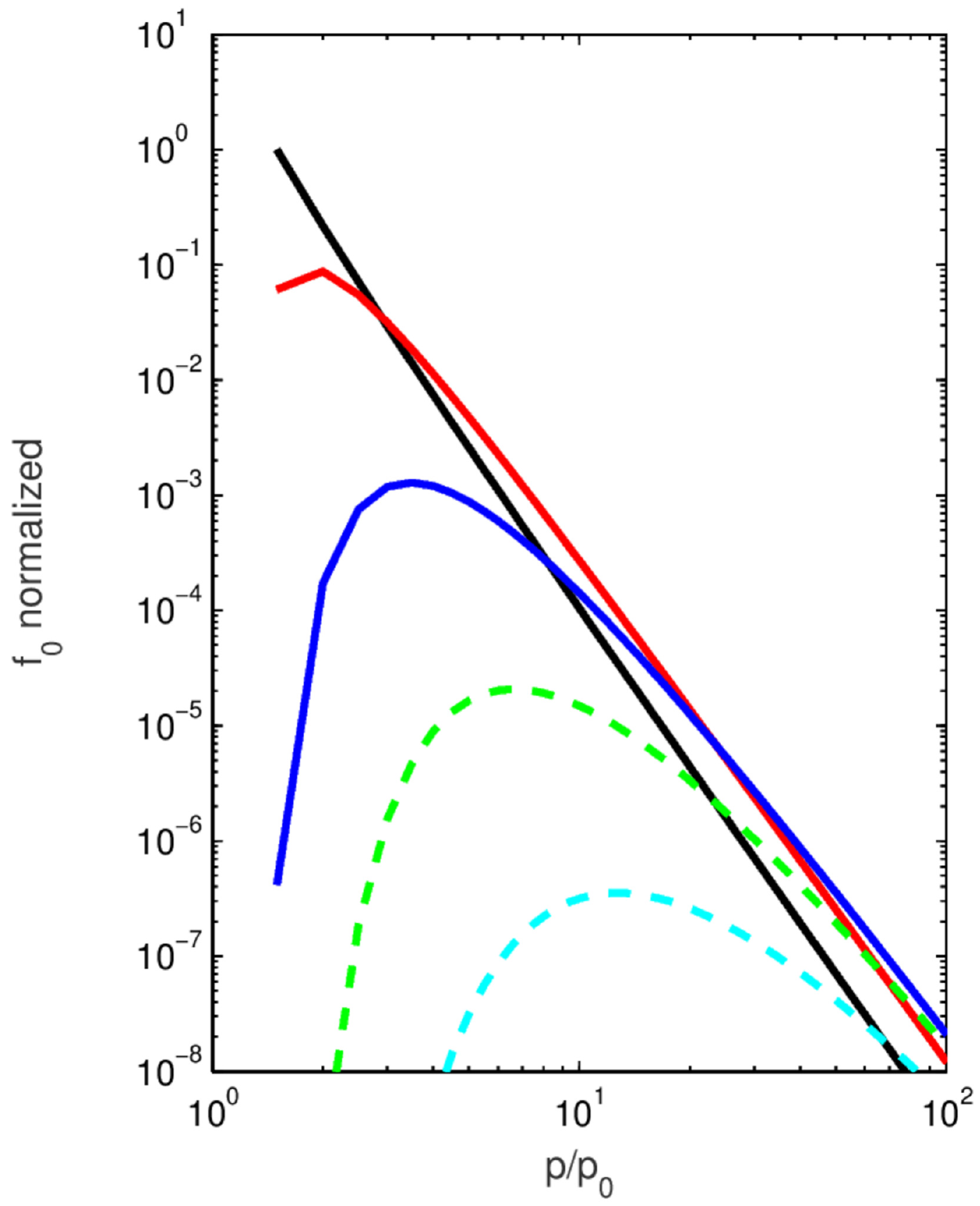}
    \caption{The 1D steady-state solution for energetic proton first order Fermi acceleration or mean SMFR compression acceleration in a uniform SMFR region in the vicinity of 1 AU (\eq{eq:leRoux11}) (reproduced from \citet{leroux2019modeling}). Left panel: The same format as \fig{fig:leRoux4}, left panel. Right panel: Same format as \fig{fig:leRoux4}, right panel.}
    \label{fig:leRoux5}
\end{figure}

Consider \fig{fig:leRoux5} where we show an analytical solution of \eq{eq:leRoux11} in the limit where first-order Fermi acceleration of energetic ions in response to the mean compression rate of SMFRs in an SMFR region at 1 AU is the dominant acceleration mechanism. Inspection of the results in the left panel for the spatial variation in the amplification factor in the accelerated particle distribution downstream of the particle injection point shows that they are remarkably similar to the results in \fig{fig:leRoux4} for second-order Fermi acceleration. In other words, with both first and second-order second Fermi SMFR acceleration the observed enhanced energetic ion flux in SMFR regions at 1 AU in Fig. 10 of \citet{khabarova2017energetic} were reproduced reasonably well. The SMFR parameters specified in the first-order Fermi solution closely follow those used in the second-order Fermi solution, except for the characteristic cross-section of SMFRs that was reduced by a factor of four. However, the reduced value falls within the range of possibility given the little that we know of the statistics of SMFR parameters in SMFR acceleration regions in the solar wind, thus accentuating the need for more detailed analysis of SMFR properties in SMFR acceleration regions. As can be seen in \fig{fig:leRoux5}, right panel, similar to the results for second-order Fermi acceleration, the modelled accelerated spectra for first-order Fermi acceleration are power laws at the particle injection point, exhibiting the same rollover trend qualitatively at lower particle energies downstream of the injection point (see also \cite{Zhao2018unusual,adhikari2019role}). Quantitatively, however, the spectral rollover trend at lower particle energies and overall increasing spectral hardening downstream of the injection location is notably stronger in the case of first-order Fermi acceleration due to the cutoff in the downstream spectrum at the injection momentum. This illuminates a key difference between the first Fermi acceleration solution, where all the particles that arrive downstream of the injection point have been accelerated systematically to momenta larger than the injection momentum to form the low-energy cutoff at the injection momentum, and the second-order Fermi solution where particles arriving downstream experienced stochastic acceleration which lowers the probability for a low-energy cutoff at the injection momentum. This predicted difference in the spectral evolution for the two acceleration mechanisms downstream of the injection point might potentially help identify the dominant operating SMFR acceleration mechanism in observations. Based on the evolution of the spectral power-law index through the SMFR region in SMFR acceleration event two of \cite{adhikari2019role}, the event with spectral power-law indices closest to our results, the less strong spectral hardening in the second-order Fermi acceleration case is closer to the observed hardening trend. More SMFR acceleration events need to be studied before conclusions can be drawn with confidence. The success of our SMFR acceleration results for both first-order Fermi acceleration (see also \cite{Zhao2018unusual, adhikari2019role}) and second-order Fermi acceleration is partially due to the term for particle escape from the SMFR region which ensured steepened accelerated particle spectra with more realistic slopes. The particle escape term reflects the need for more sophisticated solutions that specify finite boundaries for the SMFR acceleration region and allow for multi-dimensional transport. It must be pointed out that the need for steeper accelerated spectra in the solution can partly be the result of modelling particle acceleration in the test particle limit. The considerable pressure in the accelerated test particle spectra indicates that the energy exchange between the particles and SMFRs should be modelled self-consistently, thus contributing also to steeper accelerated particle spectra \citep{leRoux2016combining,leRoux2018self}.

\section{Numerical modeling of magnetic reconnection, formation of coherent structures, their dynamics and corresponding properties of turbulence. }
\label{sect:unicalturb}

Magnetic reconnection is a fundamental process that occurs ubiquitously in both space and laboratory plasmas. This process consists of an intense magnetic energy release that heats and accelerates particles \citep{drake2006electron,hesse2016theory}. One way in which reconnection may be triggered is via the tearing instability of thin current sheets. In a recent work \citep{PucciEA18}, it has been discussed the critical aspect ratios for the fastest tearing mode growth times, required to reach values comparable to ideal time-scales for current sheet equilibria that are different from the Harris current sheet.

It is now widely accepted that magnetic reconnection takes place in turbulent environments, where an energy cascade, from large to smaller spatial scales, is present \citep{matthaeus2015intermittency,bruno2016turbulence}. The interplay of magnetic reconnection and plasma turbulence is hence decisive for assessing the role of this process in natural systems and can include the effects of turbulence on magnetic reconnection and, vice-versa, the influence of reconnection on turbulence \citep{matthaeus2011needs,karimabadi2013coherent}. For example, it has been observed that turbulence can increase the reconnection rate  \citep{matthaeus1986turbulent,smith2004hall, lapenta2008self, loureiro2009turbulent, lazarian1999reconnection, kowal2011mhd, kowal2012particle, LazarianEA15}. Analogously, plasma jets generated by reconnection can also supply energy for sustaining the turbulent cascade \citep{lapenta2008self,pucci2017properties,cerri2017kinetic,pucci2018generation}.  All the above, recent studies suggest that the intense current sheets naturally produced by turbulence, and that represent the boundaries of magnetic islands, might be the preferred sites of magnetic reconnection. For this reason, reconnection is a crucial ingredient of turbulence itself \citep{matthaeus1986turbulent,servidio2010statistics}. These small-scale coherent structures are also thought to be the sites of inhomogeneous dissipation, where kinetic effects and plasma heating are concentrated \citep{Servidio2011magnetic, osman2011evidence, servidio2012local, greco2012inhomogeneous, osman2012intermittency, wu2013intermittent, servidio2015kinetic, wan2015intermittent}, together with local accelerations of energetic particles \citep{comisso2018,comisso2019interplay,Pecora18JPP}.

Due to the strong nonlinearity of complex, multi-scale plasmas, the adoption of high-resolution numerical simulations is mandatory. Therefore, fro practical reasons, considerable 
progress in the area of reconnection and turbulence has been made in the context of reduced dimensionality models that have an ignorable coordinate, or a weak dependency along the magnetic field as in the case of Reduced MHD \citep{Rappazzo17} (see \citet{oughton2017reduced} for a recent review). Much of the progress in 3D turbulent reconnection has been either experimental \citep{BrownEA06} or in a 3D numerical setup that is in effect nearly-2D \citep{KowalEA09,Daughton11}. It is obvious that the full 3D case is substantially more complex and much less understood, both theoretically \citep{Schindler:etal:1988JGR,Priest:Pontin:2009PhPl} and from the point of view of simulations \citep{Dmitruk06,BorgognoEA05, lapenta2015secondary, lapenta2016multiscale}. 

Weakly 3D cases, often studied in Reduced MHD, have interesting properties \citep{oughton2017reduced}. For weakly 3D setups, it has been confirmed that some 2D-like  effects persist \citep{RappazzoEA07,Daughton11}. There are however some fundamentally 3D effects that occur even in Reduced MHD.  An example is 
in the study of the geometry of current sheets and the trajectories of field lines that pass through them \citep{ZhdankinEA13,WanEA14,RappazzoEA17-current}.
In 2D rectilinear geometry the presence of a current sheet in one plane guarantees that a similar current sheet will be present at all other planes at different distances along the 
ignorable coordinate. Furthermore the central field line,
the X-line, pass through each of these identical current sheet.  
This is in essence a trivial restatement of two-dimensionality. 
However in the weakly 3D Reduced MHD case \citep{WanEA14} the
behavior of field lines passing through current sheets do not have this property. Instead, a field line passing through an X-point 
central to a current sheet on one plane, will likely depart nearby current sheets on nearby planes. Consequently X-points are usually found at varying distances relative to what had been their coincident current sheet. But of course reconnection in MHD only occurs when the X-point is co-located with a strong current, so this meandering has a leading order effect on where reconnection occurs and how strong it will be \citep{WanEA14}. This inherently 3D effect is further complicated by the fact that current sheets in reduced MHD have a finite extent along the guide field \citep{ZhdankinEA13}. 
Further study also showed \citep{RappazzoEA17-current} 
that reduced MHD current sheets tend to 
wander in the transverse directions {\it diffusively}, while field lines also wander randomly, but somewhat independently of
the current sheets. This is clearly a much more complex scenario 
than the 2D case, even though Reduced MHD is only weakly 3D.
This complexity will however
have potentially major impact on physical phenomena such as nanoflares~\citep{RappazzoEA07,GomezEA00} .

In general, it is well established that turbulence plays an essential role in accelerating the reconnection process
both in 2D \citep{matthaeus1986turbulent}
and in 3D \citep{lazarian1999reconnection}. 
While the physics of reconnection revealed in the 2D can be applied in some circumstances to 3D, it is likely also that there are essential physical effects that occur in reconnection
only in a strongly 3D systems, not the least of which is the 
very nature of reconnection itself in a full 3D representations (see, e.g., \cite{Priest:Pontin:2009PhPl}). 
These challenging problems have not been included in any complete way in the present review, where we concentrated, instead, on the progress in understanding reconnection in a turbulent environment which is very well magnetized and anisotropic.

\begin{figure}
    \centering
    \includegraphics[width=\textwidth]{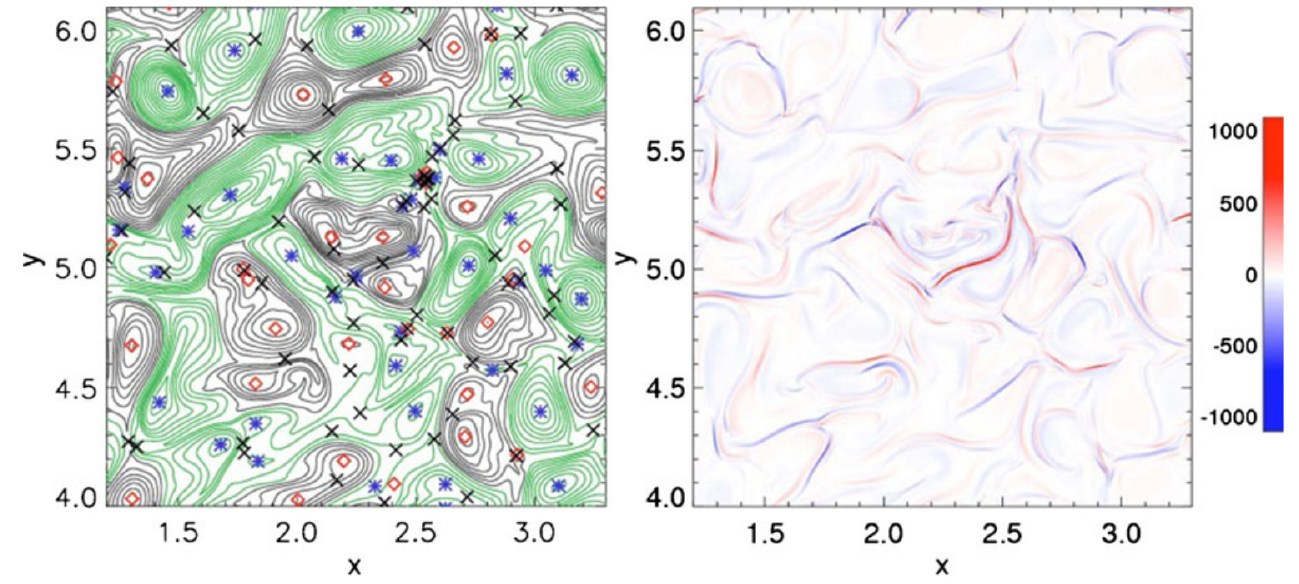}
    \caption{(Color online). 2D MHD simulation result, showing the concurrency of current sheets and magnetic reconnection sites. Left: Contour plot of the magnetic potential $a$ with the position of all the critical points: O-points (blue stars for the maxima and red open diamonds for the minima) and X-points (black), from a small section of a high-resolution 2D MHD simulation. Right: Shaded contour plot of the current density $j$ in the same region. In between magnetic islands, intense current density peaks are present. Several reconnection sites are observed associated with current sheets. (reproduced from \citet{servidio2010statistics}).}
    \label{fig:servidio2010}
\end{figure}

First numerical attempts to model magnetic reconnection were based on an MHD approach, where the plasma is treated as a collisional magneto-fluid. Turbulence tends to dynamically generate relaxed regions --where non-linearities are depleted-- separated by sheets of strong gradients (i.e. current) \citep{matthaeus2015intermittency}, where magnetic reconnection can occur \citep{Matthaeus80,carbone90, retino2007insitu, servidio2009magnetic, rappazzo2010shear, servidio2010statistics, servidio2011statistical}. Figure \ref{fig:servidio2010} shows the in-plane magnetic field lines with X-points and O-points identified (left), and the associated current sheets (right), for a 2D magnetohydrodynamics (MHD) simulation \citep{servidio2010statistics}. A sea of interacting magnetic island, with boundaries characterized by intense current sheets, is found. Current sheets are often co-located with reconnection sites. 

The {\it in-situ} observations conducted with the Cluster mission have revealed that physical ingredients beyond the pure MHD description --namely the Hall and the kinetic terms-- can have a significant role in characterizing the magnetic reconnection process  \citep{oieroset2001insitu, vaivads2004structure,  retino2007insitu, sundkvist2007dissipation}. Indeed, their effects become important when characteristic dynamical scales are comparable with the ion skin depth $d_i =c/\omega_{pi}$ ($\omega_{pi}$ being the ion plasma frequency). This motivated huge numerical efforts to include these ingredients and a variety of numerical simulations, ranging from Hall-MHD to Vlasov-Maxwell, have been performed.

\begin{figure}
    \centering
    \includegraphics[width=0.5\textwidth]{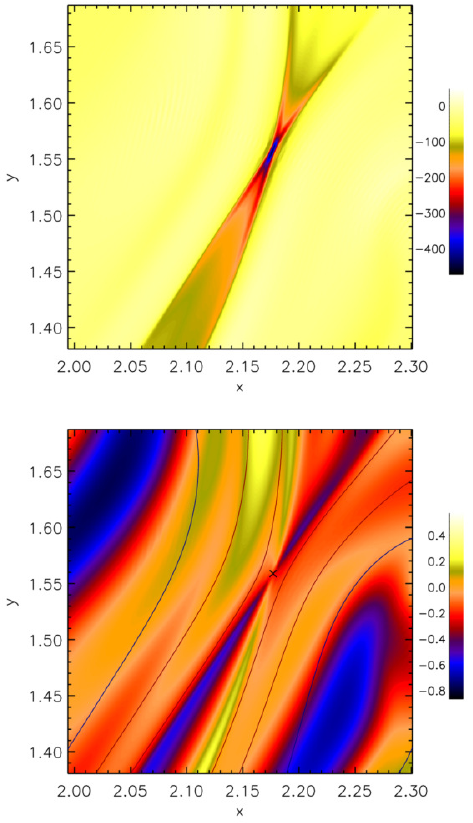}
    \caption{(Color online). 2D Hall-MHD simulation result, indicating the typical quadrupolar structure of the reconnection site due to the Hall term. Top: Contour plot of the out-of-plane component of the current $j_z$ in a sub-region of the simulation box. The bifurcation of the sheet and the typical structure of a reconnection region are clearly visible. Bottom: A contour plot of the out-of-plane component of the magnetic field $b_z$, in the same sub-region. The magnetic flux $a$ is also represented as a line contour. A quadrupole in the magnetic field can be identified, revealing the presence of Hall activity (reproduced from \citet{donato2012reconnection}).}
    \label{fig:donato2012}
\end{figure}

The Hall term, which brings the dynamics of whistler waves and dispersive effects into the system, increases the reconnection rate \citep{Ma:Bhattacharjee:2001, birn2001geospace, smith2004hall, lu2010features, papini2019hall}, while the distribution of reconnection rates become broader than the MHD case \citep{donato2012reconnection}. At variance with the pure MHD case, the magnetic energy is more frequently released through explosive, catastrophic events, that lead to fast magnetic reconnection onset \citep{cassak2005catastrophe, cassak2007onset}. The topology of the current sheets also changes and a quadrupolar structure of the out-of-plane magnetic field is observed (Fig. \ref{fig:donato2012}), as expected from theory \citep{sonnerup1979}. 

Several efforts have been performed to understand at which extent MHD and Hall-MHD models, based on a collisional closure, properly describe turbulence in almost collisionless systems where the collisional closure may not be valid (e.g., \citet{pezzi2017colliding,perrone2018fluid,gonzalez2019turbulent,papini2019hall}). In general, Hall MHD simulations retain turbulence properties (spectral features, intermittency, reconnection...) up to the sub-proton range of wavenumbers. However, together with the presence of the Hall physics, genuinely kinetic effects are also significant at scales $l\sim d_i$ \citep{valentini2008cross, valentini2011new, valentini2011short, howes2008kinetic, parashar2009kinetic, schekochihin2009astrophysical, tenbarge2013collisionless, TenBargeEA13-sheets, MatteiniEA13, franci2015solarwind, vasconez2015kinetic, valentini2016differential, parashar2016propinquity, pucci2016from, cerri2017kinetic, cerri2017reconnection, pezzi2017revisiting, pezzi2017colliding, pezzi2017turbulence, groselj2017fully, franci2018solarwind, parashar2018dependence, hellinger2019turbulence,franci2020modeling, califano2020electron}. Indeed, since magnetic reconnection onsets in a weakly-collisional plasma, the plasma is free to explore the full phase-space and can exhibit non-equilibrium features, such as temperature anisotropy, rings, beams of accelerated particles along or across the local magnetic field \citep{osman2011evidence, osman2012intermittency, servidio2015kinetic}. Both hybrid, where electrons are assumed to be fluid while protons are kinetic, and fully-kinetic models have been employed to gain insights on the magnetic reconnection onset at kinetic scales \citep{birn2001geospace, mangeney2002numerical, zeiler2002threedimensional, pritchett2008collisionless, Scudder08, lu2010features, zenitani2011new, greco2012inhomogeneous, aunai2013electron, wu2013intermittent, karimabadi2013coherent,  leonardis2013identification, valentini2014hybrid, wan2015intermittent, lapenta2015secondary, lapenta2016multiscale, shay2018turbulent, lapenta2020local}. Kinetic simulations are usually employed within two different numerical approaches: Particle-In-Cell (PIC) and Eulerian Vlasov codes. Since the computational cost of PIC algorithms is in general smaller with respect to low-noise Eulerian codes, PIC codes are able to capture the full plasma dynamics (including electron scales, although with unrealistic electron to ion mass ratio). However, at variance with noise-free Eulerian algorithms, PIC codes are affected by statistical noise and may fail in providing a clean description of small-scale fluctuations and particle distribution functions in phase space. Very recently, first Eulerian fully-kinetic codes have been implemented to describe plasma dynamics at electron scales without noise \citep{schmitz2006darwin, umeda2009twodimensional, umeda2010full, tronci2015neutral, delzanno2015multi, umeda2016secondary, umeda2017nonMHD, ghizzo2017vlasov, juno2018discontinuous,  roytershteyn2019numerical, skoutnev2019temperature, pezzi2019vida}.

The introduction of the kinetic physics leads to several novelties regarding magnetic reconnection \citep{burch2016electron, torbert2018electronscale,shuster2019MMS,chen2019electron,jiang2019role}. In the diffusion region, collisionless and collisional plasma processes affect the change of magnetic field topology. Within a general fluid framework, non-ideal mechanisms are modelled through a resistive term, which can take into account both electron-ion collisions as well as the presence of an anomalous resistivity produced by wave-particle interactions and/or turbulence. Owing to poor collisionality, non-ideal effects are induced by collisionless processes, such as electron pressure tensor gradients and electron inertia effects. It is thus decisive to describe the system through kinetic model to evaluate the role of these terms in shaping magnetic reconnection.

\begin{figure}
    \centering
    \includegraphics[width=0.8\textwidth]{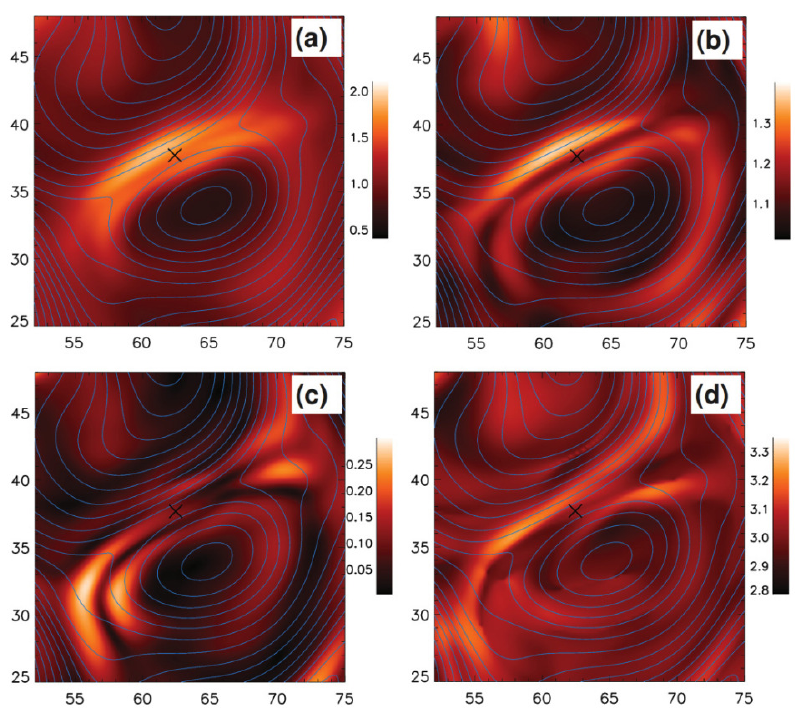}
    \caption{(Color online) Hybrid Eulerian Vlasov simulation results, indicating the concentration of kinetic effects close to current sheets. Contour plots of several indicators of non-Maxwellian features in the proton velocity distribution function, in the vicinity of an intense current sheet, identified through the PVI method (black cross in all panels): (a) $\epsilon$ parameter, that evaluates the local deviation of the distribution function from the associated Maxwellian (in per cent); (b) Proton temperature anisotropy $T_{\perp}/T_{||}$, being the parallel direction evaluated with respect to the local magnetic field; (c) heat flux, evaluated as the third-order moment of the proton distribution function; and (d) kurtosis (fourth-order moment) of the proton distribution function. (reproduced from \citet{greco2012inhomogeneous}).}
    \label{fig:greco2012}
\end{figure}

Within the hybrid framework, it has been reported that kinetic effects are concentrated close to intense current sheets \citep{valentini2007hybrid, servidio2012local,greco2012inhomogeneous, valentini2014hybrid, servidio2015kinetic, valentini2016differential, valentini2017transition}. The role of alpha particles in characterizing kinetic scales turbulence has been analyzed in the detail showing that both protons and alpha particles are not in thermal equilibrium and manifest a preferentially perpendicular heating \citep{gary2003consequences, hellinger2005alfven, ofman2007two, ofman2011hybrid, perrone2011role, perrone2013vlasov, perrone2014analysis, perrone2014generation, maneva2013turbulent, maneva2014regulation, maneva2015relative, maneva2018generation, valentini2016differential}. Figure \ref{fig:greco2012} displays four indicators of non-Maxwellian features in the proton distribution function in the proximity of a current sheet, identified through the Partial Variance of Increments (PVI) method \citep{Greco09}. Recently a similar proxy of non-Maxwellianities based on the entropy density has been adopted by \citet{liang2019decomposition,liang2020kinetic,pezzi2020dissipation}. The $\epsilon$ parameter (panel (a)), which locally quantifies deviations of the proton VDF with respect to the associated Maxwellian distribution clearly shows a broad region of significant non-Maxwellianity surrounding the X-point. Out-of-equilibrium features emerge as temperature anisotropy (b) as well as higher VDF moments such as heat fluxes (c) and kurtosis (d).  Note also that the heat flux is peaked in the exhaust region, this resembling the presence of outflows.

\begin{figure}
\begin{minipage}{0.48\textwidth}\centering
\includegraphics[width=\textwidth]{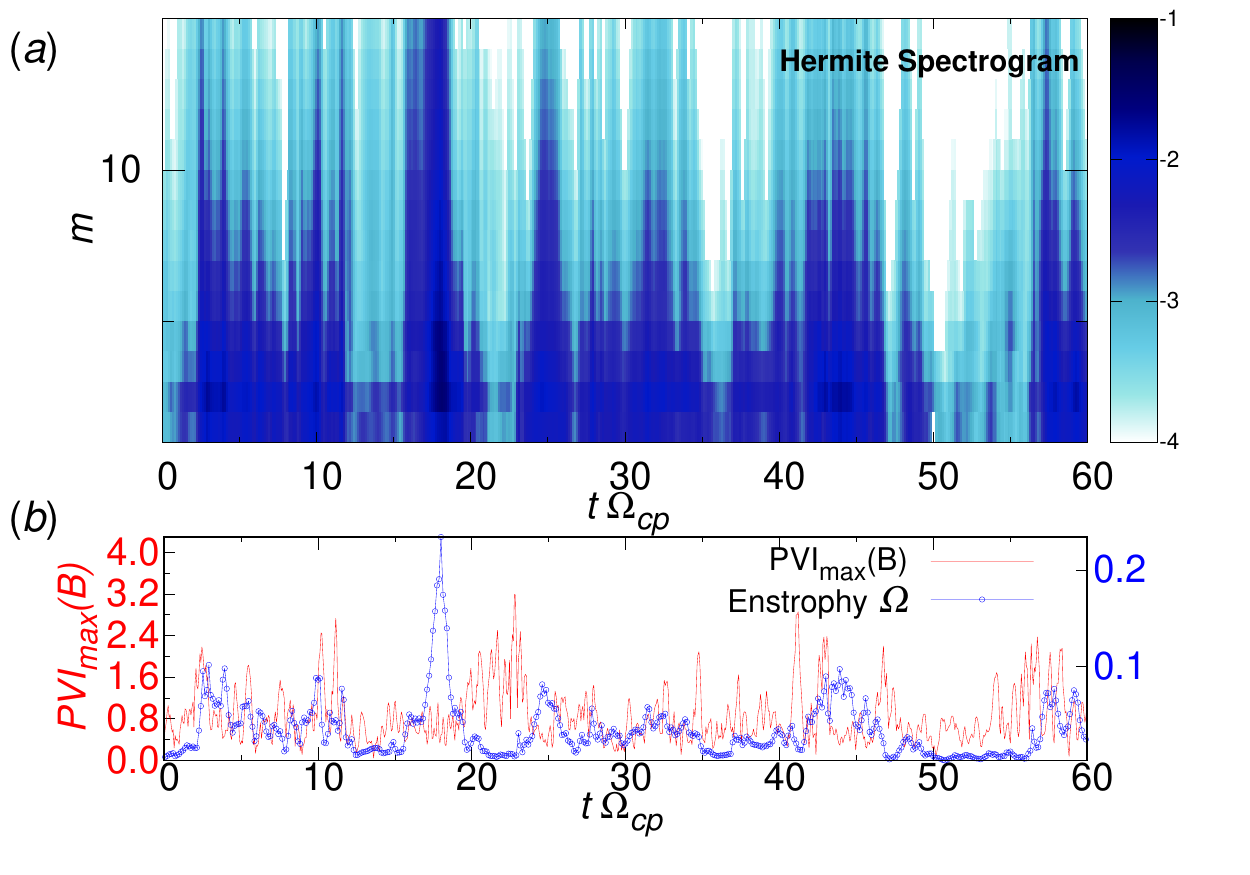}
\end{minipage}
\hfill
\begin{minipage}{0.48\textwidth}\centering
\includegraphics[width=\textwidth]{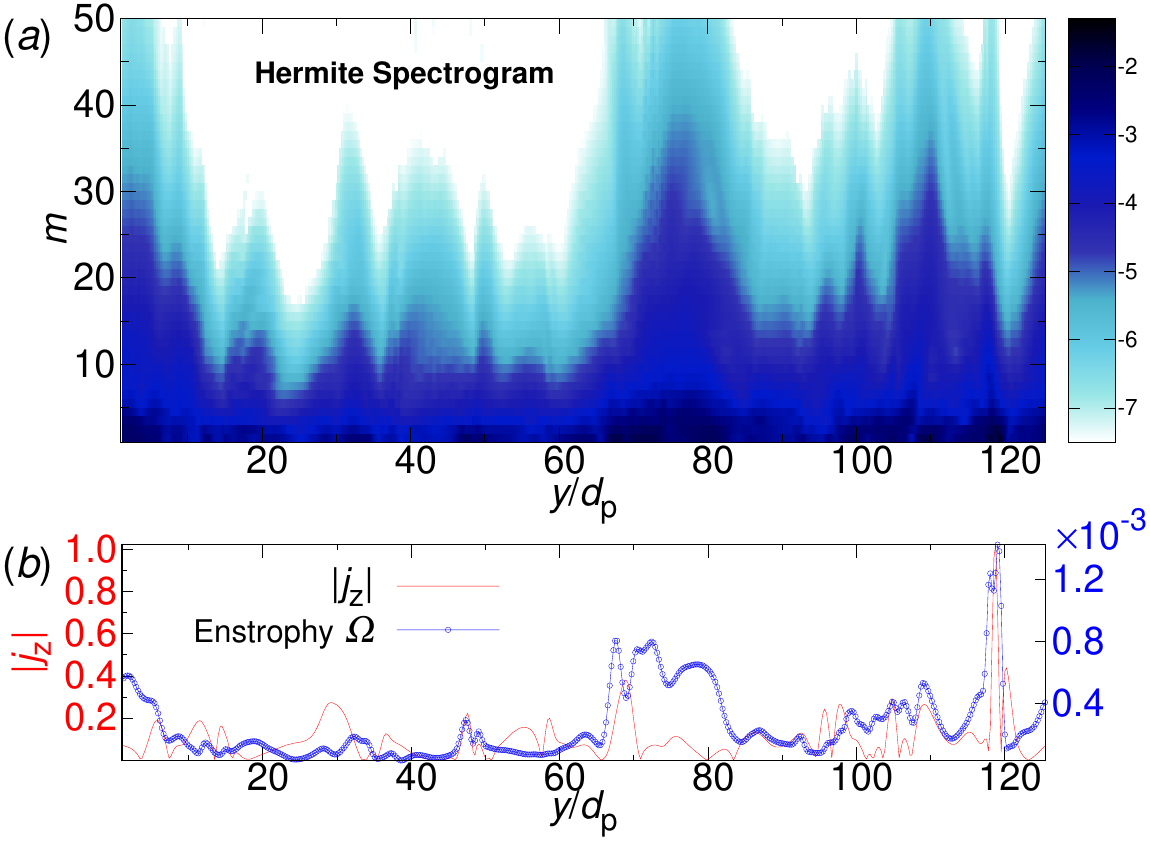}
\end{minipage}
\caption{(Color online) Magnetospheric Multi-Scale (MMS) in-situ observations (left columns) and Eulerian Hybrid Vlasov-Maxwell simulation (right columns) results, showing that the Hermite velocity-space cascade is intermittent and it is more developed close to intense current sheets. Left: Hermite spectrogram (top) and profile of the maximum PVI index for the magnetic field (i.e. magnetic field gradient, $~$ current density) bottom) as a function of time, from the MMS observations. Right: Hermite spectrogram (top) and profile of the current density $j_z$, together with the plasma enstrophy $\Omega$ (bottom), as a function of the spatial position in an one dimensional cut of the simulation (Adapted from \citet{pezzi2018velocityspace}). The Hermite spectrogram displays the evolution of the Hermite spectrum as a function of time and/or space.}
\label{fig:hermite}
\end{figure}

As suggested above, velocity-space is complexly structured in view of wave-particle interactions and turbulent cascade. Recently the Hermite decomposition of the proton distribution function has been adopted to highlight this complexity \citep{Grad49,tatsuno2009nonlinear,pezzi2016recurrence,SchekochihinEA16,servidio2017magnetospheric}. Indeed, the distribution function is decomposed in velocity-space by adopting a 3D Hermite transform, whose one-dimensional basis is:
\begin{equation}
\psi_m(v)=\frac{ H_m\!\!\left(\!\frac{v-u}{v_{th}}\!\right) }{\sqrt{2^m m! \sqrt{\pi} v_{th} }} e^{ - \frac{(v-u)^2}{2 v_{th}^2}}, 
\label{eq:psim}
\end{equation}
where $u$ and $v_{th}$ are now the local proton bulk and thermal speed, respectively, and $m\geq 0$ is an integer (we simplified the notation suppressing the spatial dependence). The above projection quantifies high-order corrections to the particle velocity DF. A highly distorted DF produces 
plasma enstrophy \citep{Knorr77}, defined as
\begin{equation}
\Omega({\bf x},t) \equiv \int_{-\infty}^{\infty} \delta f^2({\bf x}, {\bf v},t) d^3 v = \sum_{\bf m>0} \left[f_{\bf m}({\bf x},t)\right]^2, 
\label{eq:enst}
\end{equation}
where $\delta f$ indicates the difference from the ambient Maxwellian and $f_{\bf m}\equiv f_{\bf m}(m_x,m_y,m_z)$ is the Hermite coefficient, from which the enstrophy spectrum $P({\bf m})= \langle f_{\bf m}({\bf x},t)^2\rangle$ is defined. From this last quantity, reduced spectra (e.g. along perpendicular and/or parallel direction, as well as omnidirectional) can be evaluated. By taking advantage of highly accurate measurements from the Magnetospheric MultiScale mission (MMS), an enstrophy velocity-space cascade, revealed by evaluating the Hermite spectrum of the ion distribution function, has been observed in the Earth's magnetosheath \citep{servidio2017magnetospheric} and also in hybrid Vlasov-Maxwell simulations \citep{pezzi2018velocityspace, cerri2018dual}. A Kolmogorov-like phenomenology has been also proposed to interpret the observed slopes of the Hermite spectrum. Here we want to further stress that the presence of this phase-space cascade is spatially intermittent and it is more developed close to current sheets (Fig. \ref{fig:hermite}). This indicates that magnetic reconnection triggers non-thermal features in the distribution of particles, as also pointed out from {\it in-situ} observations and PIC simulations \citep{drake2006formation, shay2007two, hesse1999diffusion,pritchett2013influence, shay2014electron,drake2014structure,shay2016kinetic, pritchett2016three,lapenta2017origin, hesse2017population}. 

\begin{figure}
    \centering
    \includegraphics[width=0.5\textwidth]{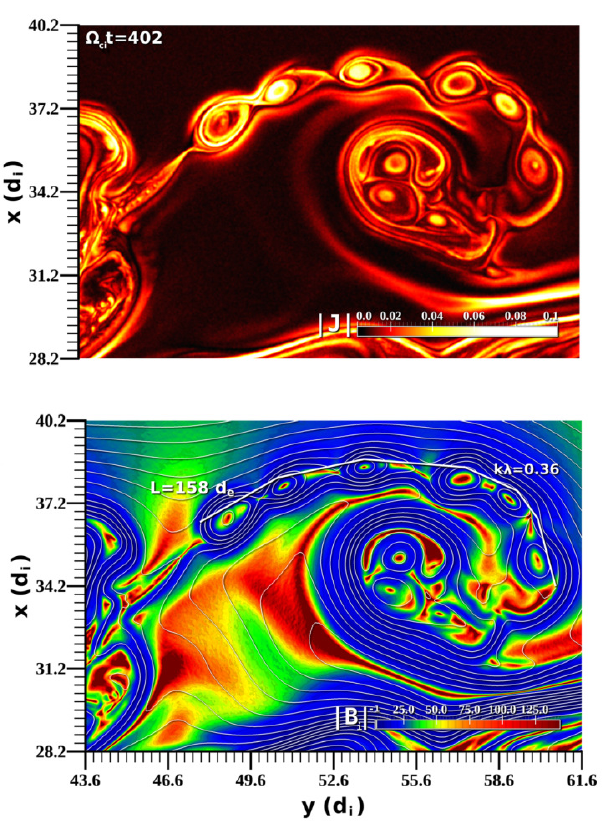}
    \caption{(Color online) 2D fully-kinetic PIC simulation results ($m_i/m_e=100$), showing the generation of secondary tearing islands at electron scales in a small sub-portion of the simulation box. Top: Contour plot of $J$ showing the formation of chains of tearing islands. Bottom: Contour plot of the in-plane magnetic field $B$, highlighting the fact that tearing modes are formed in regions where the in-plane magnetic field is weak. Contours of vector potential $a$ are also displayed. (reproduced from \citet{karimabadi2013coherent}).}
    \label{fig:karimabadi2013}
\end{figure}

Thanks to the unprecedented-resolution observations MMS \citep{burch2016magnetospheric, fuselier2016magnetospheric, torbert2016estimates, torbert2018electronscale}, electron scale reconnection has become finally accessible and have shown a complex picture where a nested set of diffusion regions, whose size range from ion to electron scales \citep{hesse2016theory}. Fully-kinetic models need to be adopted to characterize the process at such small scales. Figure \ref{fig:karimabadi2013} show the occurrence of several secondary islands, which are unstable for tearing instability. The size of these reconnection sites is comparable with electron scales.

At small scales, the dissipation of turbulent energy is thought to occur. Although it is well accepted that turbulence efficiently transfers energy from large to smaller scales \citep{verma1995,vasquez2007evaluation,sorriso2007observation,marino2008heating,hadid2018compressible,andres2019energy}, it is not clear which physical mechanisms --among a variety of proposed ones-- are controlling the dissipation process \citep{vaivads2016turbulence,matthaeus2020pathways}. Actually, several different definitions of the {\it word} dissipation have been recently proposed. One approach analyzes the dissipation associated with a peculiar phenomenon, either a kind of fluctuations (e.g. KAW and whistler) \citep{ChandranEA10, salem2012identification, gary2016whistler, vech2017nature, sorriso2018local, sorriso2019turbulence} or magnetic reconnection \citep{drake2010magnetic, Servidio2011magnetic, osman2011evidence, osman2012intermittency, osman2012kinetic, wu2013intermittent, shay2018turbulent}. 

Within the reconnection community, 
there has been a particular emphasis on 
examining 
the electromagnetic work on particles, i.e. ${\bm j} \cdot {\bm E}$ ($\bm j$ the electric current density, $\bm E$ the electric field) 
as a dissipation proxy \citep{sundkvist2007dissipation, zenitani2011new, wan2015intermittent}. 
This has been instrumental in identifying a 
large number of active reconnection sites in MMS data \citep{BurchEA16-grl,FuselierEA17,WilderEA18}.
However it has also been demonstrated in MMS analysis \citep{ChasapisEA18}
that some current sheets that are found to be 
reconnecting, with measures such as 
${\bm j} \cdot {\bm E}$ indicating conversion
of magnetic energy into particle 
energy, are actually {\it cooling} locally. 
This is not a paradox, since
fundamental Vlasov-Maxwell theory shows 
that electromagnetic work exchanges energy with microscopic {\it flows}, while the actual conversion 
into internal energy of each species $\alpha$
is accomplished by $\Pi^\alpha_{ij}\nabla_i u^\alpha_j$
the contraction of 
the pressure tensor with fluid 
velocity gradients of each species
\cite{yang2017energyPOP,yang2017energyPRE}. 
Further decomposed into a compressive part
(pressure-dilatation) and 
an incompressive part (pressure-strain, or ``Pi-D'') 
these quantities have recently been 
studied 
as direct channels of dissipation into heat
\citep{yang2017energyPOP,yang2017energyPRE,chasapis2018energy,pezzi2019energy,matthaeus2020pathways}.
It is important to note that 
within a collisional closure the pressure
strain interaction 
is approximated by a viscosity \citep{braginskii1965transport}
and becomes
irreversible; however in the Vlasov-Maxwell case, lacking collisions, these terms are sign-indefinite, just as are 
the electromagnetic work 
and the 
turbulence scale-to-scale transfer
\citep{matthaeus2020pathways}. 
Recently the statistical distribution of the Pi-D interaction
has been examined in 
in MMS data \citep{BandyopadhyayEA20}, and found to be 
consistent, in detail, with statistics from kinetic plasma simulation. 
It is encouraging that these highly fluctuating quantities in turbulent plasma 
give rise to statistical distributions 
that have reproducible properties.  

The {\it field-particle correlator} has been furthermore proposed to point out features of particular phenomena
associated with dissipation, e.g. Landau damping \citep{klein2016measuring, chen2019evidence, klein2020diagnosing}. Although in general their role is thought of being weak \citep{speiser1965particle, Kasper2008,maruca2011relative,chhiber2016solarwind}, collisional effects have been very recently also adopted to describe the ultimate dissipation, that produces an irreversible entropy growth \citep{tenbarge2013collisionless, pezzi2015collisional, navarro2016structure, pezzi2016collisional, pezzi2017solarwind, pezzi2019protonproton, vafin2019coulomb}. 

We conclude this section by highlighting that the support that numerical simulations can provide for interpreting and understanding {\it in-situ} observations is far from being exhausted. Indeed, magnetic reconnection is a multiscale phenomenon which goes ``hand-in-hand'' with turbulence evolution \citep{matthaeus2011needs}. Several issues need to be additionally addressed to clarify this puzzling picture. As introduced above, several simulations are performed in a reduced two-dimensional geometry, while the full three-dimensional case is much more complex and needs to be understood in detail. Another caveat concerns the unrealistic ion-to-electron mass ratio and the reduced box-size, that respectively affect the separation of different dynamical scales (i.e. the possibility to couple large and small scales) and the feasibility to achieve realistic Reynolds number. In this perspective, very recently, it has been proposed to couple different models, such as kinetic and global MHD \citep{Daldorff:etal:2014,drake2019computational}. Finally, based on the MHD treatment, magnetic reconnection is associated with the introduction of irreversibility in the system. However, in a weakly-collisional plasma, the transition to an irreversible dynamics is a long-standing challenge \citep{navarro2016structure,pezzi2016collisional,pezzi2019protonproton}. For a discussion of the relationship between collisional and collisionless cases, see 
\citep{matthaeus2020pathways}.

\section{Theory of particle acceleration in dynamical flux ropes/magnetic islands.}
\label{sect:unicalaccel}

Charged particle dynamics depends on the stochastic motion of magnetic field lines, whose random displacement affects the diffusion of particles both across and along the mean magnetic field \citep{JokipiiParker69}. In fact, particles gyrate along the magnetic field but, if the field is turbulent, they spread in the perpendicular direction, \quotes{jumping} from one field line to another \citep{Jokipii66,ChandranEA10,Ruffolo12}. The turbulent nature of the fields that scatter particles, makes the analytical treatment rather difficult. Because of this, we still lack an exact and universal theory to describe diffusion \citep{Bieber97, Hussein16,dundovic2020novel}.

Currently, among all diffusion theories, the nonlinear guiding centre (NLGC) \citep{Matthaeus03} and its variations
\cite{Ruffolo12,Shalchibook2009}
give a rather accurate prediction of the diffusion coefficient for systems with a three-dimensional (3D) geometry. However, a fully 3D numerical description of plasma turbulence requires huge computational efforts that can be streamlined by reducing the dimensionality of the problem. If a strong guiding magnetic field is present, turbulence becomes anisotropic and both 2D and 2.5D models become good approximations \citep{dobrowolny1980properties, dobrowolny1980fully, Shebalin83, matthaeus1986turbulent, OughtonEA94, Dmitruk2004test}. Due to this anisotropy, structures and phenomena related to turbulence, such as current sheets and reconnecting magnetic islands \citep{matthaeus1986turbulent,Greco09,Servidio2011magnetic}, magnetic field topology changes and energy conversion \citep{Parker57, Matthaeus1984particle, Ambrosiano88test, servidio2009magnetic, servidio2015kinetic}, mainly occur due to dynamics in the plane perpendicular to the main field \citep{bruno2016turbulence}. In \cite{Pecora18JPP}, a 2D version of 3D NLGC has been developed and tested with the 2.5D hybrid-PIC (Particle In Cell with kinetic ions and fluid electrons) simulation campaign described in \citep{Servidio16}. As turbulence develops, large-scale structures interact and smaller vortices and sharp current sheets appear, as shown in \fig{fig:servidio2010}. In particular, these magnetic structures represent magnetic islands (sections of three-dimensional flux tubes), and intense current sheets are associated to regions of strong magnetic gradients, in between reconnecting magnetic islands \citep{Matthaeus80,servidio2015kinetic}. Particles’ trajectories show that they can be either trapped in magnetic vortices 
\citep{Ambrosiano88test,RuffoloEA03,DmitrukEA04,TooprakaiEA07,SeripienlertEA10,TooprakaiEA16},
or scattered by magnetic discontinuities such as current sheets \citep{Rappazzo17}. 

While at short times particles can be trapped in topological structures, typically 
a stochastic transport 
regime is achieved after rather long time intervals and can be statistically described within the theory of diffusion, 
based for example, on the relation
$$\langle \Delta s^2 \rangle = 2 D \tau $$
where $\Delta \textbf{s} = \textbf{x}(t_0+\tau) - \textbf{x}(t_0)$ is the displacement of a particle, with $\textbf{x}$ being its position vector, $D$ the diffusion coefficient, and the average operation $\langle \dots \rangle$ is taken over all the particles \citep{Chandra43R,batchelor1976brownian,wang2012brownian}.
The diffusive description can be applied after 
initial transients, the simplest of which 
is referred to as the ballistic regime \citep{Servidio16}.
When applicable, particles must also escape from topological trapping in small flux tubes \cite{TooprakaiEA07} 
prior to achieving diffusive transport.
Generally speaking, 
transient regimes will last until particles sample 
uncorrelated correlated magnetic fields. 
As long as particles motion is followed for times shorter than the 
particle Lagrangian magnetic field correlation time, their motion cannot be found to be stochastic. This correlation time can be thought as the time one particle needs to move from one large 
scale magnetic vortex (island) to another. Low energy particles have longer correlation times as they cannot easily escape from vortices. Generally, the shape of trajectories suggests whether a particle has experienced trapping or scattering events, or both. Indeed, some particles have closed orbits that are a manifestation of trapping phenomena, while others show sharp turnovers that suddenly bend their trajectory when they experience strong local discontinuities \citep{drake2010magnetic, HaynesEA14}.

\begin{figure}
    \centering
    \includegraphics[width=\textwidth]{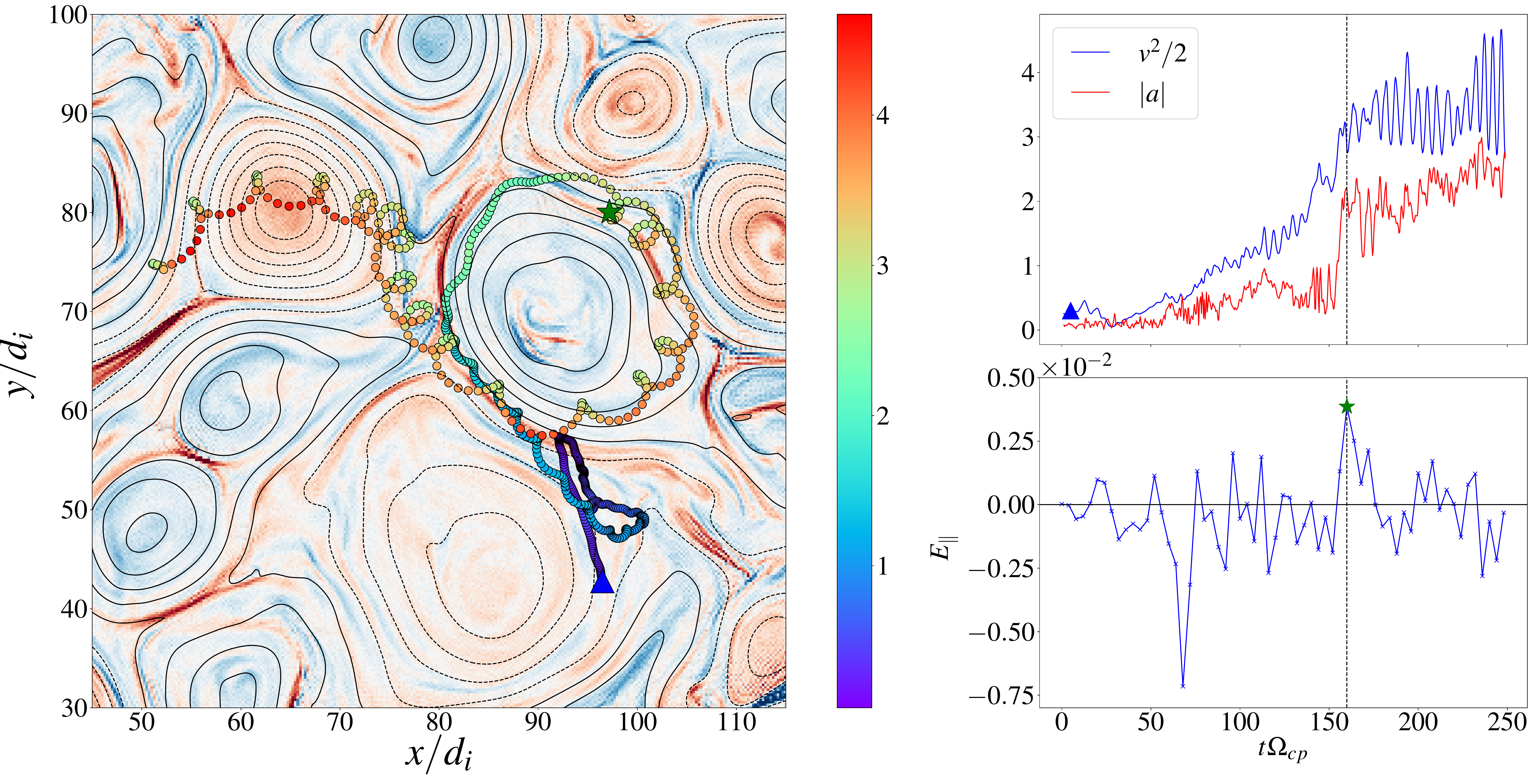}
    \caption{Snapshot of turbulence simulation when one particle is subject to a peak of the parallel electric field, located at the flank of the magnetic island in which it is trapped (adapted from \citet{Pecora18JPP}). At the same time, the particle experiences an increase in acceleration and kinetic energy and is able to escape from the island (the trajectory starts at the blue triangle and the colour code represents particle’s kinetic energy).}
    \label{fig:journey}
\end{figure}

When charged particles move in a turbulent plasma, local properties and modifications in the topology, such as island contractions and magnetic reconnection, are possible acceleration mechanisms as discussed in previous sections. Among these, one that can lead straight to particle acceleration is the electric field parallel to the local magnetic field \citep{Sturrock66,leRoux01,comisso2018}. Particles with acceleration values that exceed the variance of the distribution are found to be non-uniformly distributed in space. These particles tend to cluster where the parallel electric field is more intense, mostly on the flanks of magnetic islands. This supports the idea that accelerating particles are located close to regions where dynamical activity is occurring, notably along boundaries of interacting flux tubes, and near the associated current sheets, suggesting an association with magnetic reconnection. An example of this acceleration mechanism taking place is shown in \fig{fig:journey}.

Particles can also be accelerated by perpendicular electric fields
including betatron and curvature drift mechanisms that were discussed at some length in Section \ref{sect:leroux}. 
It should be noted that much of the elegant formal theory presented in those preceding sections was based on conservation of adiabatic invariants and correspondingly, on guiding center drift theory \cite{RossiOlbert}. While applicable 
in many situations, the basis of this set of approximations
is not always defensible; in particular 
adiabatic conservation laws such as conservation 
of magnetic moment are very sensitive to the presence of 
resonant power in the spectrum \citep{DalenaEA12-pre}.
In this regard a test particle study in a hierarchical reduced MHD model 
\cite{DalenaEA14} found that in early stages
of acceleration, parallel electric field is most important, but later, the most energetic particles are found in a so-called betatron distribution, with parallel streaming relatively 
suppressed. The affected particles are typically
trapped for extended periods of time 
in the acceleration region.
The interaction 
with the electric field is found to be in the perpendicular 
direction and resonant in the time domain as seen by the 
accelerated particles. The 
locus of this acceleration is 
near magnetic boundaries of flux tubes as they are 
compressed due to interaction 
with neighboring flux tubes.
There is a suggestion that the mechanism is limited by eventual pitch angle isotropization. 
As far as we are aware a transport 
theory describing this mechanism has not yet been developed. 

\begin{figure}
    \centering
    \includegraphics[width=.7\textwidth]{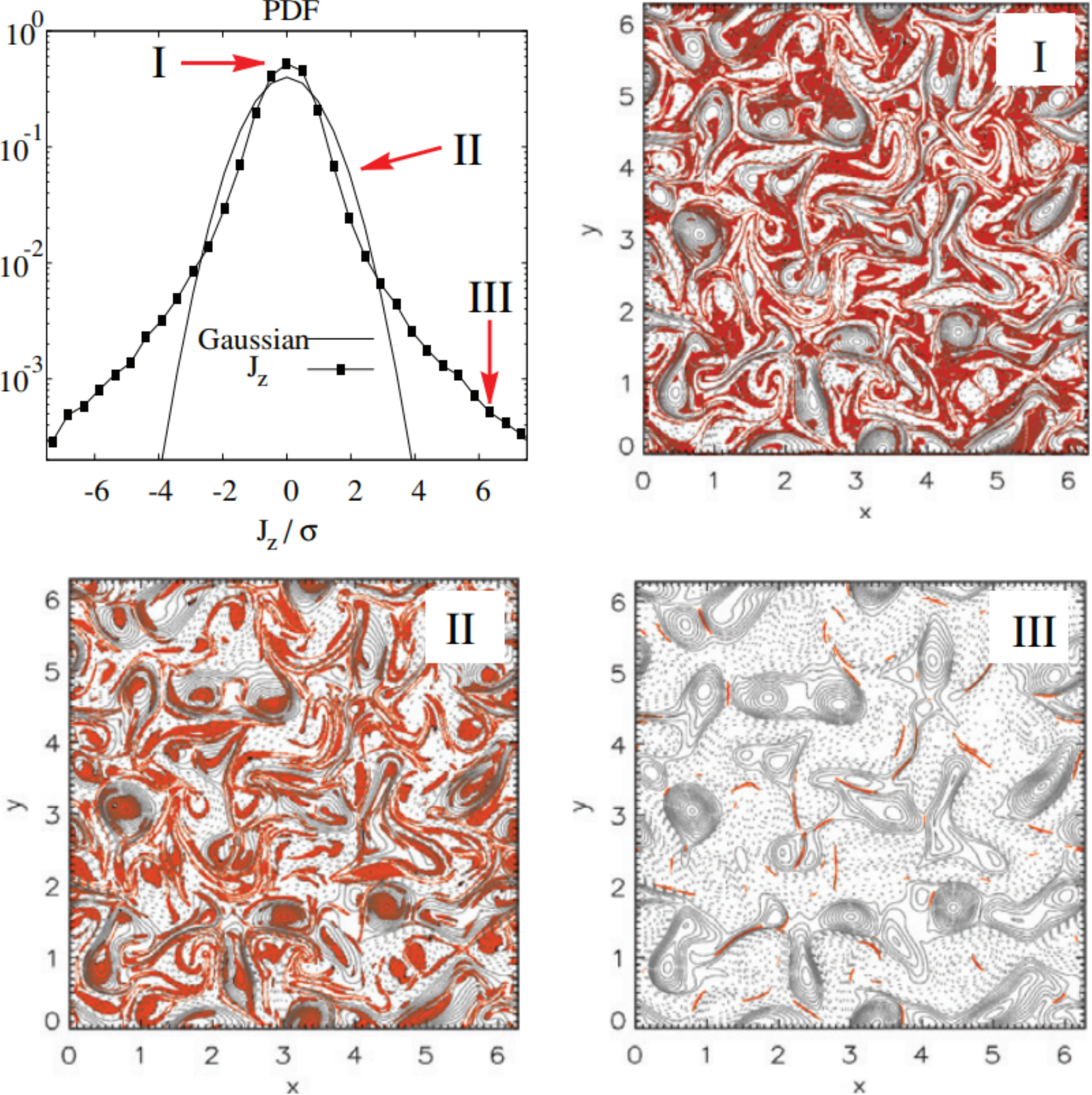}
    \caption{PDF of the out-of-plane electric current density $J_z$ from a 2D simulation, compared to a reference Gaussian (reproduced from \citet{Greco09}). For each region I, II, and III, magnetic field lines (contours of constant magnetic potential $A_z > 0$ solid, $A_z < 0$ dashed) are shown; the coloured (red) regions are places where the selected band (I, II, or III) contributes.}
    \label{fig:PVI}
\end{figure}

Coherent structures, such as current sheets, contribute to non-Gaussian statistics and therefore to intermittency which is a fundamental property of turbulence and can be associated with kinetic effects such as particle acceleration \citep{MarschandTu97, sorriso1999intermittency, osman2012intermittency, greco2012inhomogeneous, servidio2015kinetic, wan2015intermittent}. A well-established proxy to locate regions in which such structures appear is the Partial Variance of Increments (PVI) technique \citep{Greco09}. In turbulence simulations, it has been found that the regions of large parallel electric field occur in correspondence of magnetic discontinuities rather than in smooth regions, at the boundaries of the magnetic islands as previously shown in \fig{fig:PVI} of \citet{Greco09}.

Along with the parallel electric field, another feature has emerged to be extremely relevant in the process of acceleration and energization of particles. For energization to be effective, there must be a sort of spatial resonance between particle and diffusive characteristic scales. In particular, the gyroradius has to be comparable with Taylor dissipation length, $\lambda_T = \sqrt{\langle \delta b_\perp^2 \rangle / \langle j_z^2 \rangle}$, (qualitatively, the in-plane length of the largest current sheet).
In \citet{Pecora19SolPhys} this resonance is achieved in low-$\beta$ simulations and particles are more effectively energized with respect to those in large-$\beta$ plasmas. Indeed, changing plasma $\beta$ does not change turbulence scales such as correlation and Taylor lengths. but it modifies plasma's thermal energy and hence particles' gyroradii. These local spatial resonances, at low $\beta$, make the magnetic moment no longer a constant of the motion, leading to acceleration and eventually to energization, reminiscent of the nonadiabatic betatron mechanism described above 
\citep{DalenaEA14}. 

. On the other hand, in larger-$\beta$ plasma simulations, a consistent percentage of particles have a Larmor radius that is larger than the Taylor length, meaning that particles barely notice the current sheets during their gyrating motion and no resonance is possible, resulting in lower to none energization features.

Energization processes are ubiquitous in the whole universe and are not confined within the heliospheric description. High-energy particles, cosmic rays, black holes jets, all require a slightly different description that takes into account relativistic effects and strongly magnetized environments. As for non-relativistic plasmas, observations are tightly related to numerical simulations, especially when dealing with far unfathomable places. Magnetically dominated regions are a source of strong reconnection events and simulations show that acceleration of particles takes place as a two-stage process. First, the electric field acts as a primary source of acceleration in what can be considered an injection phase. Subsequently, many different processes, such as stochastic interaction with turbulent fluctuations, curvature drift and magnetic islands contraction, can take place and result in a Fermi-like acceleration \citep{guo2015particle, guo2016efficient, guo2016particle, huang2017development, ball2018electron, comisso2019interplay, trotta2020fast}.

There is a lot of observational evidence about the structured texture, previously depicted with simulations, of space plasmas at different spatial scales \citep{Schatten71,Bruno01,Borovsky08, khabarova2017energetic, malandraki2019current, verscharen2019multi}. As discussed also in Part I of this review, cross-sections of 3D structures such as elongated plasmoids, flux ropes and blobs of different origins, are 2D magnetic islands. The presence of these structures, on-average aligned with the Parker spiral, was suggested by \citep{Borovsky08}, following the spaghetti solar wind paradigm developed in 1970\textit{th} (see Section 2.2 of Part I). There is no clue if such structures originate close to the Sun and then advected in the interplanetary space, or generate locally in the HCS and other SCSs filling the solar wind, due to magnetic reconnection and numerous instabilities (e.g., \cite{khabarova2015small, khabarova2016small, khabarova2017energetic, malandraki2019current, khabarova2020counterstreaming}). Particularly relevant is the contribution given by the MMS mission \citep{curtis1999magnetospheric}, whose measurements help unravelling the fine-scale structure of the near-Earth interplanetary space. Lots of effort have been dedicated into correlating spacecraft measurements to topological properties of the magnetic field and kinetic properties of plasma, showing also that reconnection sites can generate secondary structures such as reconnection-generated flux ropes within their exhausts \citep{lapenta2015secondary, burch2016electron, burch2016magnetic, phan2016mms, eastwood2016ion, lapenta2018nonlinear, stawarz2018intense, stawarz2019properties}.

\begin{figure}
    \centering
    \includegraphics[width=\textwidth]{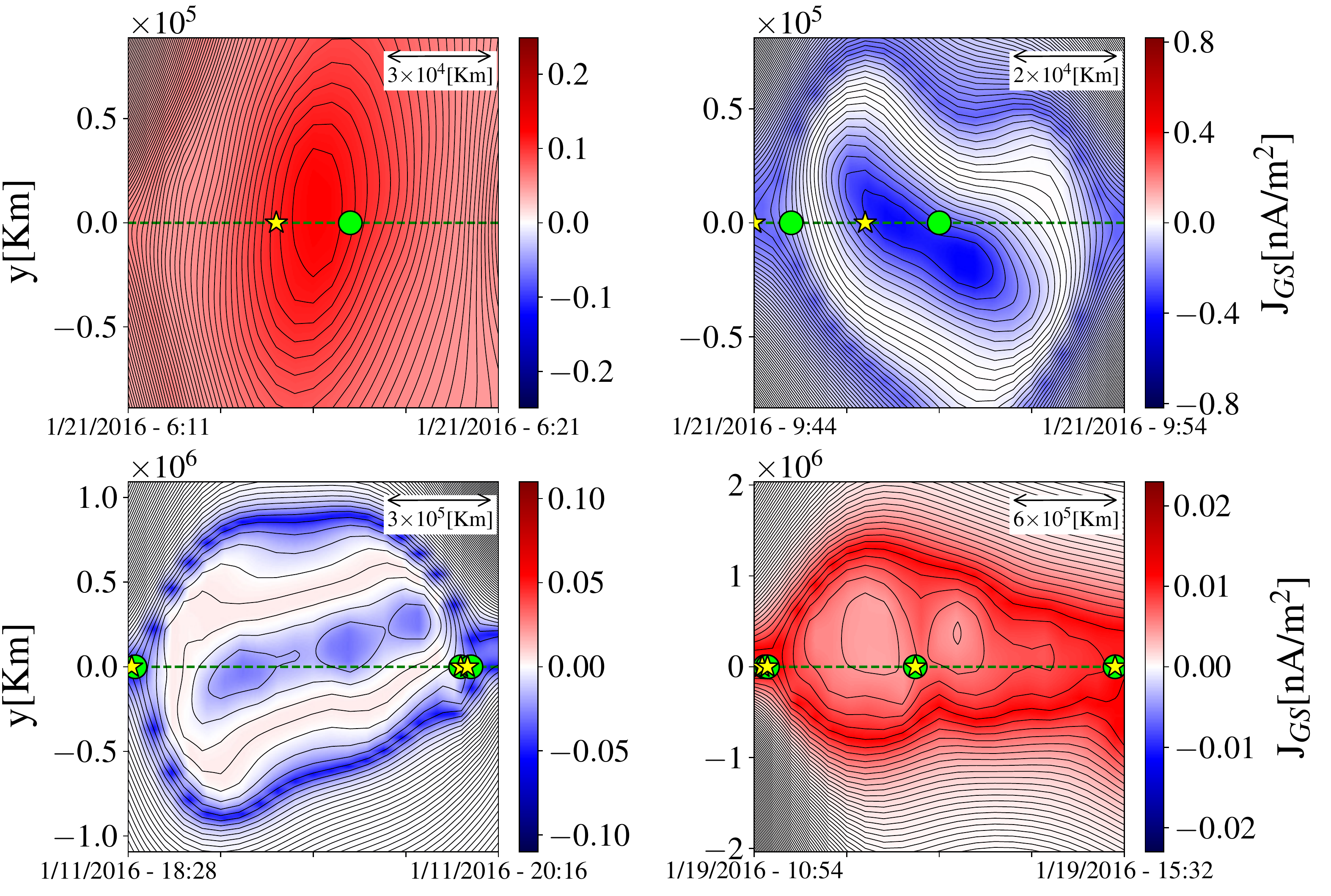}
    \caption{Reconstructed flux ropes, adapted from \citet{Pecora19GS}. The colour represents the out-of-plane current density, the black solid lines are the isocontour of the magnetic potential and the boundaries of a PVI event are indicated with a star (beginning) and a circle (end).}
    \label{fig:FR}
\end{figure}

Of course, it is not possible to obtain a clear picture, like that provided by simulations, of the surroundings of a spacecraft, but there are powerful tools that can be used to get 2D and 3D information when specific conditions are met. A three-dimensional perspective of magnetic field lines can be recovered using the First Order Taylor Expansion (FOTE) method \citep{fu2015how} that can be applied in the vicinity of a null point. This method requires multispacecraft measurements and the null point to be enclosed within the spacecrafts configuration volume but, despite the restricted regions of applicability, it can provide important information such as the identification of a 3D reconnection site where kinetic effects can be studied locally and linked to topological properties \citep{fu2016identifying, wang2019electron}. On the other hand, the Grad-Shafranov (GS) reconstruction method \citep{hu2002reconstruction} is able to reveal the 2D magnetic field texture around one single spacecraft when it passes through an MHD stationary structure with cylindrical symmetry (e.g. a flux rope, whose 2D section in the plane perpendicular to the symmetry axis is a magnetic island). This technique has been also adopted to study the effect of magnetic clouds on galactic cosmic ray intensity \citep{benella2019grad}. Very recently, \citet{Pecora19GS} enhanced the GS method by synergising it with the PVI technique providing additional observational evidence of the complex structure of the solar wind. In fact, the large-scale texture of the solar wind, reconstructed with the GS method at about $10^5 - 10^6$ km, shows flux tubes that are filamentary or \quotes{spaghetti-like}. The additional information provided by the PVI technique reveals where strong small-scale gradients of the magnetic fields are located. 
These regions are found at the boundaries of flux ropes/plasmoids where, possibly, reconnection, heating and particle acceleration are taking place. A few examples are shown in \fig{fig:FR}.
These structured flux tubes, hence, provide conduits for energetic particle transport and possible trapping and acceleration (see \citet{Tessein13,khabarova2016small,Pecora18JPP} and Section \ref{sect:leroux}).

The possibility to visualise 2D and 3D maps of the magnetic field and locate reconnection sites is extremely valuable when it comes to relating particle properties with the surrounding environment, as simulations already allow to do (\fig{fig:journey}). Eventually, one may envision numerous applications in which the structures revealed by the combined GS/PVI method can be relevant to understand complex physics and turbulent interplanetary dynamics.

\section{Summary and conclusions}
\label{sect:concl}

In this review we have depicted the dynamics of magnetic and plasma structures, such as CSs, FRs/MIs and plasmoids/blobs, and their impact on the surrounding medium, especially related to particle energization. The complexity involved within such structures makes the adoption of the fully 3D approach essential since several important pieces of information are inevitably lost by adopting models with explicit symmetries (planar and/or spherical). Observations and numerical simulations in the last decades support these ideas (see, e.g., \citet{lapenta2015secondary,lapenta2018nonlinear,adhikari2019role,khabarova2020counterstreaming,lazarian20203D}). At the same time, the 2D approach can be employed in some cases. For instance, because of the spectral anisotropy, this is generally applicable for studying turbulence. Indeed, 2D/2.5D models of plasma turbulence are able to describe most of the features of both turbulence dissipation and magnetic reconnection, providing a valid tool for investigation of the related processes and understanding of particle heating and energization in astrophysical plasmas (e.g., \citet{matthaeus2015intermittency}).

Modern theories and observational studies describe the HCS and similarly strong CSs as essentially non-planar complex plasma structures surrounded by a plasma sheet in which numerous small-scale reconnecting CSs separated by plasmoids occur \citep{khabarova2015small,malova2018structure,adhikari2019role,mingalev2019modeling}. CSs are unstable in natural plasmas. Owing to the constantly changing environment, CSs are subject to different instabilities, including the tearing instability \citep{zelenyi1998multiscale, zelenyi2004nonlinear, tenerani2015tearing}. These instabilities may impact CSs simultaneously with various fluctuations, destabilizing, triggering nonlinear processes at CSs and even destroying them. Therefore, strong quasi-stable CSs, such as the HCS and CSs at leading edges of ICMEs and CIRs/SIRs, represent a well-known source of turbulence and intermittency. On the other hand, numerous thin and unstable CSs are generated in turbulent and intermittent regions (e.g, \citet{servidio2009magnetic,matthaeus2015intermittency}). This dualism reflects an intrinsic tie between instabilities, wave processes,  magnetic reconnection, turbulence, and intermittency in space plasmas.     

Summarizing the results of studies of particle acceleration associated with magnetic reconnection, we would like to stress out the fact that, besides the obvious role of the reconnection-induced electric field, charged particles can be energized by the first- and second- order Fermi mechanisms and the so-called anti-reconnection electric field operating during contraction and merging of FRs/plasmoids/MIs \citep{Zank2014particle,LeRoux2015kinetic,leRoux2016combining,xia2018particle,xia2020particle,leroux2019modeling}. Numerical simulations of processes occurring in turbulent plasmas successfully describe energization of charged particles by several acceleration mechanisms acting simultaneously. An analysis of observations confirms that theoretical predictions and results of numerical simulations discussed above correspond to reality and show that the ubiquitous occurrence of magnetic reconnection in the heliosphere makes this phenomenon essential for  local particle acceleration \citep{khabarova2015small,khabarova2016small,khabarova2020counterstreaming}. Comprehensive studies of the formation, evolution and dynamics of CSs, FRs/plasmoids, MIs in the heliosphere are necessary for better understanding of the transport of energetic particles in the heliosphere, including galactic cosmic rays \citep{E19}.

CSs, FRs/MIs, and plasmoids/blobs of various origins and scales exist in the turbulent solar wind. Therefore, we provide an analysis of the main properties of solar wind turbulence, from large (MHD) to smaller (kinetic) scales for convenience of the readers interested in numerical simulations. Here, we particularly focus on the most recent advances in kinetic numerical simulations (e.g., \citet{servidio2015kinetic}). Owing to the weak collisionality, plasma turbulence naturally evolves in the fully 6D phase-space and excites a variety of genuinely kinetic effects in the particle VDF \citep{valentini2016differential, servidio2017magnetospheric, pezzi2018velocityspace}, ranging from beams along the background magnetic field to a highly-structured velocity-space, as recently found via the Hermite decomposition of the particle VDF. 

To conclude, one may offer numerous applications in which the structures revealed by the combined GS/PVI method can be relevant in understanding the complex physics and turbulent dynamics of both the interplanetary magnetic field and the solar wind plasma. It should be emphasized that most of contemporaneous theoretical works describing solar-wind properties have been made in a frame of 2D modeling and reconstructions. Very often, the available methods do not allow complete understanding the real shape of 3D plasma structures, as discussed in Part I of this review. Talking about 2D modeling, one should take into account a certain degree of uncertainty in the interpretation of the results, which is similar to problems of the interpretation of observations. In particular, the shapes and corresponding dynamics of magneto-plasma structures in the solar wind may only be suggested since one considers a cross-section of real 3D structures which may be both FRs and/or finite shaped plasmoids/blobs. That is why we talk about 2D magnetic islands above. We admit that the lack of 3D models due to an obvious complexity of their building as well as the insufficiency of multi-spacecraft data to restore 3D structures are disadvantages of the modern theoretical and observational approaches. Meanwhile, we can conclude that employing contemporary 2D models and corresponding simplified methods widely used in space science allows general understanding of the numerous complex processes if one initially considers various possible 3D topologies occurring in the real space plasmas. The transition from the 2D to 3D models to describe complex 3D processes and non-planar structures in the solar wind is the next step of the development of heliospheric physics.

\begin{acknowledgements}
This work is supported by the International Space Science Institute (ISSI) in the framework of International Team 405 entitled ``Current Sheets, Turbulence, Structures and Particle Acceleration in the Heliosphere.''. 
O.P. thanks Dr. D. Trotta and Dr. F. Catapano for friendly and precious conversations on some of the topics discussed in this review. \\ \\
\textbf{Funding} R.K., H.M. and O.K. are partially supported by Russian Foundation for Basic Research (RFBR) grant 19-02-00957. H.M. acknowledges the partial support of Volkswagen Foundation grant Az90 312. J. A. le Roux acknowledges support from NASA Grants NNX15AI65G, 80NSSC19K027, NSF-DOE grant PHY-1707247, and NSF-EPSCoR RII-Track-1 Cooperative Agreement OIA-1655280. S.S. acknowledges the European Union’s Horizon 2020 research and innovation programme under Grant Agreement No. 776262 (AIDA,www.aida-space.eu). \\ \\
\textbf{Conflicts of interest/Competing interests} The authors declare no conflict of interests. \\ \\
\textbf{Availability of data and material (data transparency)} All data and material used are from public open-access data depositories and archives (see Acknowledgements for details).  \\ \\
\textbf{Code availability} Not applicable

\end{acknowledgements}

\bibliographystyle{spbasic}  

\end{document}